\documentclass{aa}  
\usepackage[multidot]{grffile}
\usepackage{graphicx}
\usepackage{txfonts}
\usepackage{lscape}
\usepackage{natbib}
\usepackage[colorlinks,allcolors=blue,bookmarks=false]{hyperref} 
\pdfoutput=1
\usepackage{lscape}

\begin{document} 

   \title{Properties of the dense cores and filamentary\\ structures in the Vela C molecular cloud \thanks{Table~\ref{coreobs},~\ref{corederive}, and ~\ref{filamentpara} are only available in electronic form
at the CDS via anonymous ftp to cdsarc.cds.unistra.fr (130.79.128.5)
or via https://cdsarc.cds.unistra.fr/cgi-bin/qcat?J/A+A/}}

   \author{Xue-Mei Li \inst{1,2}
          \and
          Guo-Yin Zhang \inst{2}
          \and 
          Alexander Men'shchikov\inst{3}
          \and
          Jin-Zeng Li \inst{2} 
          \and
          Chang Zhang \inst{2}
          \and
          Zhong-Zu Wu \inst{1}
          }

   \institute{College of Physics, Guizhou University, Guiyang 550025, PR China
         \and
         National Astronomical Observatories, Chinese Academy of Sciences, Beijing 100101, PR China\\
             \email{zgyin@nao.cas.cn; ljz@nao.cas.cn}
        \and 
        Universit{\'e} Paris-Saclay, Universit{\'e} Paris Cit{\'e}, CEA, CNRS, AIM, F-91191, Gif-sur-Yvette, France\\
            \email{alexander.menshchikov@cea.fr}
        }

\date{Received / Accepted }

\titlerunning{Properties of cores and filaments in the Vela C molecular cloud}
 
  \abstract{
   The initial and boundary conditions of the Galactic star formation in molecular clouds are not well understood.
   In an effort to shed new light on this long-standing problem, we measured properties of dense cores and
   filamentary structures in the Vela C molecular cloud, observed with $\it Herschel$.
   We used the \textit{hires} algorithm to create high-resolution surface densities (11.7{\arcsec}) from the $\it Herschel$
   multiwavelength dust continuum. We applied the {\it getsf} extraction method to separate the components of sources and
   filaments from each other and their backgrounds, before detecting, measuring, and cataloging the structures. The cores and filamentary structures constitute $40${\%} of the total mass of Vela C, most of the material is in the low-density molecular background cloud. We selected 570 reliable cores, of which 149 
   are the protostellar cores and 421 are the starless cores. Almost 78{\%} 
   (329 of 421) of the starless cores were identified with the gravitationally bound prestellar 
   cores. The exponent of the CMF ($\alpha=1.35$) is identical to that of the Salpeter IMF. 
   The high-resolution surface density image helped us determine and subtract backgrounds and measure sizes of the structures. We
   selected 68 filaments with at least one side that appeared not blended with adjacent structures. The filament widths are in the
   range of 0.15 pc to 0.63 pc, and have a median value of $W = 0.3\pm 0.11$ pc. The surface densities of filaments are well correlated
   with their contrasts and linear densities. Within uncertainties of the filament
   instability criterion, many filaments (39) may well be both supercritical and subcritical. A large fraction of filaments (29)
   may definitely be considered supercritical, in which are found 94 prestellar cores, 83 protostellar cores, and only 1 unbound 
   starless core. Taking into account the uncertainties, the supercritical filaments contain only prestellar and protostellar 
   cores. Our findings support the idea that there exists a direct relationship between the CMF and IMF and that filaments play a key role
   in the formation of prestellar cores, which is consistent with the previous \emph{Herschel} results.}
   
 \maketitle

\section{Introduction} 

Star formation takes place within molecular clouds \citep{Andre2014,Heyer2015} that appear as blends of (round) sources,
filamentary structures, large-scale background, small-scale fluctuations, etc. Separation of the structural components,
measurement, and analysis of the sources and filaments are thought to improve our understanding of the initial conditions of star
formation \citep[cf. ][]{Bergin2007,Andre2010}.

Photometric surveys of the nearby ($\lesssim $ 500 pc) molecular clouds and Galactic plane with \emph{Herschel}, covering thermal
far-IR and sub-millimeter continuum in five bands between 70 and 500 $\mu $m, probe the peak of the dust spectral energy
distribution (SED), allowing determination of the temperature and mass of the clouds
\citep{Andre2010,Molinari2010,2022RAA....22e5012Z}. Multiband images from the \emph{Herschel} telescope probe molecular clouds in
a wide range of spatial scales. The orbital telescope has revolutionized our understanding of the link between the structure of the
cold interstellar medium (ISM) and the process of star formation. In particular, it revealed filamentary structures with a wide
range of sizes, densities, and morphologies that are indeed widespread in the molecular clouds \citep{Men2010,Li2012,Schisano2020}.
Most dense cores are located on the supercritical filaments, which suggests that the filaments can fragment into the cores which
eventually collapse into stars \citep{Konyves2015}. The detailed fragmentation manner may be controlled by the linear density,
geometrical bending, continued accretion of gas, and magnetic fields in the filaments \citep{Zhanggy2020}. The mass distribution of
stars at their birth is known as the stellar initial mass function (IMF) \citep{Salpeter1955,Chabrier2005}. Since stars are born in
the dense cores, studies of the core mass function (CMF) might help to understand the IMF \citep{Andre2000,Ward2007}. A lot of
observational evidence supports the idea that CMFs are similar with the IMF in shape \citep[e.g.
][]{Motte1998,Alves2007,Konyves2015}. A mass conversion efficiency (from the cores to the stars) of approximately $1/3$ was
proposed by \citet{Alves2007}.

The Vela molecular ridge (Galactic longitudes $l \approx 260$ to $275^{\circ}$ and latitudes $b \approx \pm 3^{\circ}$), was
initially identified from low-resolution ($5^{\circ}$) CO emissions in the direction of Vela, and roughly outlined as four
molecular clouds named as Vela A, B, C and D \citep{May1988,Murphy1991}. Vela A, C, and D have a kinematic distance of 0.7$\sim$1
kpc, whereas Vela B has a kinematic distance of 2 kpc \citep{Murphy1991}. Vela C, the most massive molecular cloud among Vela A, C,
and D, spans roughly 3.5 degrees (south to north), and there exist nest structures dispersed with a large amount of low-density
gas, in which low-mass stars form, and tight linear ridges in which massive or intermediate-mass stars form
\citep{Liseau1992,Lorenzetti1993,Yamaguchi1999,Massi2003,Baba2006,Netterfield2009,Hill2011}. In previous studies
\citep{Molinari2011,Giannini2012,Minier2013,Massi2019}, a distance of $\sim 700$ pc to Vela C was adopted \citep{Murphy1991}. The
mean distance of Vela C, estimated by \cite{Zucker2020} from Gaia DR2 \citep{Gaia2018} stellar parallaxes, is $905\pm45$ pc. The
different distance measurements bring deviations of about 20\% in the size and 60\% in the mass of Vela C. Considering that GAIA
has measured distances with unprecedented precision, we adopted the distance of 905 pc in this work.

A decade ago, \cite{Giannini2012} detected dense cores in \emph{Herschel} multiband images of Vela C with the CuTeX source
extraction algorithm \citep{Molinari2011} and \cite{Hill2011} traced skeletons of filaments with the DisPerSE algorithm
\citep{Sousbie2011}.
In this work, we chose to apply the multiscale source and filament extraction method \emph{getsf}
\citep{Men2021method} that separates the structural components of sources, filamentary structures, and their backgrounds before
applying a consistent approach to extracting both sources and filaments. The method has no free (unconstrained) parameters, which
is a property that sets \textsl{getsf} apart from other extraction methods. Its single user-definable parameter, the maximum size
of the sources or filaments of interest, is constrained from the observed images and used just to place a reasonable upper limit on
the range of scales in a spatial decomposition of the observed images. Interested readers are referred to \cite{Men2021method} for
a full description of \textsl{getsf} and to \cite{Men2021benchmark} for its benchmarking that showed improved detection
completeness and measurement accuracy with respect to the older \emph{getsources} and \emph{getfilaments} extraction methods
\citep{Men2012getsources,Men2013getfilaments}.

The outline of the present paper is as follows. In Sect. \ref{sec:observations}, we describe the {\it Herschel} sub-millimeter dust
emission data. The data analysis and results are presented in Sect. \ref{sec:analysis}. In Sect.~\ref{sec:discussion}, we discuss
the extracted sources at different resolutions, their core mass function, and the filamentary structures. We summarize
our conclusions in Sect.~\ref{sec:conclusions}.

\section{The \emph{Herschel} images}\label{sec:observations}

The Vela C molecular cloud was observed on May 18, 2010 within the frame of the HOBYS imaging survey (PI: F. Motte, observation
IDs: 1342196657, 1342196658) \citep{Motte2010}. Using the PACS/SPIRE parallel mode \citep{Poglitsch2010,Griffin2010} with a
scanning speed of 20{\arcsec}\,s$^{-1}$, an area of $\sim 1.4\times3.2$ deg$^{2}$ was imaged simultaneously at 70, 160, 250, 350,
and 500 $\mu$m, with the angular resolutions of 5.9, 11.7, 18.2, 24.9, and 36.3{\arcsec}, respectively. We retrieved the images
from the {\it Herschel} Science Archive (HSA) \footnote{http://archives.esac.esa.int/hsa/whsa/}. The SPIRE extended-source
calibrated map and PACS Unimap \citep{Piazzo2015} products at the level 2.5 were used, because the level 3 images did not undergo
quality control at the \emph{Herschel} Science Centre.

    \begin{figure}
       \centering
       \includegraphics[width=0.45 \textwidth]{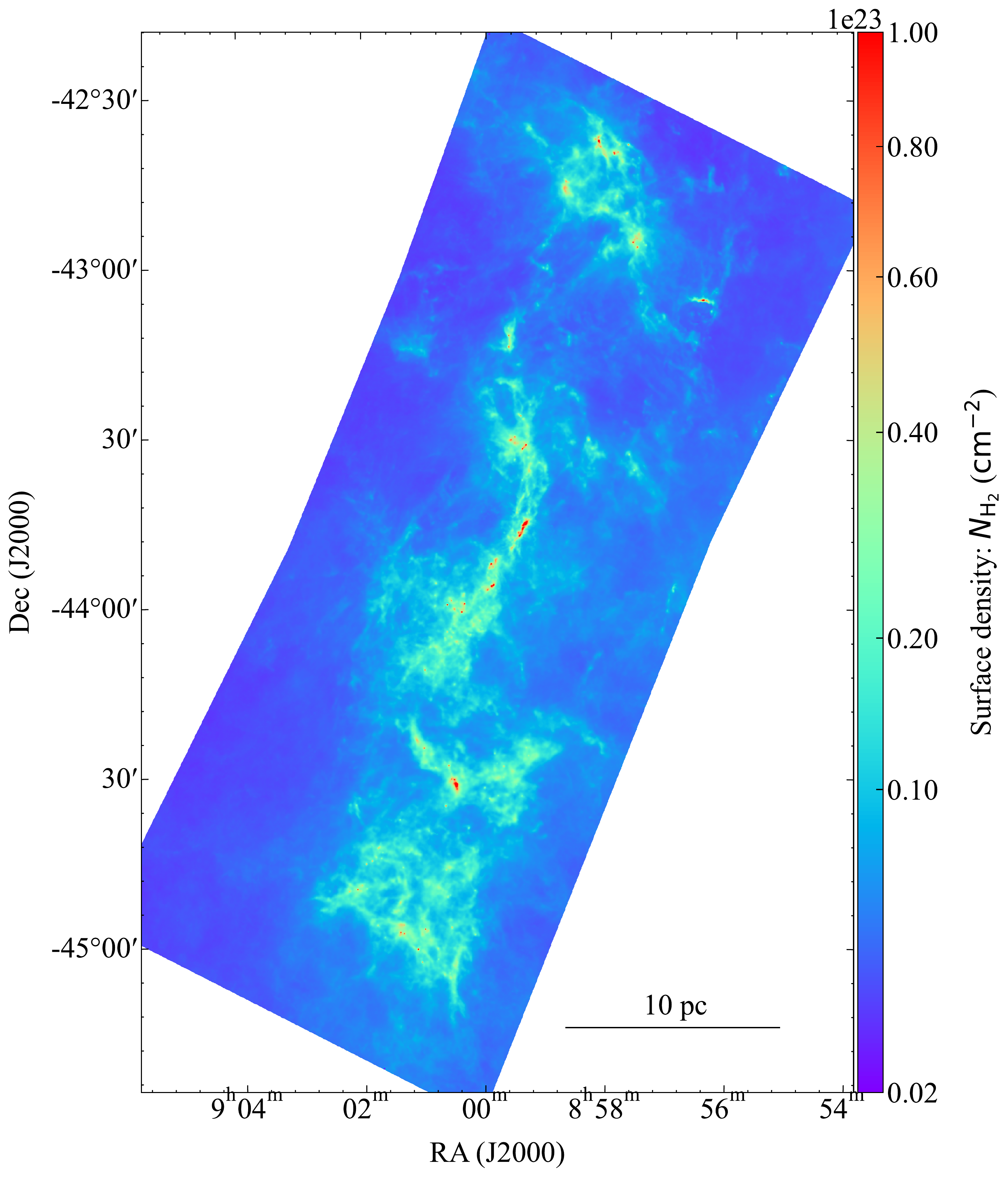}
       \caption{A high-resolution (11.7{\arcsec}) surface density map of Vela C, obtained from the \emph{Herschel} images.}
       \label{nh2map}
    \end{figure}

    \begin{figure*}
       \centering
       \includegraphics[width=0.85 \textwidth]{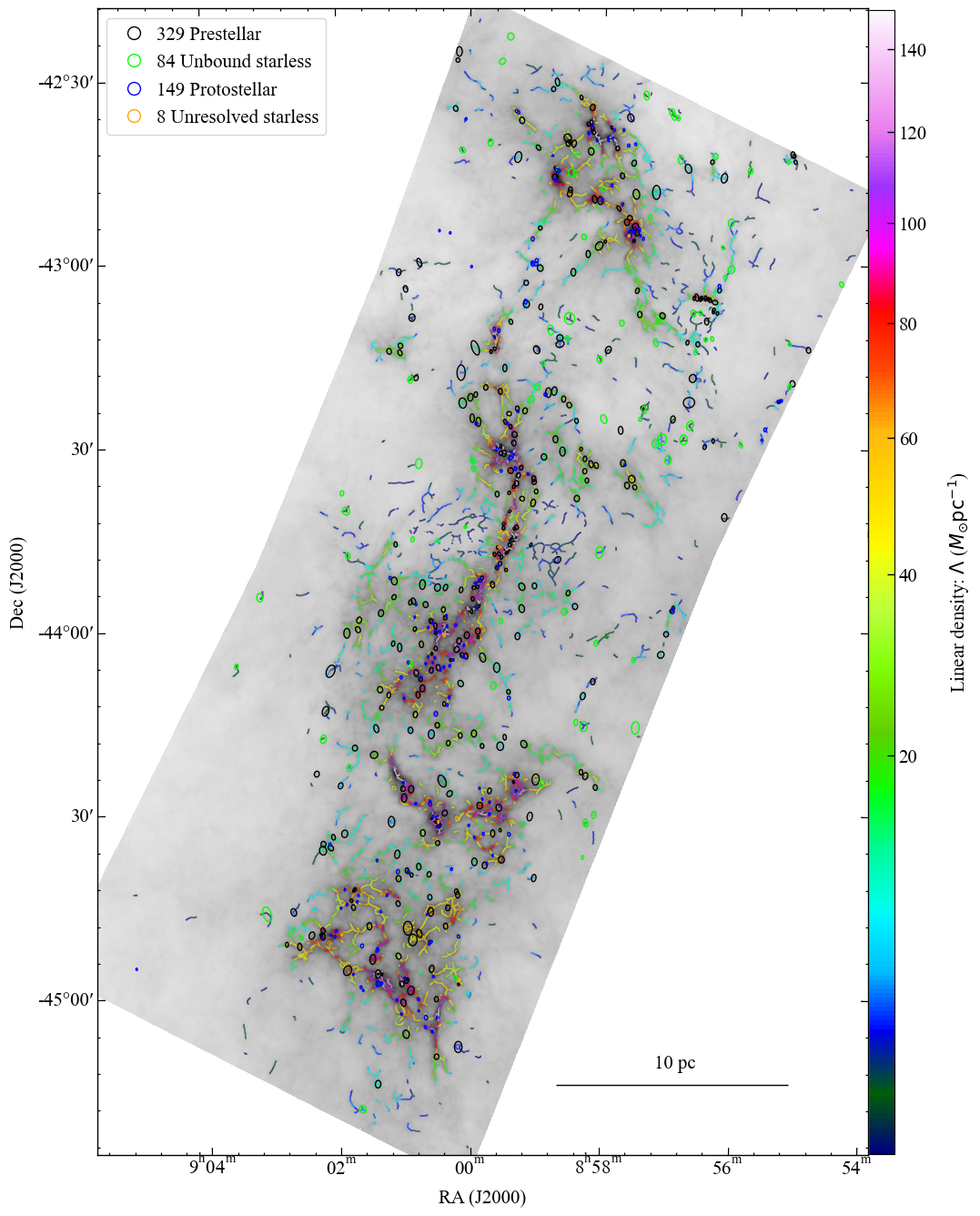}
       \caption{Extracted cores and filaments in the Vela C molecular cloud on the 11.7{\arcsec}-resolution surface density image. 
       The locations and sizes of the cores are measured from the surface density map. The black, green, and blue ellipses mark the 
       329 prestellar cores, 84 unbound starless cores, and 149 
       protostellar cores, respectively. The full axis of the ellipses is 2 times the Gaussian half-maximum size (FWHM). 
       Filamentary networks are shown by the global, scale-independent skeletons of the filamentary structure. The filament linear 
       density $\Lambda^{\rm W}$ in each point along the skeletons was estimated from Eq.~(\ref{mlineP}) assuming 
       a width of $W = 0.3$ pc (Fig.~\ref{filamentshist}).}
       \label{coresinmap}
    \end{figure*}

\section{Data analysis and results}\label{sec:analysis}

\subsection{The high-resolution surface density image}

To properly compute the surface densities, it was necessary to derive zero-level offsets for the five \emph{Herschel} images. 
We assume that the images probe the optically thin dust emission following the empirical fit (e.g., \citealt{Planck2014}),
\begin{equation}
I_\nu = \tau_{\nu_0} \, B_\nu(T) \, \left( \frac{\nu}{\nu_0} \right)^{\beta},
\label{emissionlaw}
\end{equation}
where $\nu_0$ is the reference frequency, at which the optical depth $\tau_{\nu_0}$ is estimated, $I_{\nu}$ is the specific
intensity, and $B_{\nu}(T)$ is the Planck function for the emission of dust at temperature $T$ and frequency $v$. Using a modified
blackbody fit to the \emph{Planck} 353, 545, and 857 GHz images, and IRAS 100 $\mu$m data, \cite{Planck2014} presented all-sky maps
of the dust optical depth at $\nu_0 = 353$\,GHz (850 $\mu$m) and the dust temperature map with a resolution of 5{\arcmin}. Using
$\beta = 2$, adopting $\nu_0 = 353$\,GHz as the reference frequency, and substituting the optical depth at 353 GHz in
Eq.~(\ref{emissionlaw}), we used the \emph{Planck} dust temperature map to derive the images at 70, 160, 250, 350 and 500 $\mu$m of
Vela C, corresponding to the \emph{Herschel} images. Convolving each \emph{Herschel} image to the \emph{Planck} resolution, we
obtained the offsets of each \emph{Herschel} image by a comparison with the derived intensity maps \citep{Bernard2010}. They were
found to be 23, 320, 146, 59.8, and 18.7 $\mathrm{MJy}\,\mathrm{sr^{-1}}$ at 70, 160, 250, 350 and 500 $\mu$m, respectively.

We used \emph{prepobs} from \emph{getsf} to reproject the \emph{Herschel} images to the same size and number of pixels (3{\arcsec},
$2820 \times 4050$) and the \emph{hires} method \citep{Men2021method} to compute the temperature and surface density images at the
\emph{Herschel} resolutions. An advantage of the \emph{hires} algorithm over other similar methods is that it can
use the shortest-wavelength highest-resolution images, assuming that their emission is dominated by the normal dust grains in
radiative equilibrium. Such high-resolution surface density images are helpful in our detailed study of the complex structural
diversity in Vela C. The pixel-to-pixel SED fitting was performed by the \emph{fitfluxes} utility \citep{Men2016fitfluxes}. The
images at 160, 250, 350, and 500 $\mu$m were used to derive the temperature and surface density maps at 36.3{\arcsec} resolution,
the images at 160, 250, and 350 $\mu$m were used to obtain those maps at 24.9{\arcsec} resolution, and the images at 160 and 250
$\mu$m were used to compute the maps at 18.2{\arcsec} resolution. These three sets of temperatures and observed images (from 70 to
500 $\mu$m) were used to compute the surface density and temperature images (Fig.~\ref{nh2map}) with the improved resolutions of
5.9, 11.7, and 18.2{\arcsec} \citep{Men2021method}. To reduce the noisiness of the 5.9{\arcsec} surface densities,
induced by the \emph{Herschel} 70 $\mu$m image, we smoothed the map to a 8.5{\arcsec} resolution and used the latter in our work.

\subsection{Source and filament extraction}

We applied the multiscale, multiwavelength extraction method \emph{getsf} \citep{Men2021method} to extract sources and filaments
from the \emph{Herschel} multiwavelength dust continuum images. The only \emph{constrained} parameter that it
requires is the maximum sizes of the structures to be extracted, determined from the observed images as the footprint radii of the
largest source and widest filament \citep[Sect. 3.1.3 in][]{Men2021method}. The footprint of a source or filament is defined as
the full extent of the structures at zero (background) level, i.e., as the area, beyond which their contribution becomes
negligible.

In our multiwavelength extraction of the sources and filaments with \emph{getsf} in Vela C, we used the five
\emph{Herschel} images at 70\,--\,500 $\mu$m and three surface density images with the angular resolutions of 8.5, 11.7, and
18.2{\arcsec}. In extracting sources, the detection images combined information from all the images, whereas to extract filaments,
the 70 $\mu$m image was not used. Having inspected the images, we chose the following sets of maximum sizes: 40, 60, 80, 100, and
120{\arcsec} for the \emph{Herschel} wavebands and 25, 60, and 80{\arcsec} for the surface densities.

\subsection{Mass distribution in Vela C}

Molecular clouds are complex structures that consist of various components \citep{Bergin2007}. Sources are the
round-shaped intensity peaks, corresponding to the individual fragments within molecular clouds.
Filaments are the significantly elongated structures, whose shapes are often significantly curved and properties depend on the
position along their crests. Both type of structures are characterized by a local overdensity and minimum in the gravitational
potential of their parent cloud.

\begin{table} 

\caption{Masses of the structural components of the Vela C molecular cloud, separated by \textsl{getsf} in the surface density
images with different angular resolutions. The value of $M_{\mathrm{S}}$ at $8.5${\arcsec} is likely significantly overestimated,
because it includes the protostellar sources with strong temperature gradients. The values of $M_{\mathrm{B}}$ and $M_{\mathrm{C}}$
are given after subtracting the global background $N^{\rm b}_{\rm H_{2}} = 2 \times 10^{21}$ cm$^{-2}$ of the image (or $3.7 \times
10^{4}$ $M_{\odot}$), taken from the lower-left corner of the image in Fig.~\ref{nh2map}.
} 
\begin{tabular}{ccccc}
\hline\hline
\noalign{\smallskip}
\!Resolution & \!\!Sources & \!Filaments & \!\!Background & \!Total \\
\!(\arcsec) & \!\!$M_{\mathrm{S}}$ ($M_{\odot}$) & \!$M_{\mathrm{F}}$ ($M_{\odot}$) & \!\!$M_{\mathrm{B}}$ ($M_{\odot}$) & 
 \!$M_{\mathrm{C}}$ ($M_{\odot}$) \\
\noalign{\smallskip}
\hline
\noalign{\smallskip}
\,\,$8.5$ & \!\!$6.3 \times 10^{3}$ & \!$3.0 \times 10^{4}$ & \!$5.3 \times 10^{4}$ & \!$9.0 \times 10^{4}$ \\
 \!$11.7$ & \!\!$2.2 \times 10^{3}$ & \!$2.5 \times 10^{4}$ & \!$4.6 \times 10^{4}$ & \!$7.3 \times 10^{4}$ \\
 \!$18.2$ & \!\!$2.4 \times 10^{3}$ & \!$2.6 \times 10^{4}$ & \!$4.4 \times 10^{4}$ & \!$7.2 \times 10^{4}$ \\
 \!$24.9$ & \!\!$-$ & \!$-$ & \!$-$ & \!$7.2 \times 10^{4}$ \\
 \!$36.3$ & \!\!$-$ & \!$-$ & \!$-$ & \!$7.2 \times 10^{4}$ \\
\noalign{\smallskip}
\hline
\label{masses}
\end{tabular}

\end{table} 

The hierarchical structure of Vela C (Figs.~\ref{nh2map} and \ref{coresinmap}) includes the filamentary network,
sources, and their backgrounds, all of them are separated from each other by \emph{getsf} in the extraction process
\citep{Men2021method,Zhanggy2020}. Assuming the cloud is optically thin at the observed wavelengths, the molecular cloud mass can
be derived from its surface densities,
\begin{equation} 
M_{\rm C} = \delta \mu_{\mathrm{H}_2} m_{\mathrm{H}} \sum_i \left(N_{\mathrm{H}_2} - N^{\rm b}_{\mathrm{H}_2} \right), 
\label{molecularcloudmass} 
\end{equation} 
where $\delta$ is the pixel area, $\mu_{\mathrm{H}_2} = 2.8$ the mean molecular weight per hydrogen molecule, $m_{\mathrm{H}}$ the
mass of the hydrogen atom, $N^{\rm b}_{\mathrm{H}_2}$ the global background of the image, and the summation is done over all
pixels. In Eq.~(\ref{molecularcloudmass}), we adopted the value $N^{\rm b}_{\rm H_{2}} = 2 \times 10^{21}$ cm$^{-2}$ as the image
background, corresponding to an integrated mass of $3.7 \times 10^{4}$ $M_{\odot}$.

Comparisons with a simulated star-forming region \citep[Fig.~A.1 in][]{Men2021method} suggested that the
highest-resolution surface densities are expected to be more accurate for the structural components than the standard maps at a
36.3{\arcsec} resolution. One exception is the unresolved protostellar peaks that can be strongly overestimated, because of the
steep temperature gradients in the protostellar cores, created by the accretion luminosity. To give readers a sense of possible
uncertainties in the derived masses of the structural components, Table~\ref{masses} presents the estimates, obtained from the
surface densities at each of the three angular resolutions used in the extraction, along with the total masses at the lower
resolutions. The bulk of the cloud is the low-density background with $N_{\rm H_{2}}\lesssim 5\times 10^{21}$ cm$^{-2}$, accounting
for $60${\%} of the total mass $M_{\rm C}$, whereas the components of filaments and sources constitute $35$ and $3${\%} of $M_{\rm
C}$, respectively.

    \begin{figure}
       \centering
       \includegraphics[width=0.45 \textwidth]{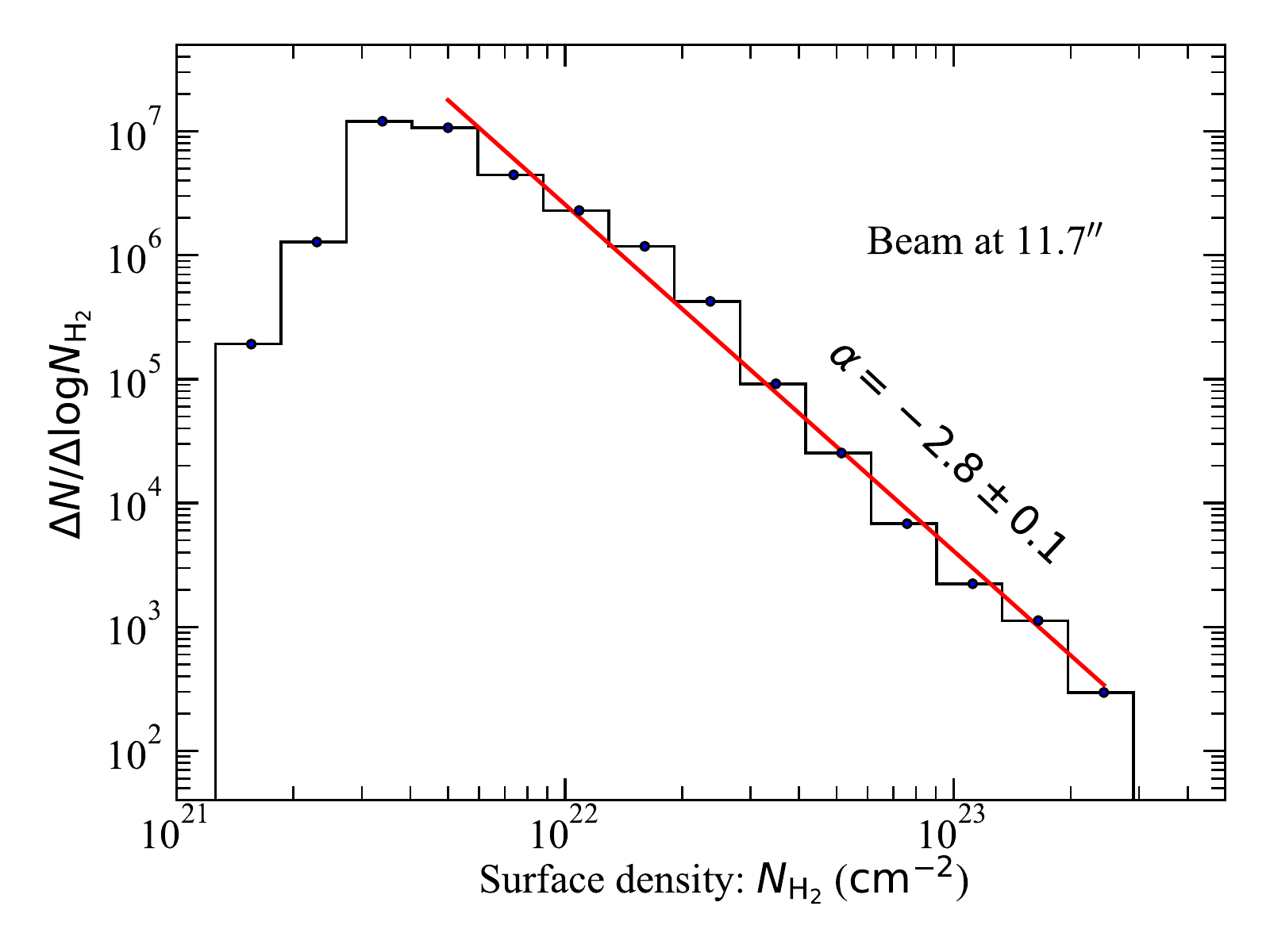}
       \caption{Probability density function PDF$_{N}$ for Vela C. The vertical axis gives the actual
       number of pixels per bin in the 11.7{\arcsec} resolution surface density image. At $N_{\mathrm{H}_2} \gtrsim
       7\times10^{21}$ cm$^{-2}$, corresponding to $N_{\mathrm{H}_2} \gtrsim 7\ A_V$, it can be well fitted with a power-law.}
       \label{PDF11p7}
    \end{figure}
  
      \begin{figure*}
       \centering
       \includegraphics[width=1.0 \textwidth]{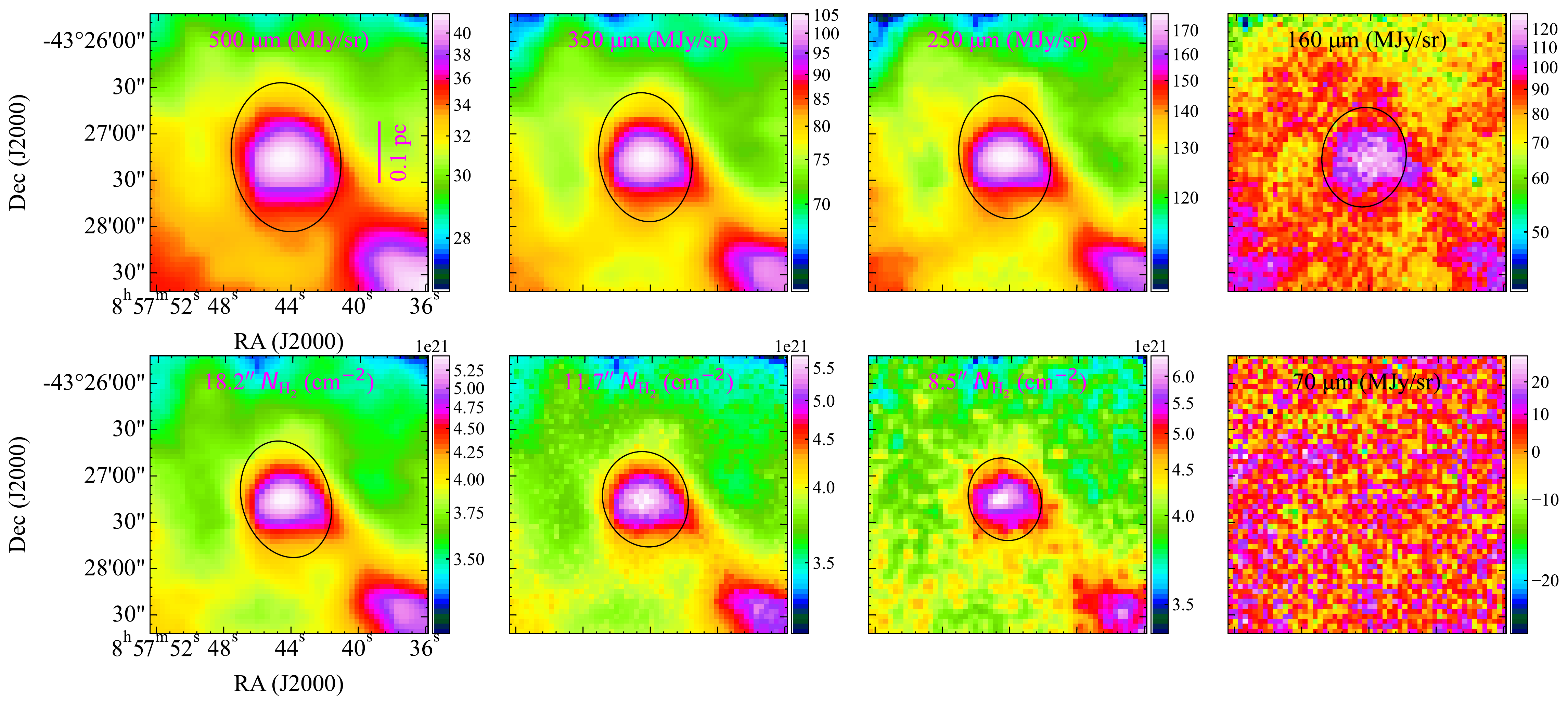}
       \caption{   
       Illustration of the selection criteria for the starless cores in the five \emph{Herschel} images and in the high-resolution 
       surface density maps at the 8.5, 11.7, and 18.2{\arcsec} resolutions (Sect.~\ref{sec:core_selection}). 
       Black ellipses represent the measured major and minor FWHM sizes.}
       \label{starlesscore}
    \end{figure*}

A surface density probability distribution function (PDF$_{N}$) is useful in getting insights into the physical processes acting in
the molecular cloud
\citep{Vazquez1994,Passot1998,Federrath2008,Kainulainen2009,Tassis2010,Kainulainen2011,Kritsuk2011,Ballesteros2011}.
Figure~\ref{PDF11p7} shows the PDF$_{N}$ of the Vela C surface densities at 11.7{\arcsec} resolution. At high surface densities
$N_{\mathrm{H}_2} \gtrsim 7\times10^{21}$ cm$^{-2}$, PDF$_{N}$ can be well fitted by a power law: ${\rm d}N / {\rm d}\log
N_{\mathrm{H}_2} \propto N_{\mathrm{H}_2}^{-2.8 \pm 0.1}$. Surface densities are convertible to visual extinction, assuming $A_V$
(mag) = $0.94\times10^{21}N_{\rm H_{2}}$ cm$^{-2}$ \citep{Bohlin1978}. The high-density tail corresponds to $N_{\rm H_{2}} \gtrsim
$ 7 $A_V$. The $\textrm{PDF}_{N}$ of the interstellar medium, and in particular of the molecular clouds, is also a rapidly
decreasing function, unless a special arrangement of masses occurs. Most of the space in the molecular cloud is filled with a
low-density background and only a small fraction of the dense structures, as can be inferred from the rapid decline in PDF$_{N}$
\citep{Kainulainen2009}.

\subsection{Selection of reliable cores}\label{sec:core_selection}

Having applied \textsl{getsf} for source extraction, we adopt its definition of sources \citep[cf. Sects.~1 and
3.2.2 in][]{Men2021method}: sources are the relatively round emission peaks that are significantly stronger than the local
surrounding fluctuations (of background and noise), suggesting the presence of the physical objects in space that produced the
observed emission. In total, 3711 sources were detected by \emph{getsf}.

Only good enough sources are then selected by \emph{getsf} at each wavelength $\lambda$, using the following conditions
\citep[Sect.~3.4.6 in][]{Men2021method}: (1) $\Xi_{{\lambda}} > 1$, where $\Xi_{{\lambda}}$ is the monochromatic
goodness (combining detection significance and signal-to-noise ratio); (2) $\Gamma_{{\lambda}} > 1$, where $\Gamma_{{\lambda}}$ is
the detection significance from monochromatic single scales; (3) $F_{{\rm P}{\lambda}}/\sigma_{{\rm P}{\lambda}} > 2$, where
$F_{{\rm P}{\lambda}}$ is the peak intensity and $\sigma_{{\rm P}{\lambda}}$ the peak intensity error; (4) $F_{{\rm
T}{\lambda}}/\sigma_{{\rm T}{\lambda}} > 2$, where $F_{{\rm T}{\lambda}}$ is the total flux and $\sigma_{{\rm T}{\lambda}}$ the
total flux error; (5) $A_{{\lambda}}/B_{{\lambda}} < 2$, where $A_{{\lambda}}$ and $B_{{\lambda}}$ are the major and minor sizes at
half-maximum; (6) $A_{{\rm F}{\lambda}}/A_{{\lambda}} > 1.15$, where $A_{{\rm F}{\lambda}}$ is the full major axis of the
elliptical footprint of a source.
    
    \begin{figure*}
       \centering
       \includegraphics[width=1.0 \textwidth]{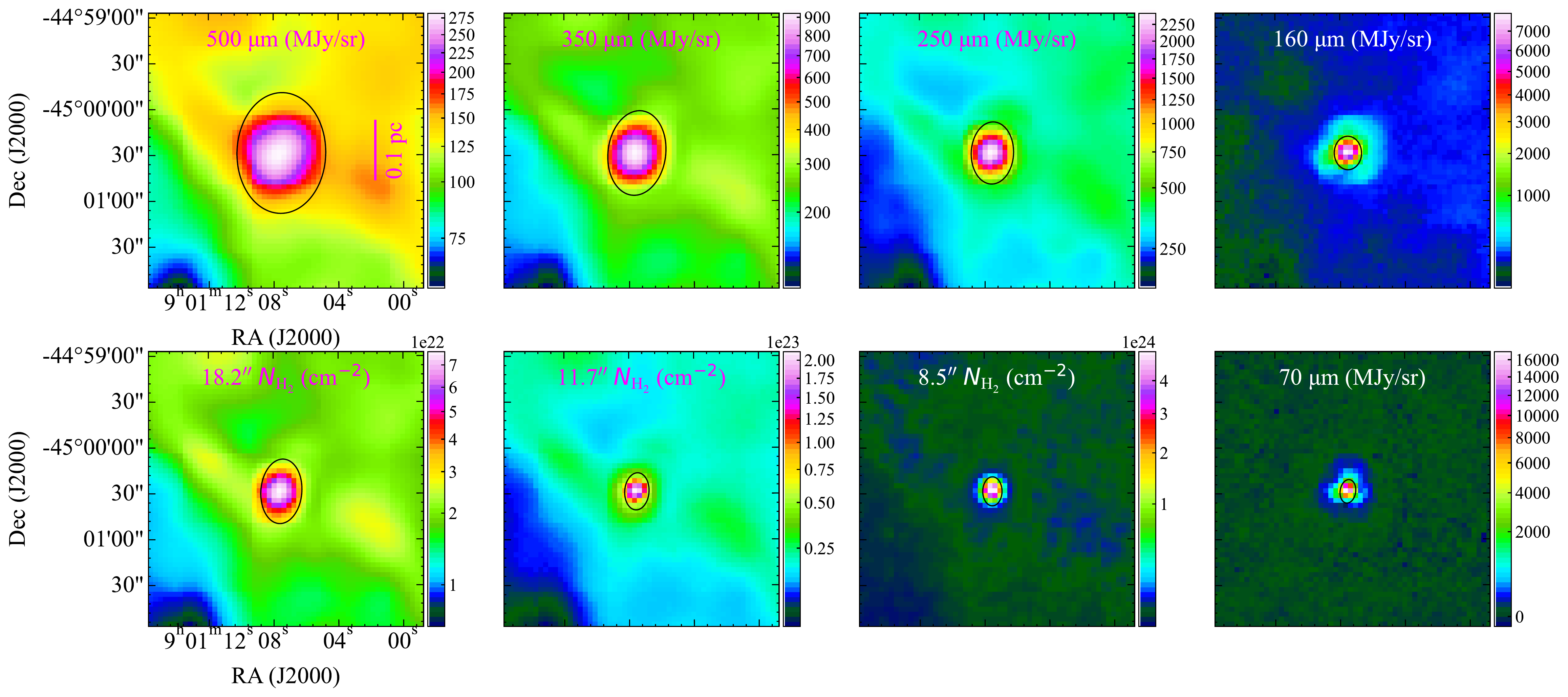}
       \caption{
      Illustration of the selection criteria for the protostellar cores in the five \emph{Herschel} images and in the high-resolution 
      surface density maps at the 8.5, 11.7, and 18.2{\arcsec} resolutions (Sect.~\ref{sec:core_selection}). Black 
      ellipses represent the measured major and minor FWHM sizes.}
       \label{protocore}
    \end{figure*}

The above criteria clean up the extraction catalog, removing spurious detections and selecting for further analysis only reliable
and well-measurable sources, as verified by \cite{Men2021benchmark}. The conditions (1)\,--\,(4) ensure that the detected source
peaks are sufficiently strong to be distinguished from the background and noise fluctuations. The condition (5) ensures that
sources are circular or elliptical, excluding too elongated structures that are unlikely to be sources. The condition (6) is used
to discard the sources with unrealistically small ratios of their footprint and half-maximum sizes. The criteria excluded a
substantial number of detected peaks that are likely spurious and removed the sources with unreliable (inaccurate) measurements,
leaving 1755 selected good sources.

Extragalactic sources are physically different from the starless cores or YSOs of our interest and must be excluded. Based on their
peak positions, we cross-matched the sources with the SIMBAD database\footnote{https://simbad.cds.unistra.fr/simbad/}
\citep{Wenger2000} and the NASA/IPAC Extragalactic Database\footnote{https://ned.ipac.caltech.edu} (NED) within a 6{\arcsec} radius.
As a result, we removed one source based on SIMBAD and nine sources based on NED, which left us with 1745 acceptable sources.
  
The conditions (1)\,--\,(6) are necessary for a source to be acceptable in a single image or in a surface density map. In order to
improve the reliability, we further selected as starless cores only those sources that are acceptably good in at least two of the
four wavebands of 160, 250, 350, and 500 $\mu$m and also in the surface density map at the 11.7{\arcsec} resolution, but not
detectable in the 70 $\mu$m map. Figure~\ref{starlesscore} illustrates the selection process that resulted in
421 acceptable starless cores.

The protostellar cores differ from the starless cores in that they produce accretion luminosity at their centers,
which creates a strong unresolved peak at short wavelengths ($\lambda \la 100$ $\mu$m), where the prestellar cores do not radiate
any noticeable amount of energy. Following the standard approach, we identified as protostellar cores those sources that are
detected and acceptably good at 70 $\mu$m, not much resolved with a $5.9${\arcsec} beam and not too elongated, that is their major
and minor half-maximum sizes $A$ and $B$ obey the relations $(A B)^{1/2} < 1.5 \times 5.9${\arcsec} and $A/B < 1.3$. In addition,
the protostellar cores must be acceptably good in at least two of the four images at 160, 250, 350, and 500 $\mu$m and in the
surface density image at the 11.7{\arcsec} resolution. Figure~\ref{protocore} illustrates the selection process that resulted in
149 acceptable protostellar cores.

    \begin{figure*}
    \centering
    \includegraphics[width=0.33 \textwidth]{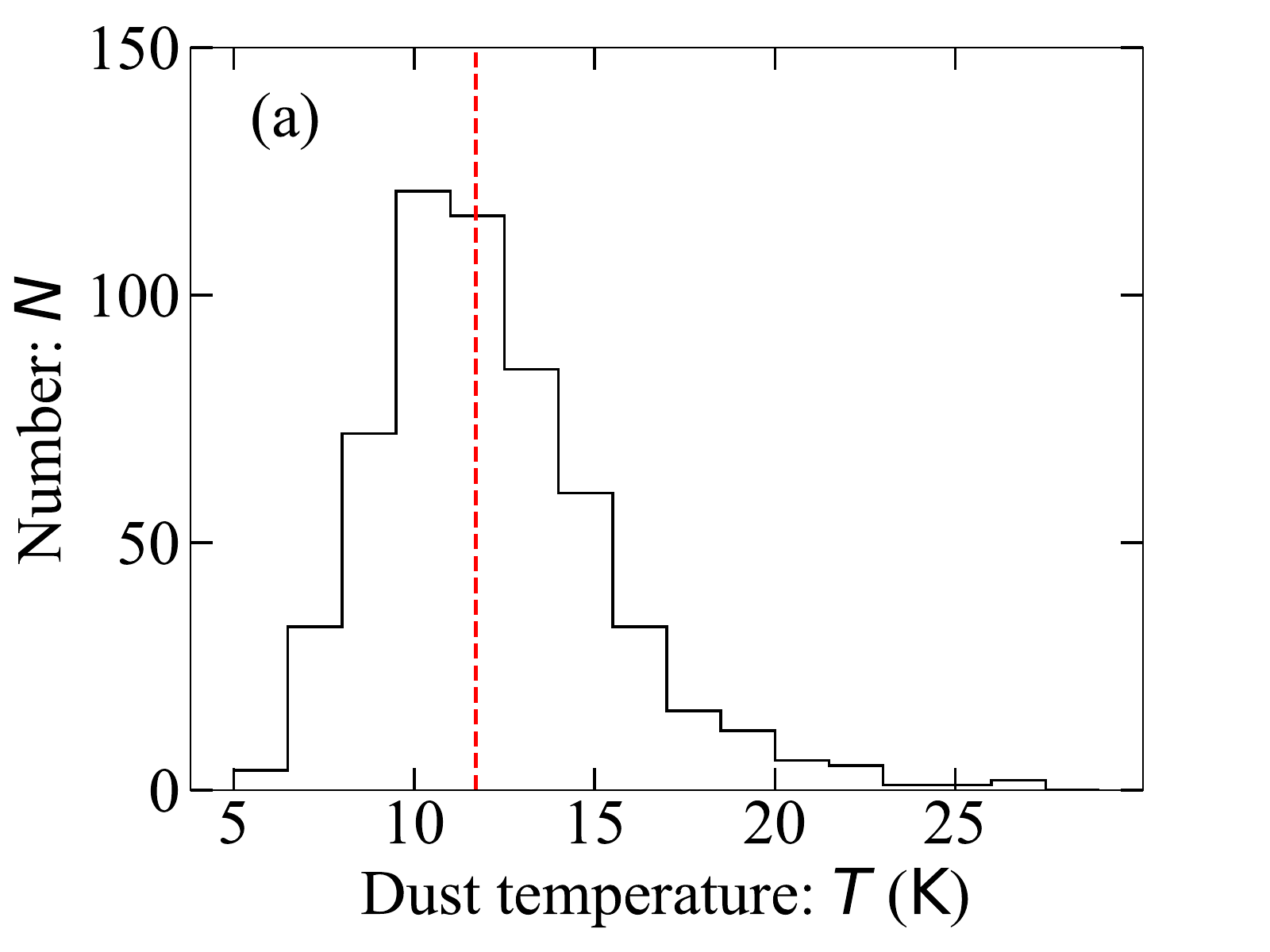}
    \includegraphics[width=0.33 \textwidth]{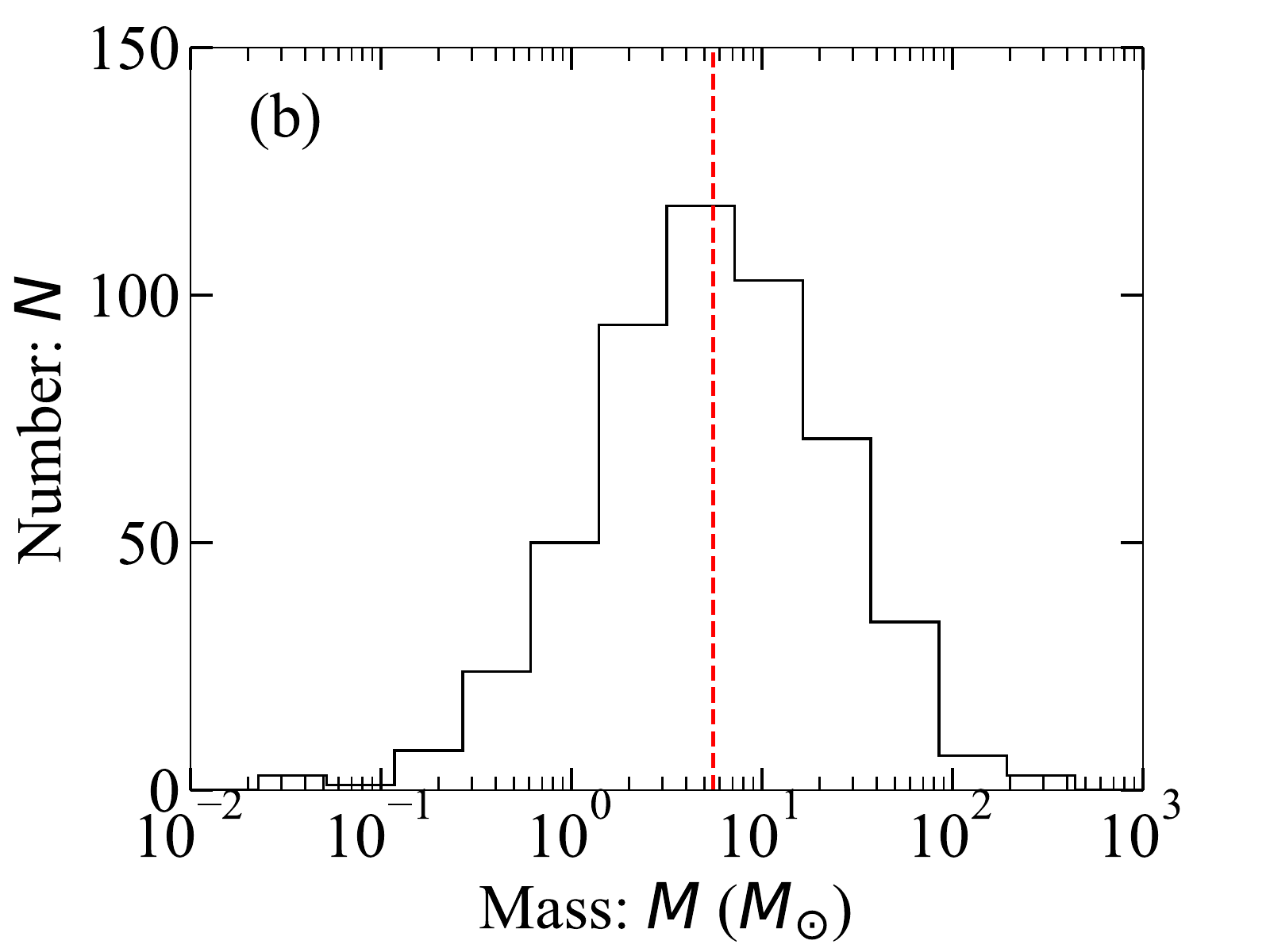}
    \includegraphics[width=0.33 \textwidth]{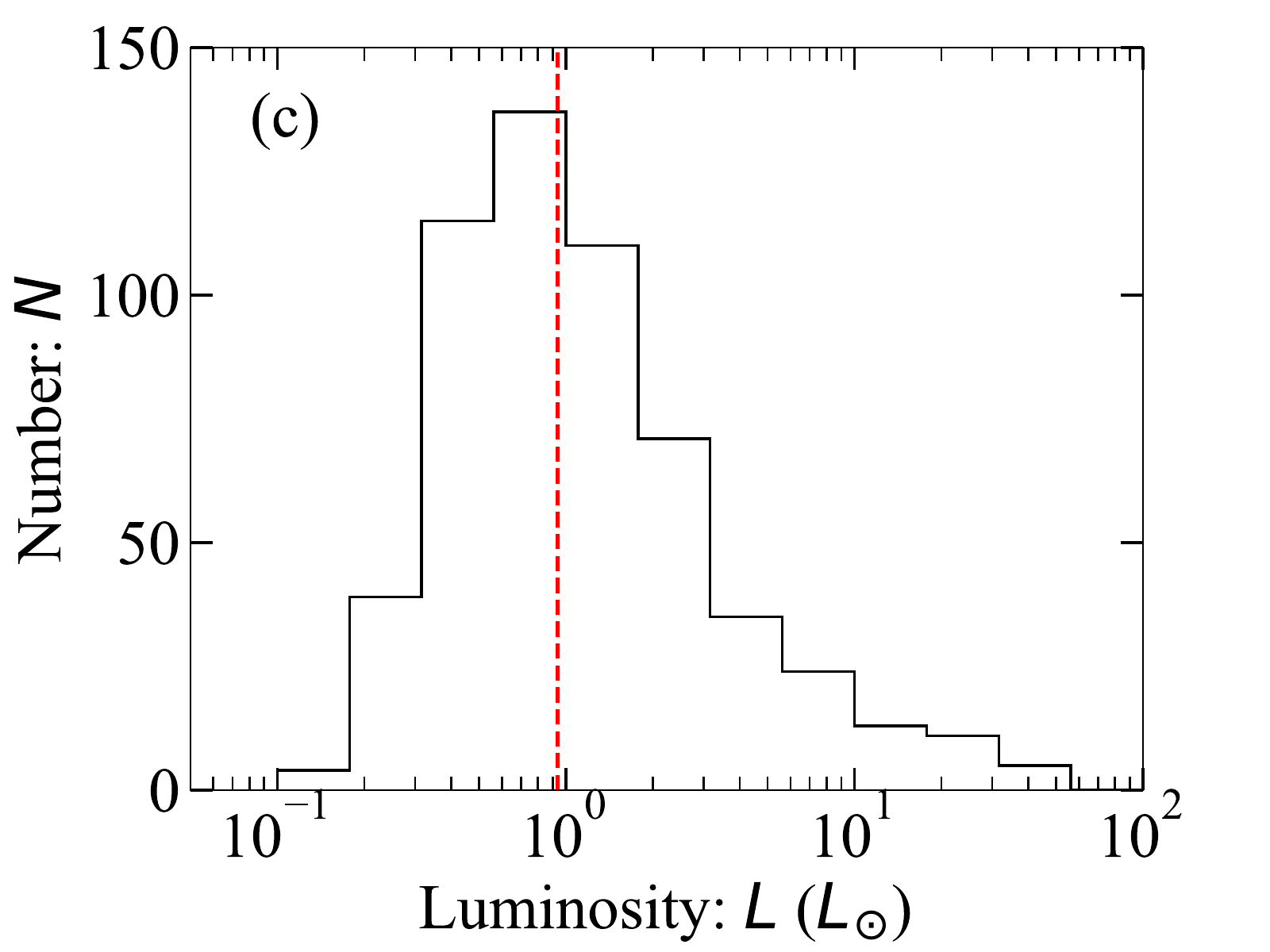}
        \caption{Histograms of the physical parameters of the 570 reliable cores obtained from fitting SEDs of the extracted 
        sources. Panel (\emph{a}): distribution of their average temperatures (median of 11.7 K), 
        Panel (\emph{b}): distribution of their masses (median of 4.8 $M_{\odot}$), Panel (\emph{c}): distribution of their 
        luminosities (median of 0.9 $L_{\odot}$). The red dashed lines indicate the median values.}
    \label{MTLRhist}
    \end{figure*}

\subsection{Classification of extracted cores}\label{sec:core_class}

\begin{figure*}
    \centering
    \includegraphics[width=0.33 \textwidth]{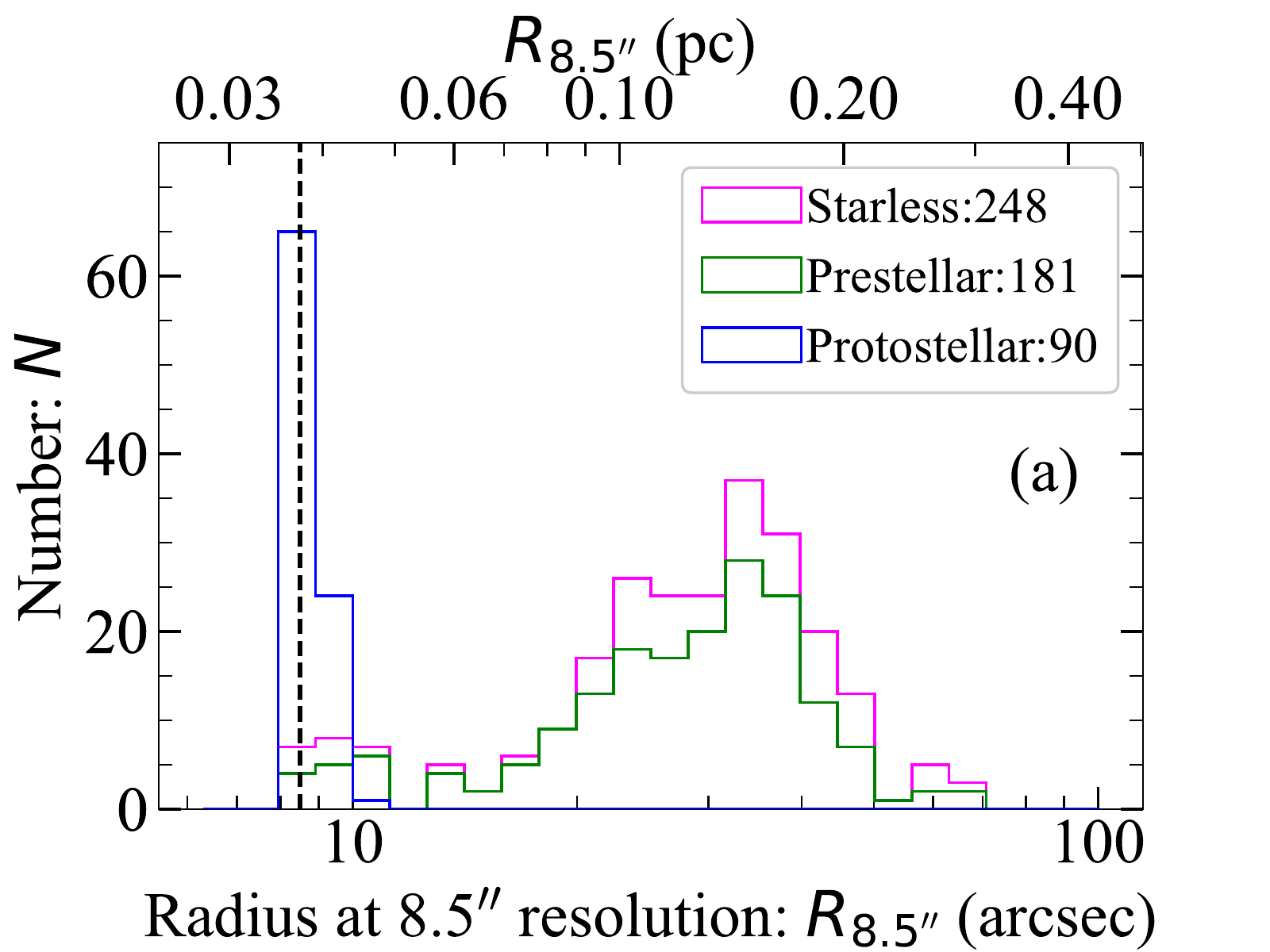}
    \includegraphics[width=0.33 \textwidth]{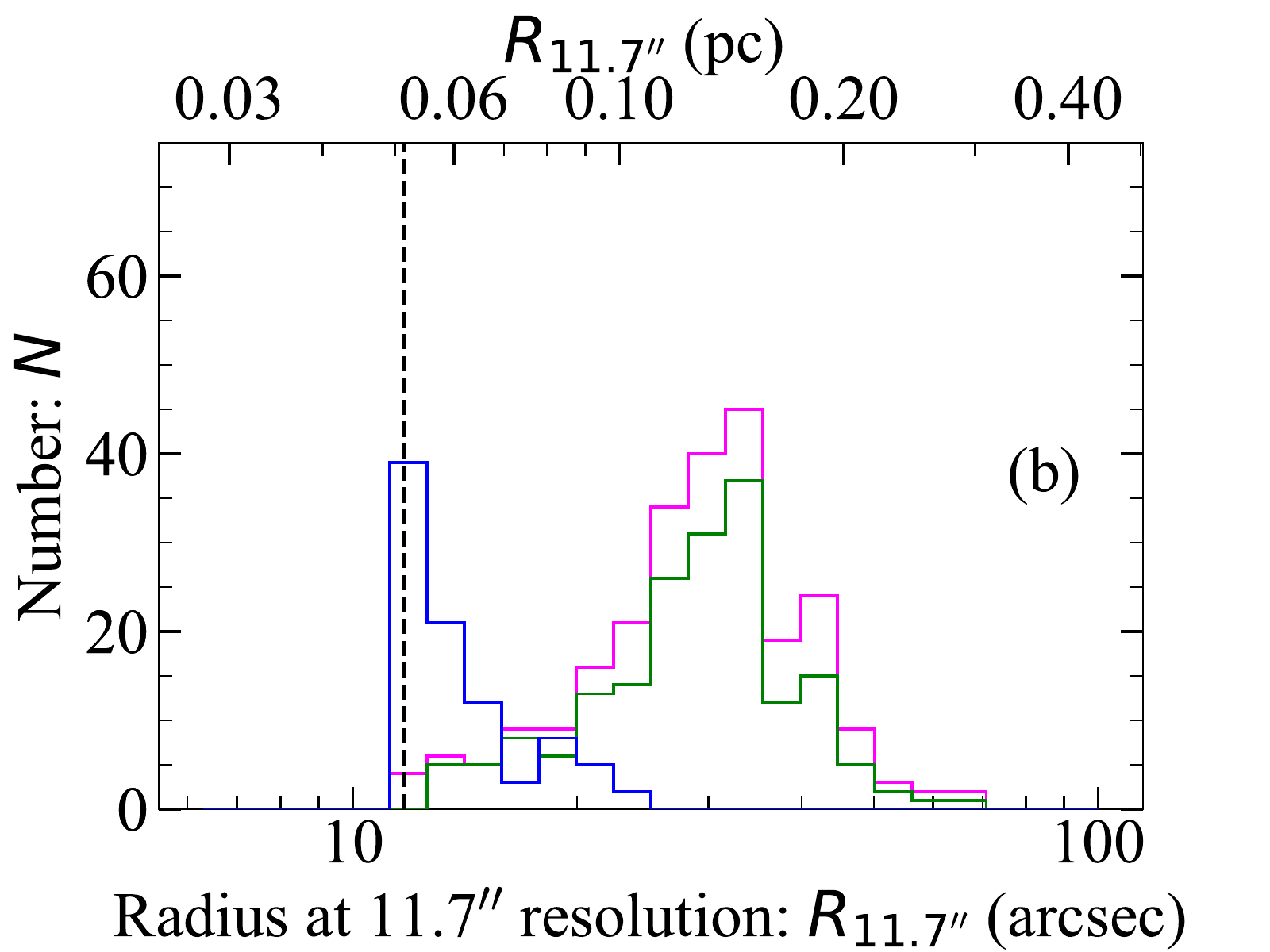}
    \includegraphics[width=0.33 \textwidth]{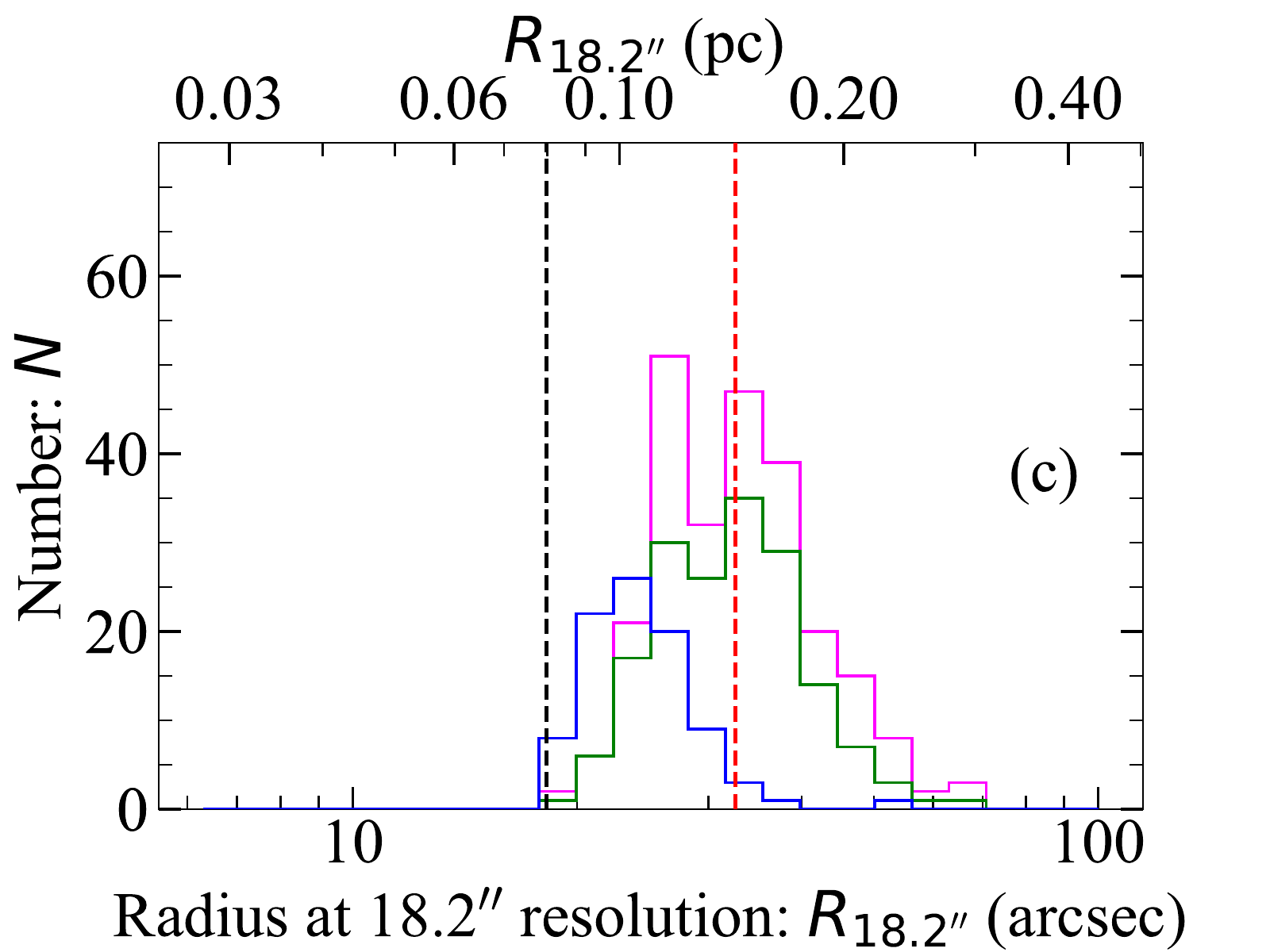}
    \caption{
    Histograms of the sizes of isolated cores, not blended with any other source in any 
    waveband. The core radius is the arithmetic mean of the major and minor half-maximum sizes, measured in the surface density
    images with three different resolutions (indicated by the vertical dashed lines). Panel (\emph{a}): distribution of source radii
    at the 8.5{\arcsec} resolution. Panel (\emph{b}): distribution of source radii at the 11.7{\arcsec} resolution. Panel (\emph{c}):
    distribution of source radii at the 18.2{\arcsec} resolution.}
    \label{coreR}
    \end{figure*}

Stars form in the process of the gravitational collapse of dense prestellar cores of a typical size of $\sim 0.1$ pc that represent
the local density maxima of the parent molecular cloud, producing the local minima of the gravitational potential
\citep[e.g.,][]{Bergin2007, Andre2014, Zhanggy2020}. Prestellar cores are the gravitationally bound starless cores with the
potential to give birth to stars \citep{Ward2007,Bergin2007,Andre2014,Konyves2015}. The prestellar cores, supported by the pressure
of the ambient gas of the parental cloud \citep[e.g.][]{Alves2001, Kirk2005}, can be approximated by the equilibrium,
self-gravitational, isothermal Bonnor-Ebert (BE) spheres \citep{Ebert1955, Bonnor1956}. The BE model is useful even if the real 
core is not strictly isothermal or not in the hydrostatic equilibrium \citep[e.g.,][]{Galli2002,Ballesteros2003}. The mass of the
critical BE sphere can be expressed as
\citep{Bonnor1956}
  \begin{equation}
 M_{\rm BE} \approx 2.4\, R_{\rm BE}\, c_{\rm s}^{2}/G,
 \label{bemass}
  \end{equation}
where $G$ is the gravitational constant, $c_{\rm s}$ the isothermal sound speed, and $R_{\rm BE}$ is the radius 
of the BE sphere, approximated in our work by the deconvolved radius,
\begin{equation}
R_{\rm dec}=(A B-O^2)^{1/2},
\label{deconv}
\end{equation}
where $A$ and $B$ are the major and minor half-maximum sizes, measured in the surface densities with the angular resolution of $O =
11.7{\arcsec}$. The sound speed in Eq.~(\ref{bemass}) was computed for the actual SED temperature $T_{F}$ of each 
extracted core. Only the partially-resolved and unresolved sources (resolvedness $(A B)^{1/2}\,O^{-1} > 1.1$) were used in 
Eq.~(\ref{deconv}) to avoid unacceptably large size deconvolution errors \citep{Men2023}, hence 8 unresolved starless cores
were ignored.
The starless cores are deemed
self-gravitating (bound) and classified as prestellar cores, when the ratio $\alpha_{\rm BE} = M_{\rm BE} / M_{\rm core} \leqslant
2$, where the core masses (and luminosities) are estimated by fitting the spectral energy distribution of the total fluxes
$F_{\lambda}$ of the sources over the \emph{Herschel} wavebands. The resulting total number of prestellar cores is
329 and the number of the unbound starless cores is 84 (Fig.~\ref{coresinmap}).
The distributions of the source radii measured from the surface density maps at the angular resolutions of 
8.5, 11.7, and 18.2{\arcsec} are shown in Fig.~\ref{coreR}.

Table \ref{coreobs} presents a template of the online catalog of our multiwavelength source extraction and Table \ref{corederive} is a template of the online catalog of the derived parameters for the 570 reliable cores from Table \ref{coreobs}, which includes their dust temperatures $T_{F}$, masses $M_{F}$, and luminosities $L_{F}$ (Fig.~\ref{MTLRhist}), obtained from fitting the spectral energy distributions (SED) of each core.

\begin{figure}
    \centering
            \includegraphics[width=0.49 \textwidth]{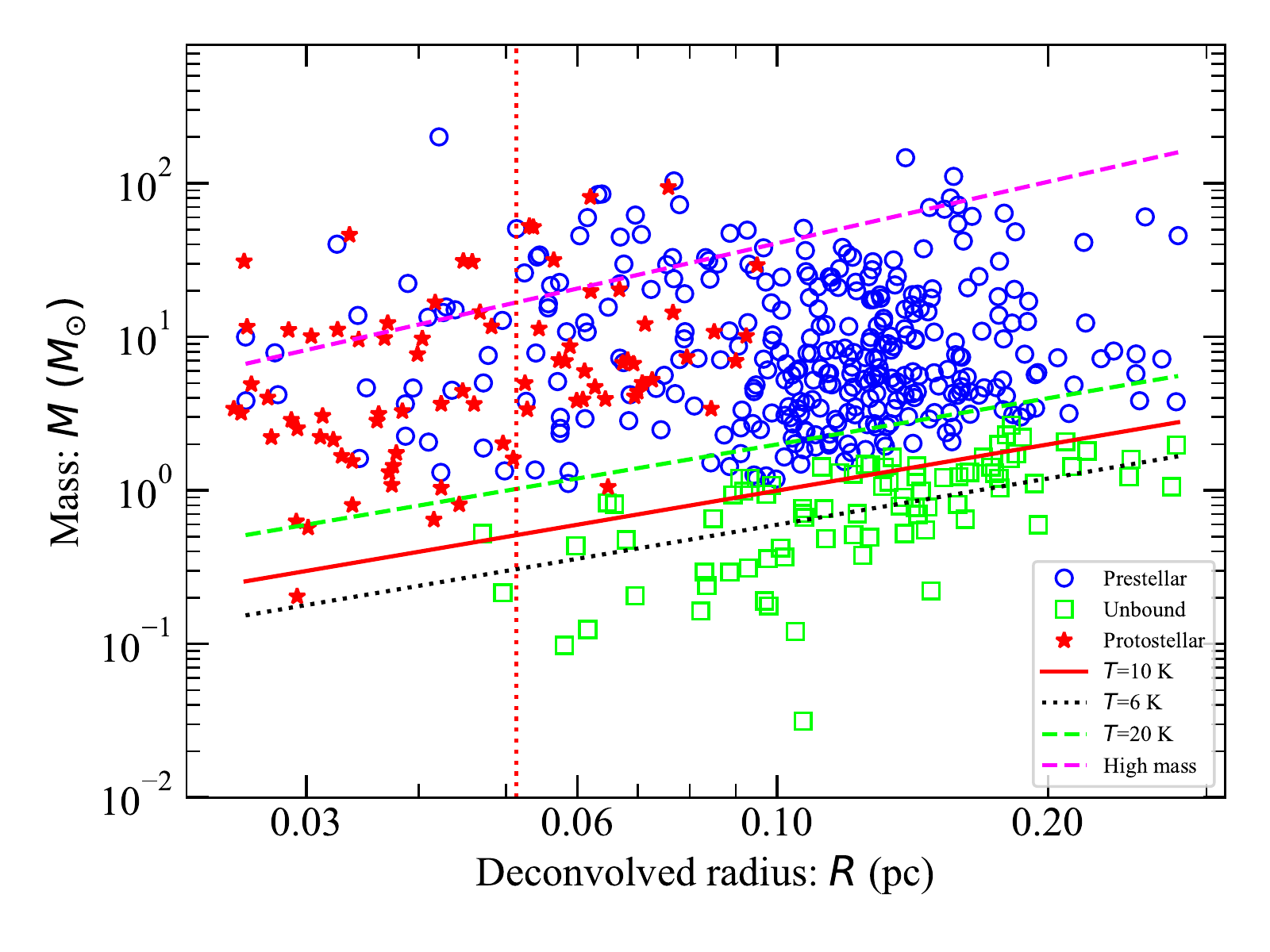}
        \caption{
        The mass-size diagram for Vela C for the unbound cores (green squares), prestellar cores (blue circles), 
        and protostellar cores (red stars). Only the partially-resolved and resolved cores are shown (with sizes by $10${\%} larger
        than the $11.7${\arcsec} beam) to ensure acceptably accurate deconvolution results using Eq.~(\ref{deconv}). The critical
        Bonnor-Ebert spheres with temperatures of $6$, $10$, and $20$ K are displayed by the dashed black, dashed green, and solid
        red lines, respectively. The dashed magenta line indicates an empirical relationship proposed by \citet{Kauffmann2010}.
        The physical scale of $0.05$ pc at the $11.7${\arcsec} resolution is indicated by the vertical line.}
    \label{masssize}
    \end{figure}

Figure~\ref{masssize} presents the mass-size diagram for the reliable cores in Vela C, where the deconvolved sizes
were obtained from Eq.~(\ref{deconv}) using the measured sizes from the surface density image at $11.7${\arcsec} resolution. The 
prestellar cores with $\alpha_{\rm BE}\leq 2$ are more
massive than the starless cores not bound by gravity ($\alpha_{\rm BE}>2$).
\citet{Kauffmann2010} proposed an empirical formula $M\propto 870\,M_{\odot}\,(R/{\rm pc})^{1.33}$ that provides a lower limit for
the mass of massive stars that can form in molecular clouds, derived from observations of the distribution of masses of young stars
in different environments.
The presence of prestellar and protostellar cores above this line suggests that they likely form massive stars.
 
\subsection{Core mass function}

\begin{figure}
        \centering
        \includegraphics[width=0.45 \textwidth]{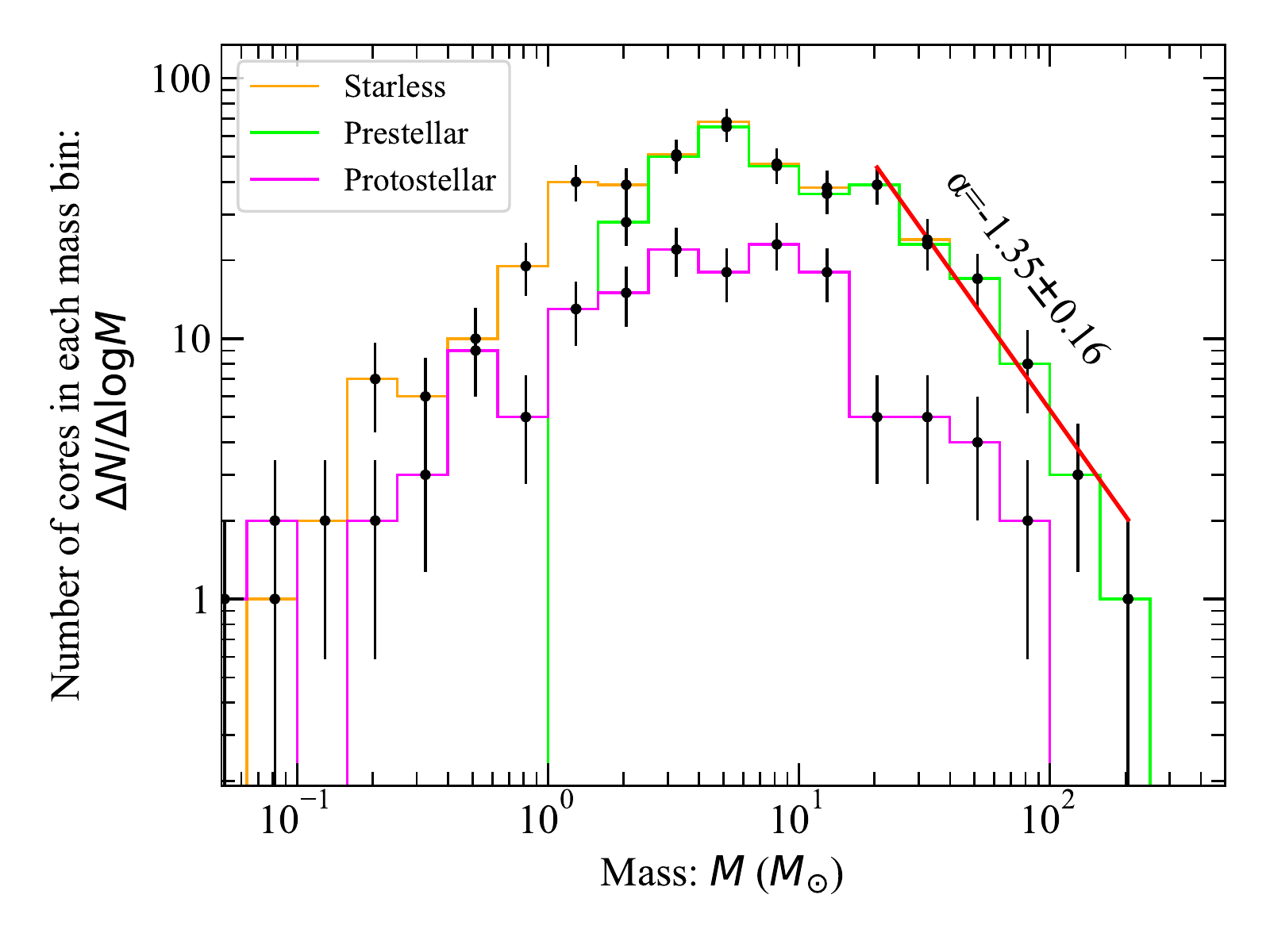}
        \caption{
        Differential core mass function (${\rm d}N/{\rm dlog}M$) for the 329 prestellar cores (green)
        421 starless cores (orange) and 149 protostellar cores (magenta). The error bars
        correspond to the $\sqrt{N}$ statistical uncertainties.}
        \label{CMFsed}
    \end{figure}

A fundamental property of molecular clouds is their mass distribution \citep{Javier2020}. Investigations of their CO emission
\citep{Sanders1985,Stutzki1990,Williams1994,Heyer1998,Kramer1998} and later studies of their dust continuum emission
\citep{Motte1998,Johnstone2000,Alves2007}, have shown that the molecular clouds and their substructures do not have any
characteristic masses. Typically, they display a power-law distribution over masses,
    \begin{equation}
\frac{{\rm d}N}{{\rm dlog}M}\propto M^{-\alpha}.
    \end{equation}
Using the masses derived from the SED fitting, we obtained the mass functions for the 421 starless,
329 prestellar cores and 149 protostellar cores (Fig.~\ref{CMFsed}). The CMFs for
the starless and prestellar cores are indistinguishable at $M > 2\ M_{\odot}$ (all starless
cores are bound) and they can be well fitted with ${\rm d}N/{\rm dlog}M \propto M^{-1.35 \pm 0.16}$ between 20 and 200
$M_{\odot}$. The power-law exponent is the same as that of the stellar IMF ($\alpha = 1.35$) \citep{Salpeter1955}.
The CMFs also reveal that the population of protostellar cores has a deficit of massive cores with respect to the
prestellar cores.

    \begin{figure*}
        \centering
        \includegraphics[width=0.45 \textwidth]{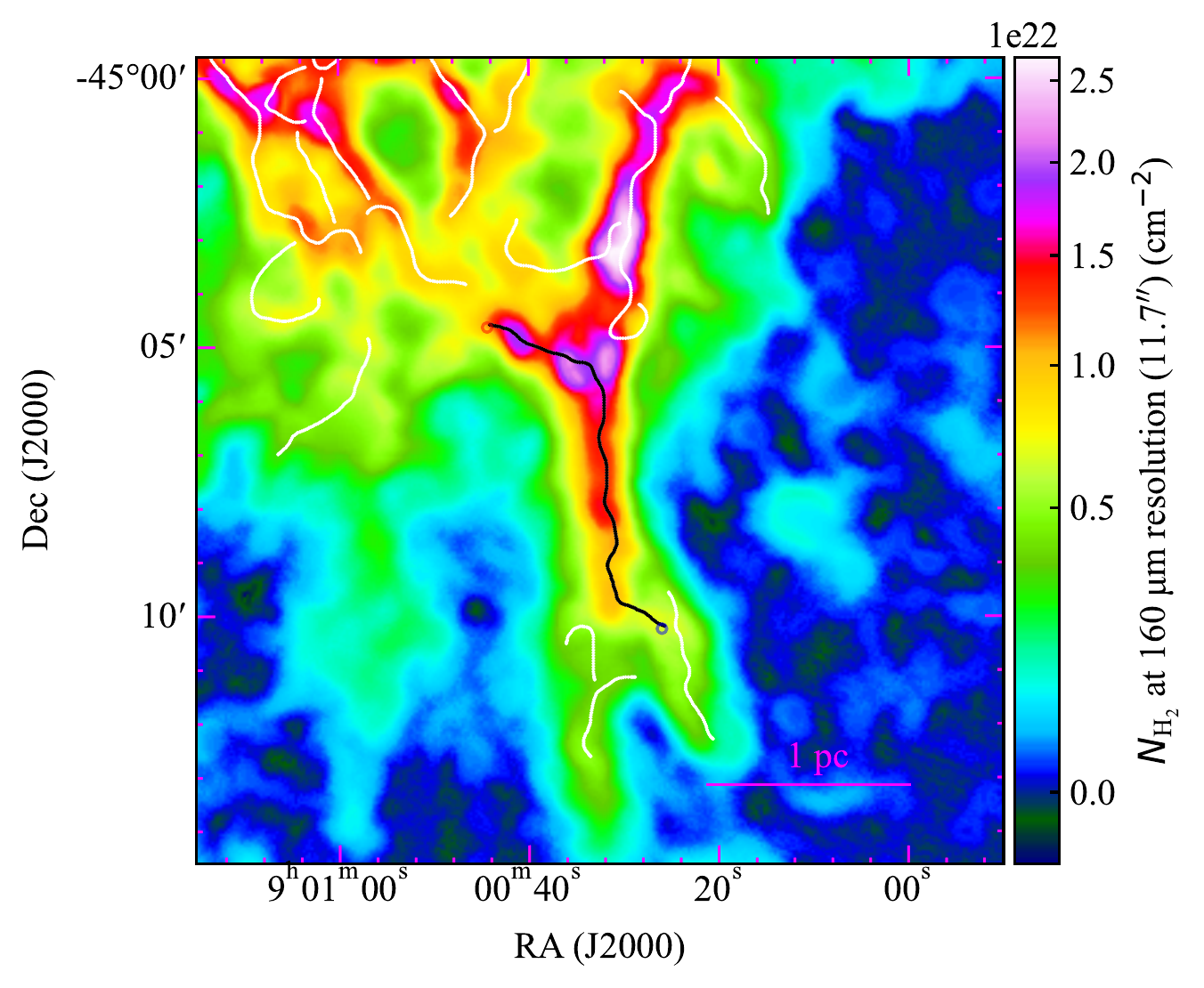}
        \includegraphics[width=0.40 \textwidth]{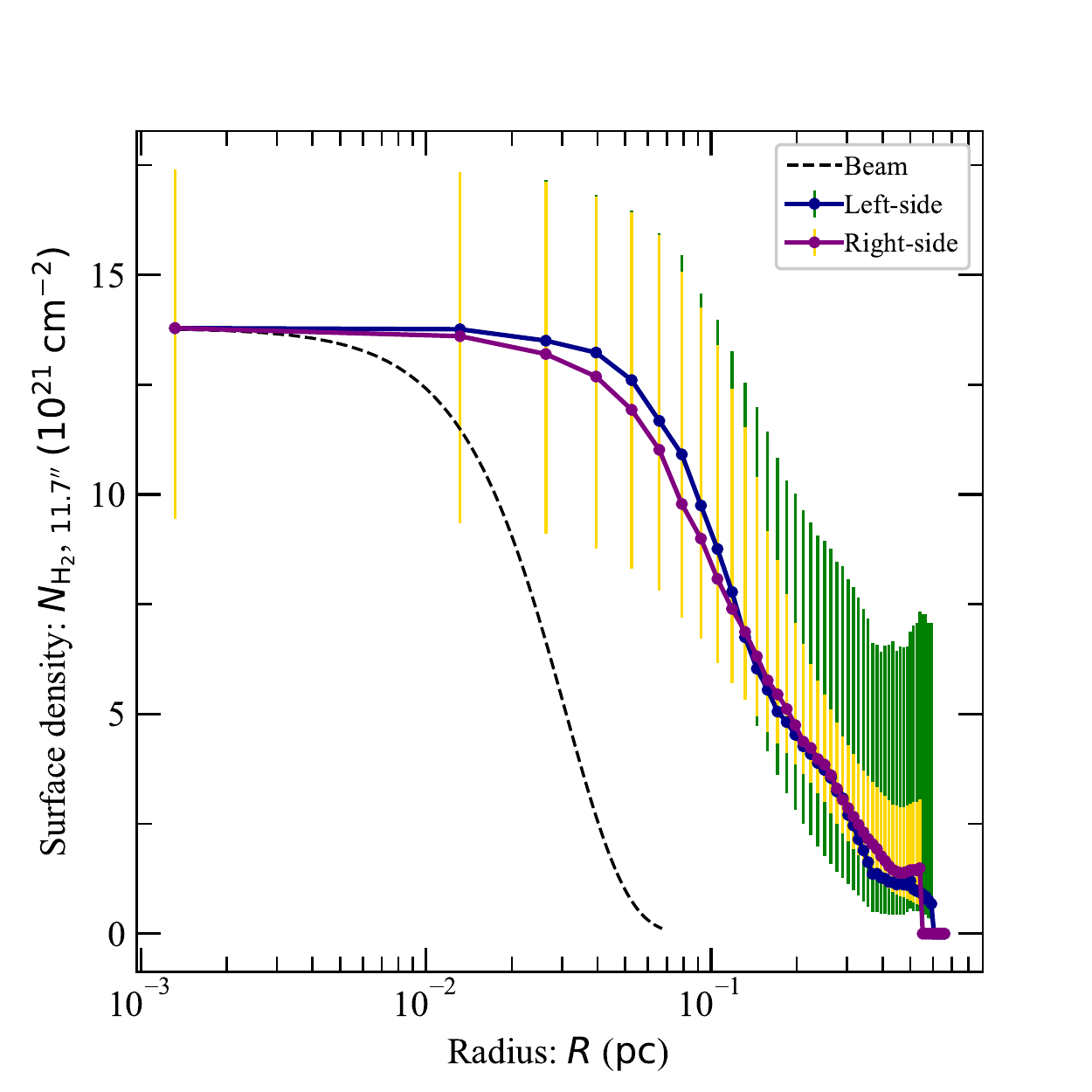}
        \caption{
        Illustration of a detected and measured filament in the high-resolution surface density image at 11.7{\arcsec}. 
        \emph{Left}: skeletons of the detected filaments with a contrast $C > 0.5$ is shown by the white curves and the black 
        curve is the skeleton of a filament whose measured radial profile is shown in the right panel. \emph{Right}: median 
        one-sided surface density profiles measured on the two sides of the filament (blue and purple solid lines). The black 
        dashed curve shows the beam profile. The median half-maximum width of the filament is $0.33\pm0.19$ pc.}
        \label{filamentexample}
    \end{figure*}
   
    \begin{figure*}
        \centering
        \includegraphics[width=0.3 \textwidth]{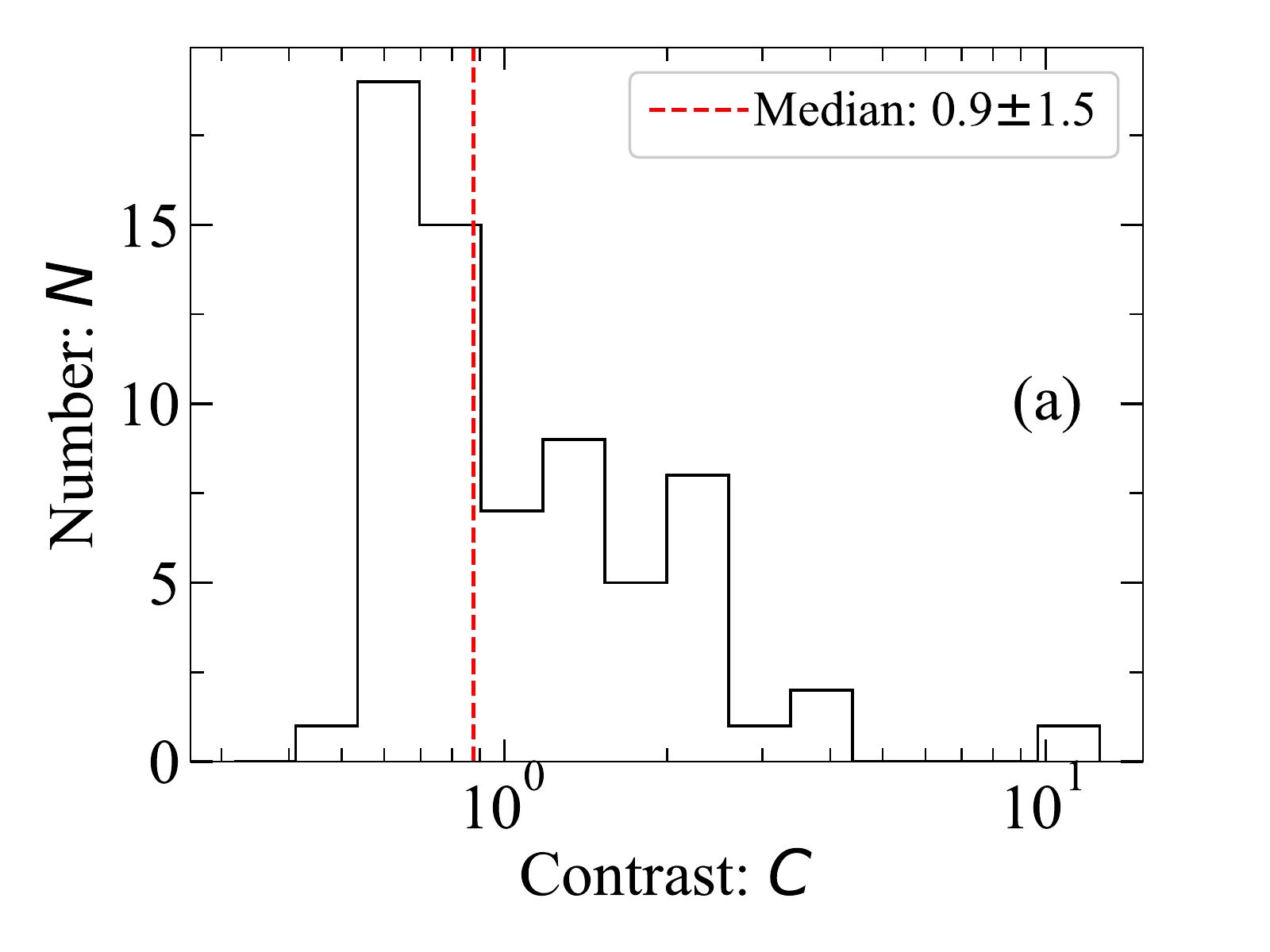}
        \includegraphics[width=0.3 \textwidth]{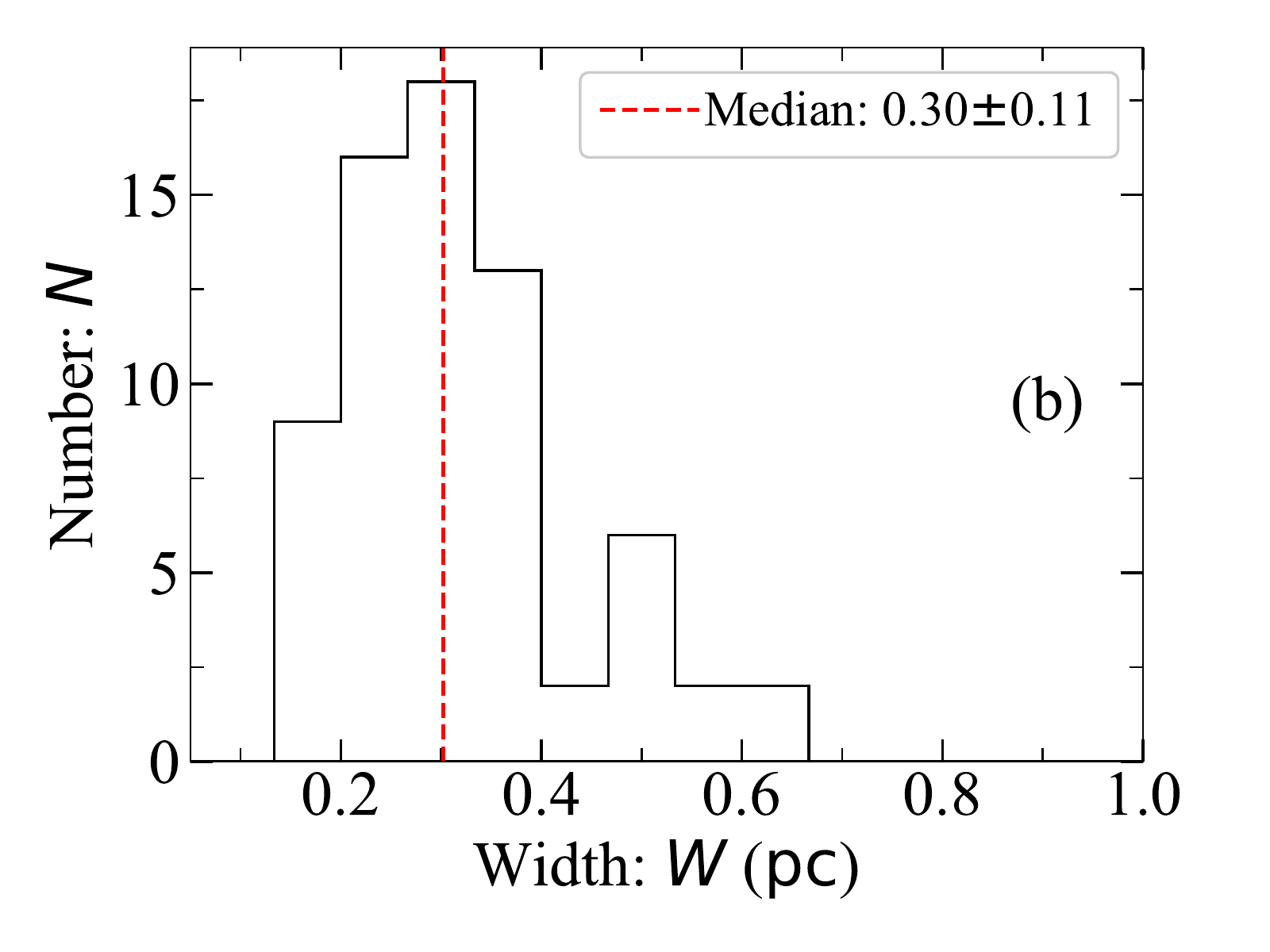}
        \includegraphics[width=0.3 \textwidth]{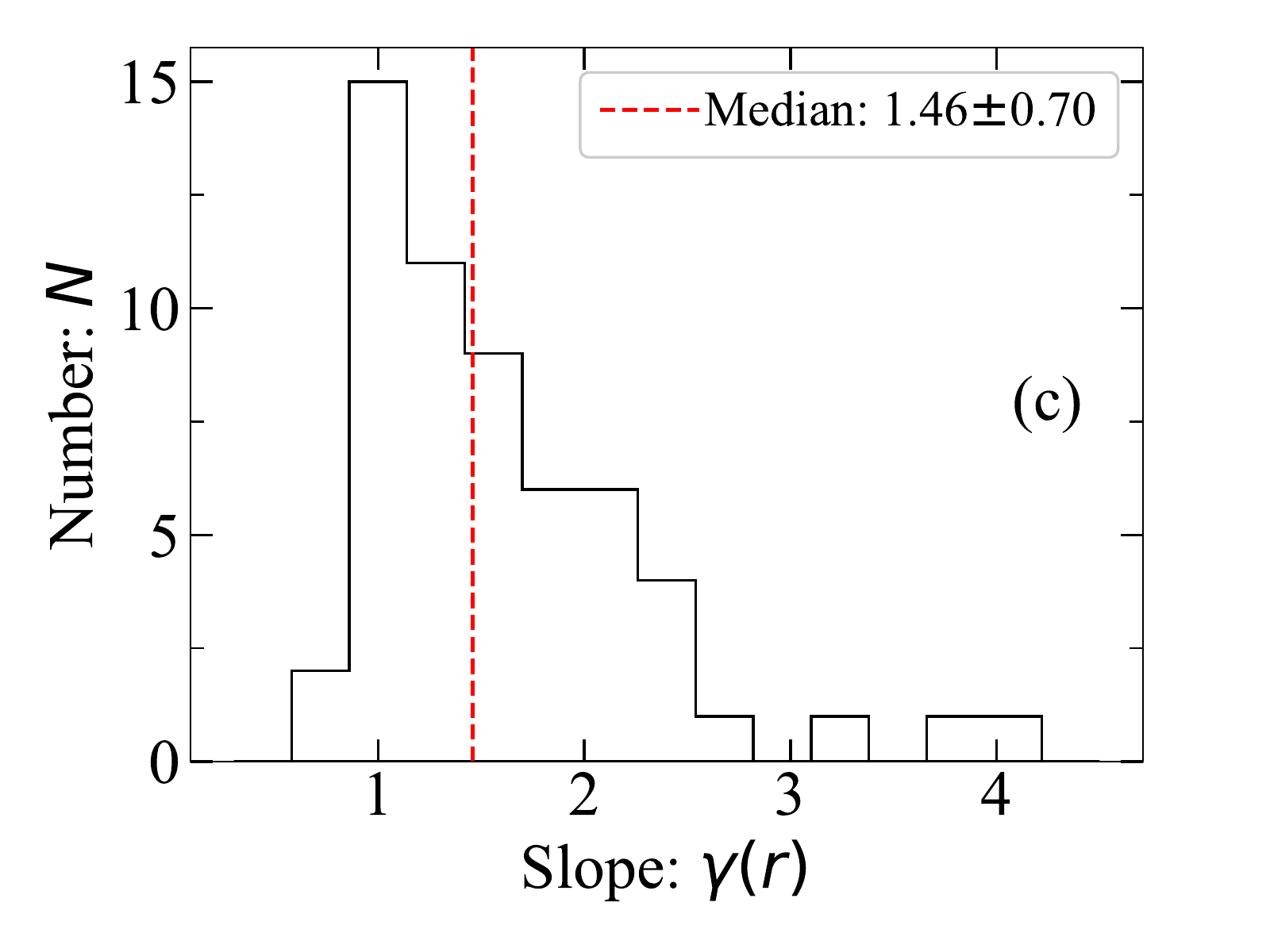}
        \includegraphics[width=0.3 \textwidth]{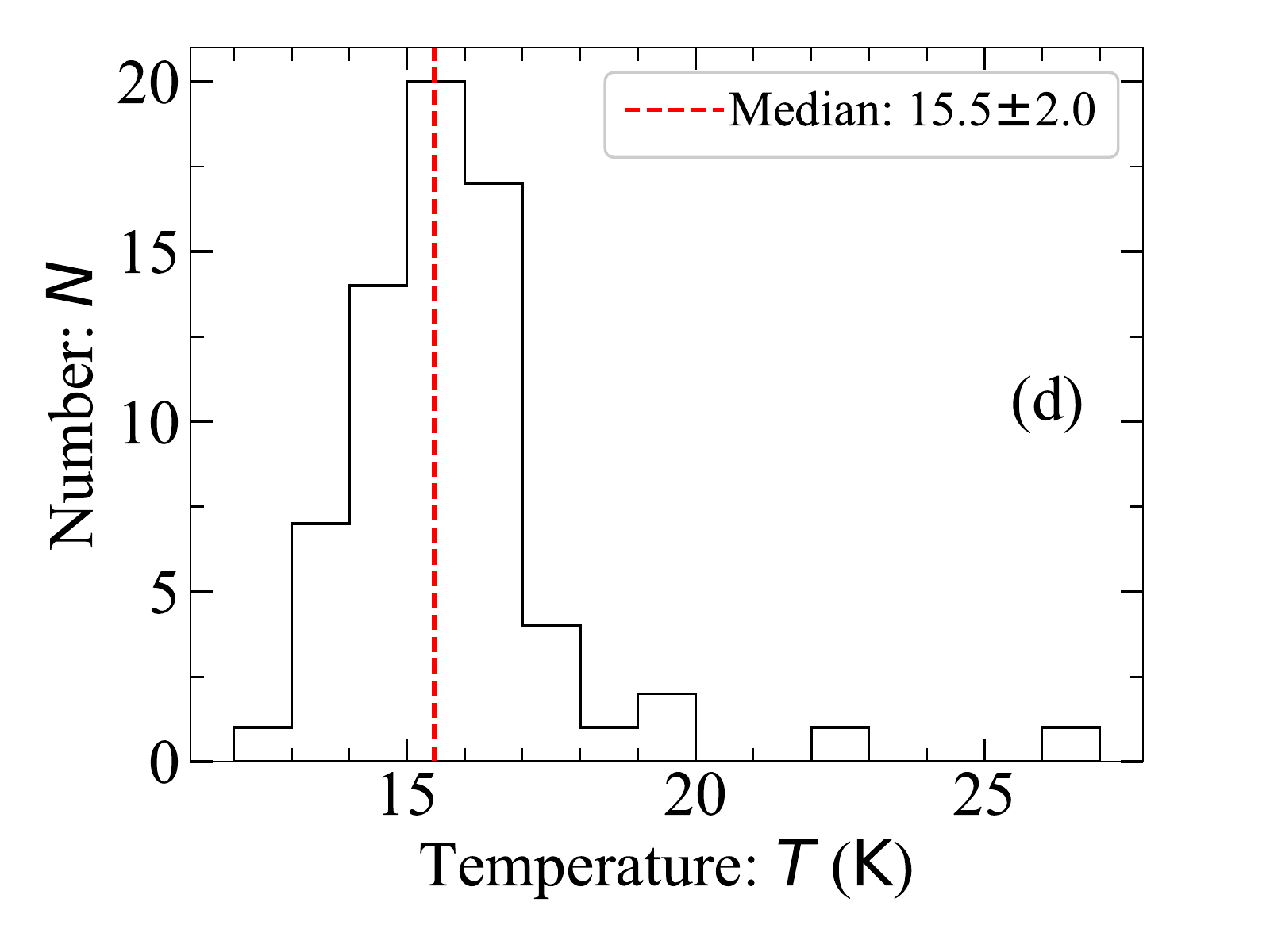}
        \includegraphics[width=0.3 \textwidth]{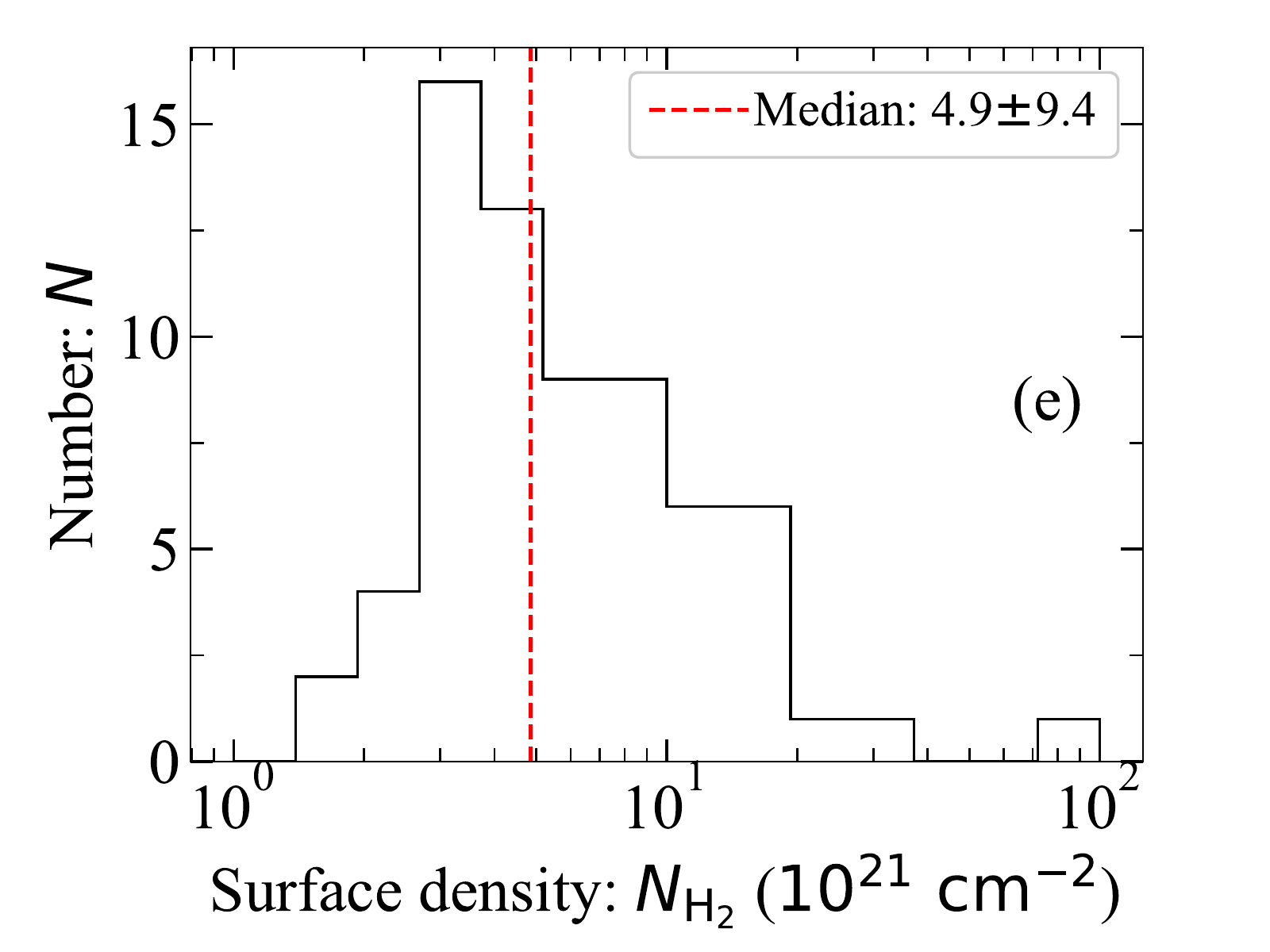}
        \includegraphics[width=0.3 \textwidth]{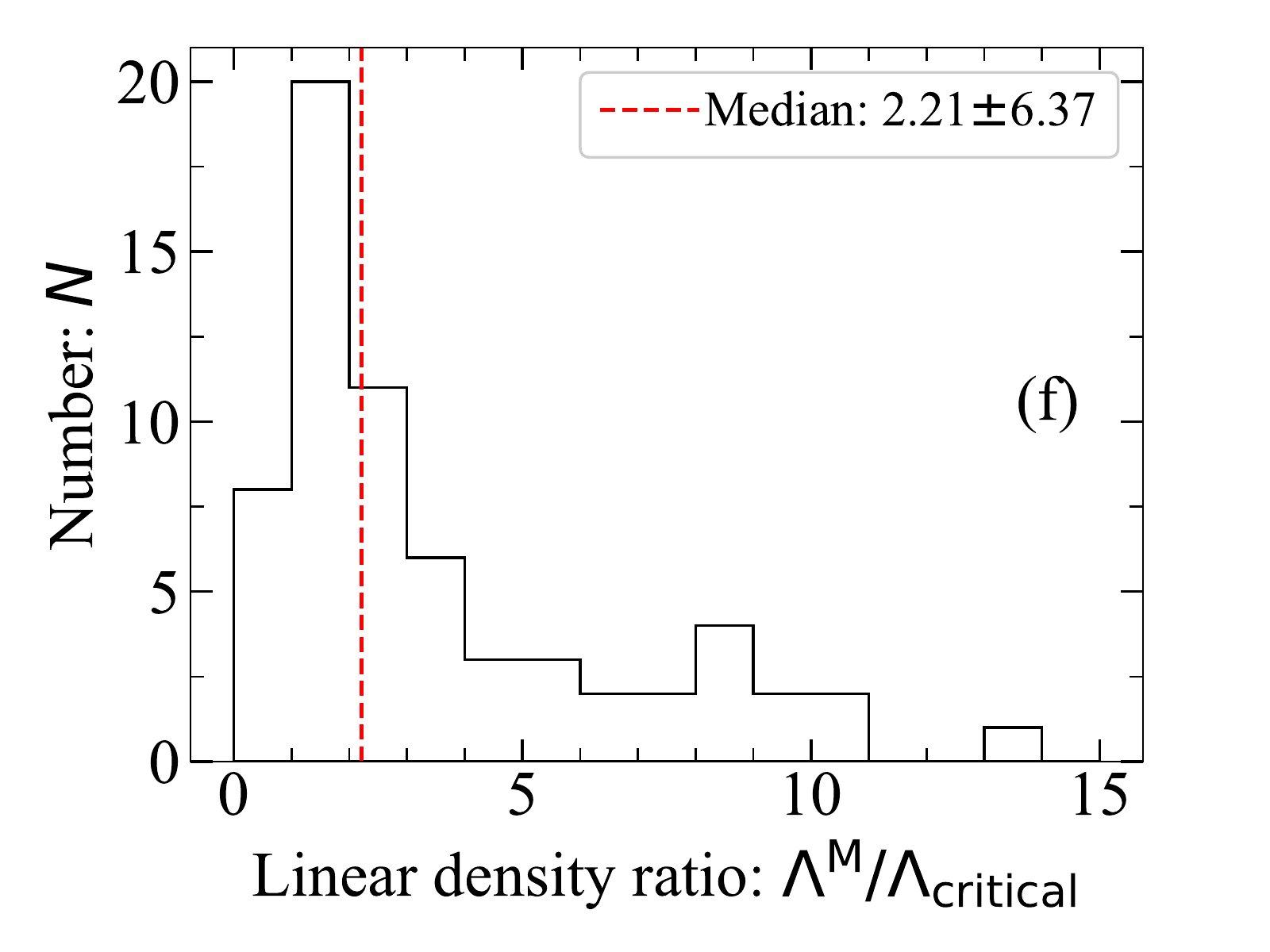}
        \caption{        
        Histograms of the physical parameters of the 68 selected filaments in Vela C. (\emph{a}): filament 
        contrasts, (\emph{b}): filament widths, (\emph{c}): logarithmic slope of profile, (\emph{d}): dust temperatures 
        along the filament crests, (\emph{e}): surface densities along the crest; (\emph{f}): ratio of linear density to the 
        critical value. The red dashed line indicates the median value, and the error is the rms of the data.}
        \label{filamentshist}
    \end{figure*}
    
    \begin{figure*}
        \centering
        \includegraphics[width=0.3 \textwidth]{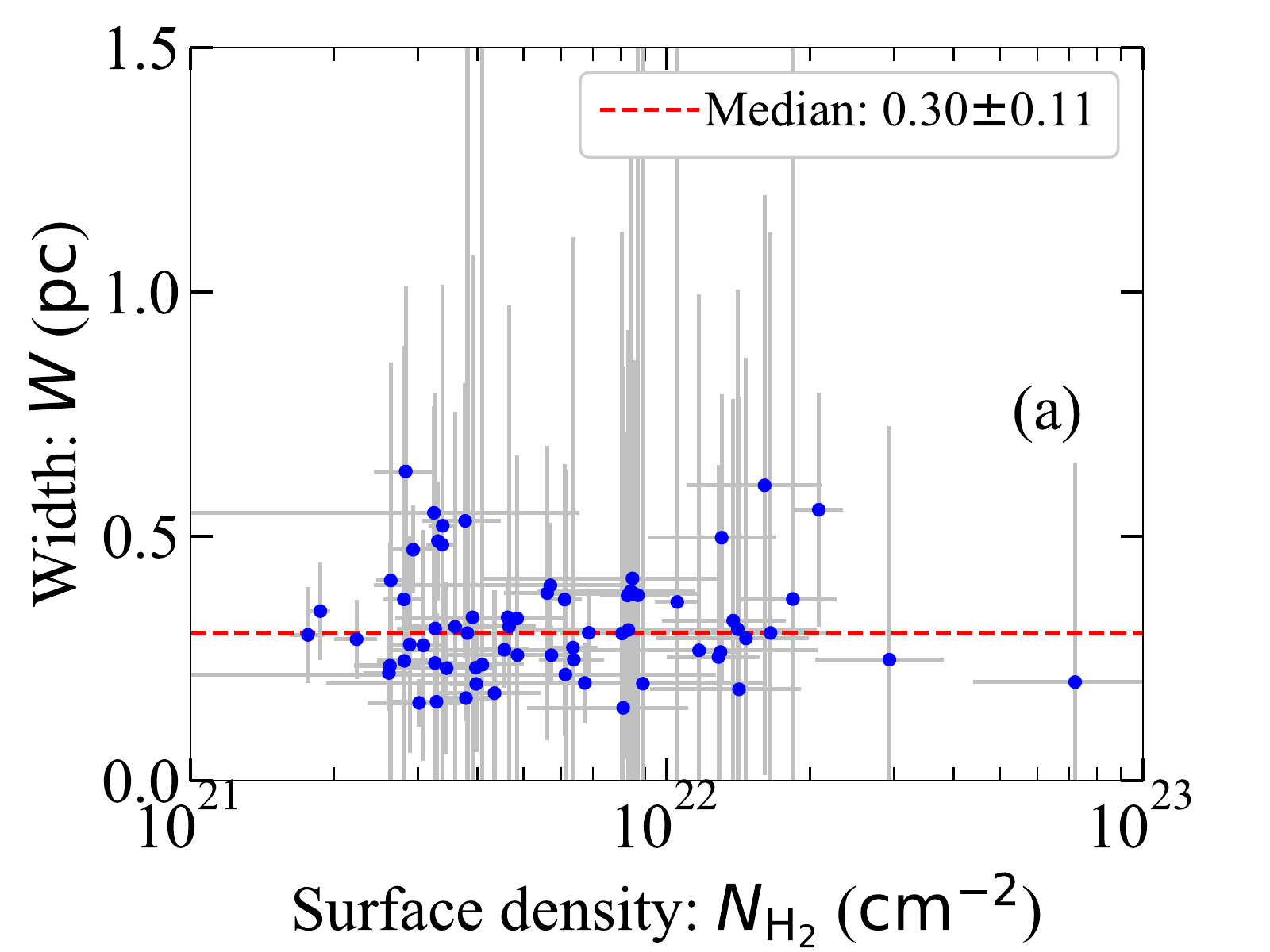}
        \includegraphics[width=0.3 \textwidth]{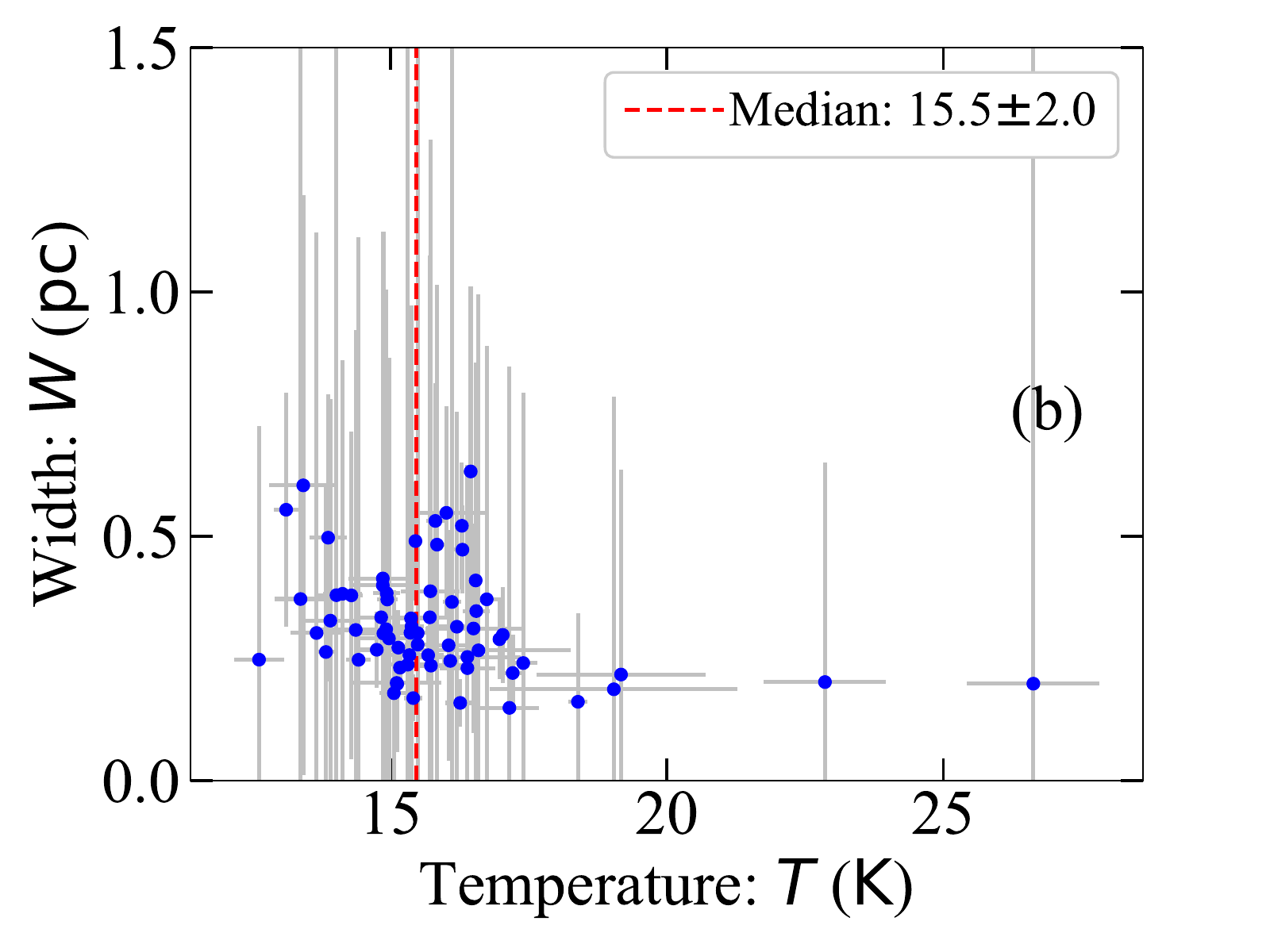}
        \includegraphics[width=0.3 \textwidth]{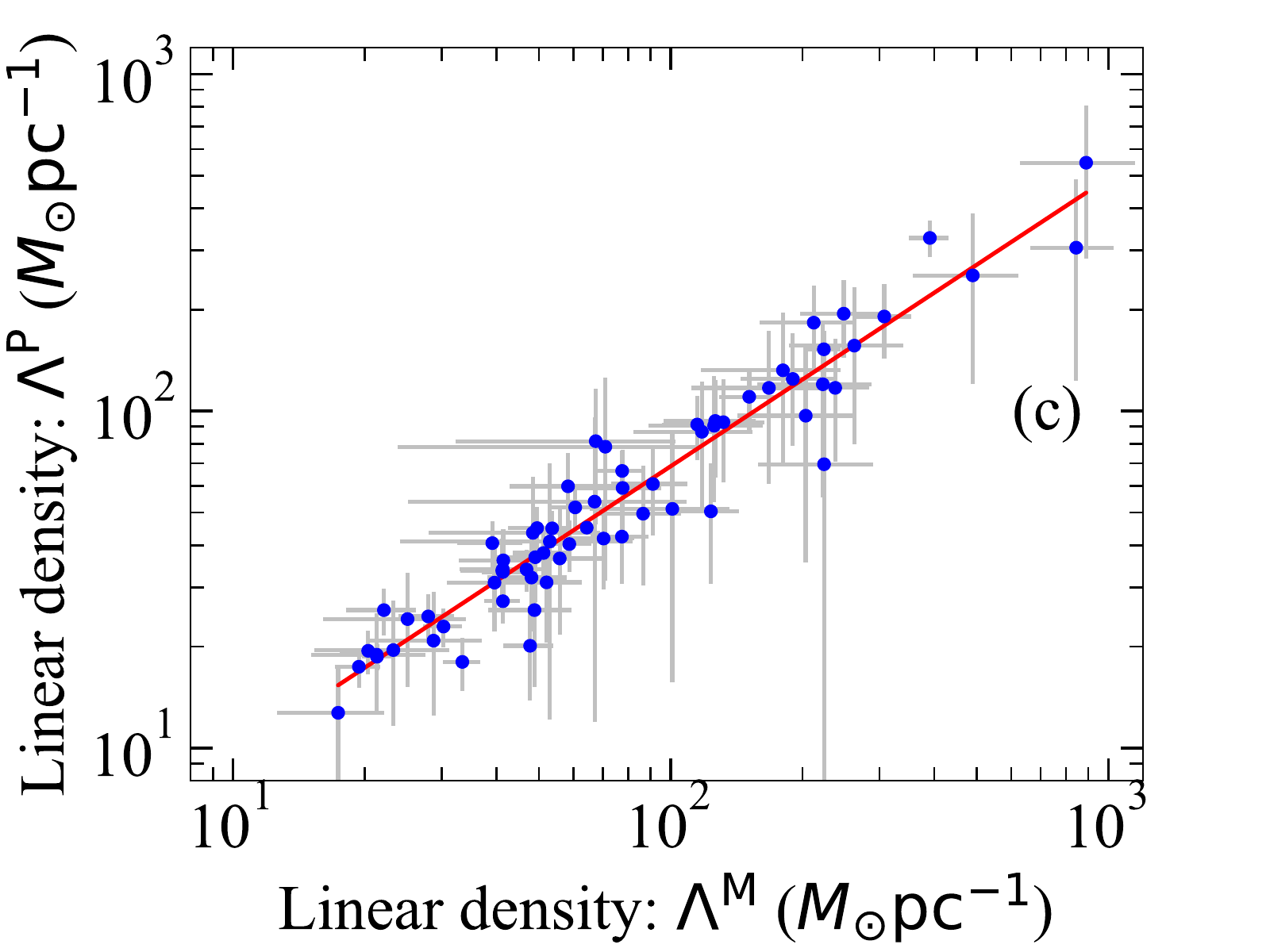}
        \includegraphics[width=0.3 \textwidth]{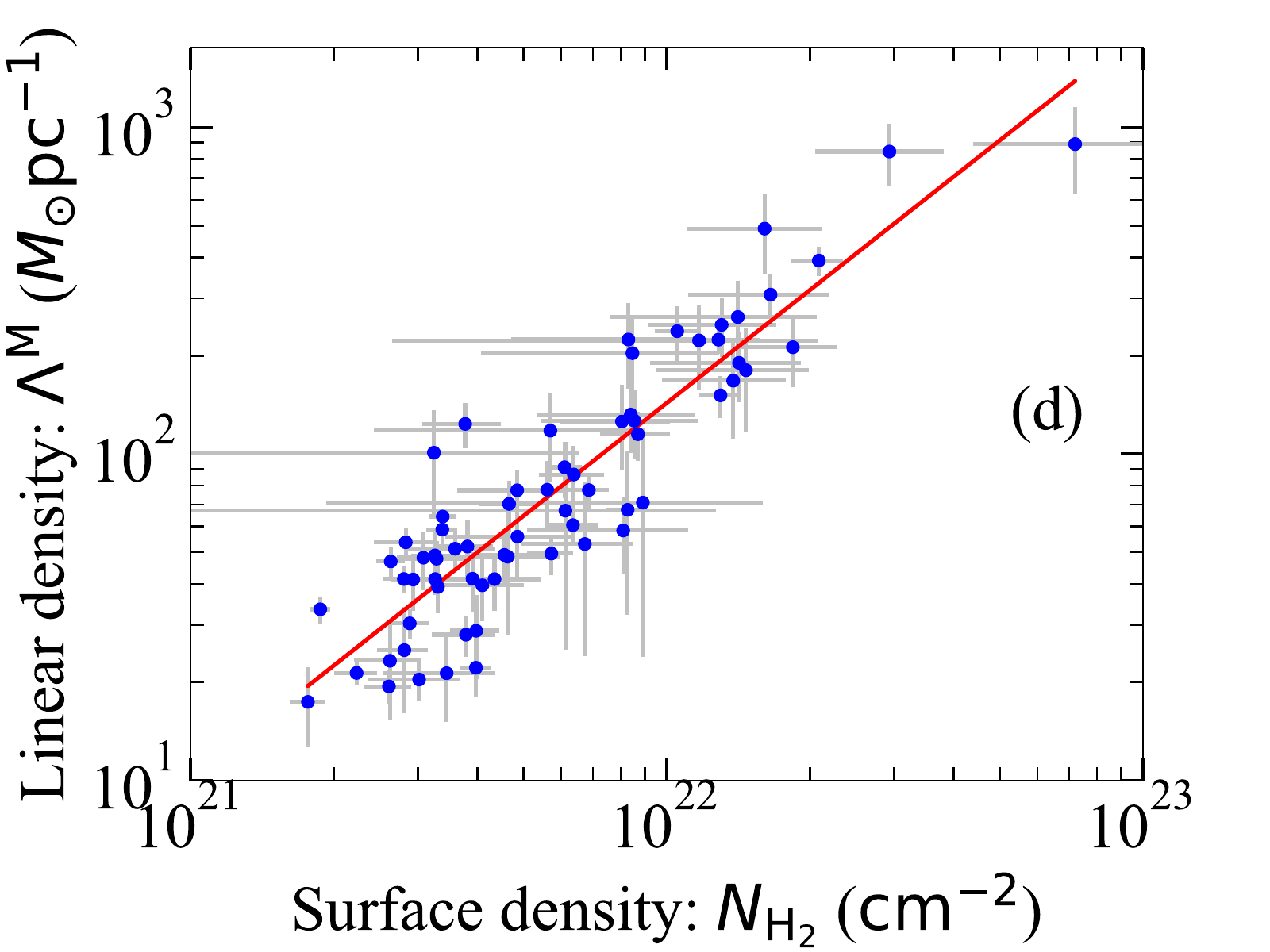}
        \includegraphics[width=0.3 \textwidth]{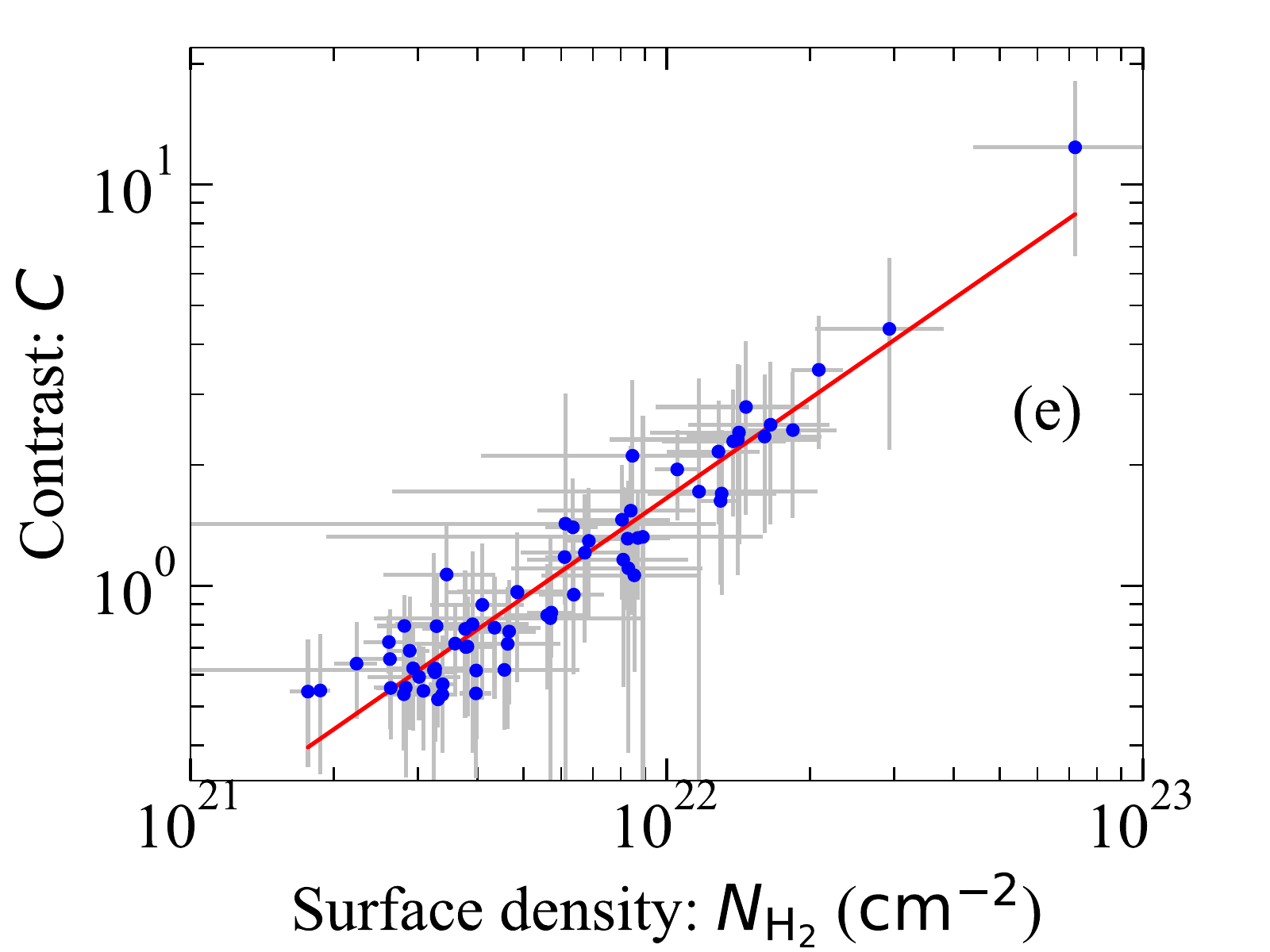}
        \includegraphics[width=0.3 \textwidth]{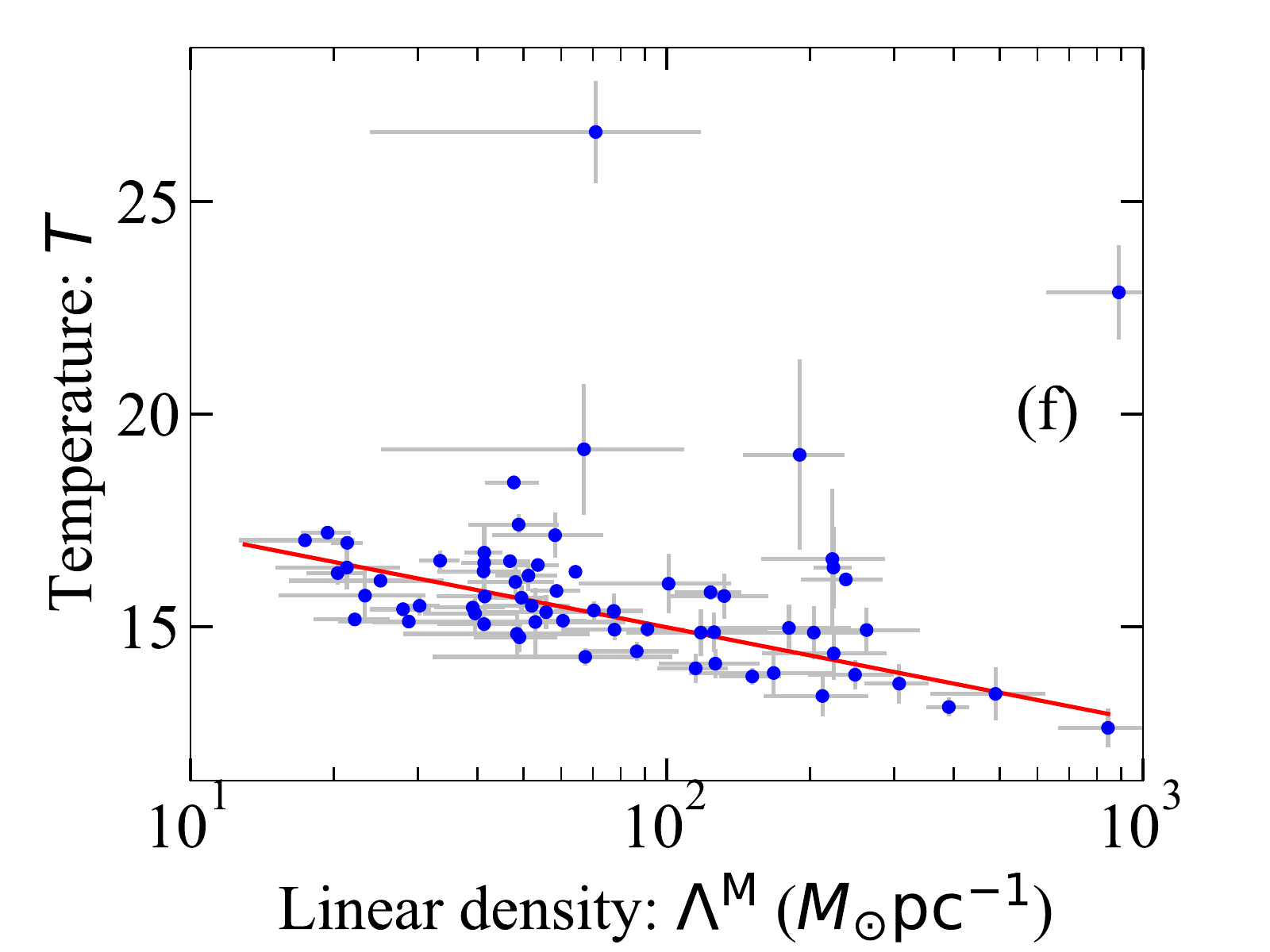}
        \caption{
        Correlations of the physical parameters of the 68 filaments in Vela C. The red dashed lines indicate the 
        median values and the red solid lines represent fits to the data: $\log \Lambda^{\rm P} = 0.86\log \Lambda^{\rm M} + 0.13$ 
        ($r = 0.96$), $\log \Lambda^{\rm M} = 1.15 \log N_{\rm {H}_2} - 23.16$ 
        ($r = 0.91$), $\log C = 0.82 \log N_{\rm {H}_2} - 17.92$ ($r = 0.95$), $T = - 2.20 \log \Lambda^{\rm M} + 19.38$ 
        ($r = 0.77 $), where $r$ is the Pearson product moment correlation coefficient.}
        \label{filamentcorrelation}
    \end{figure*}

\subsection{Physical properties of filaments}\label{sec:fil_extraction}

Molecular filaments are the strongly elongated structures in molecular clouds \citep[e.g.,][]{Men2010,Andre2014}. The omnipresent
filaments are usually blended with each other, sources, and fluctuating background clouds and the degree of blending becomes more
severe at lower angular resolutions \citep{Men2021method}. Unfortunately, there are no reliable methods to accurately subtract
backgrounds and deblend filaments from other structures, in order to properly measure their physical parameters. Therefore, it
would be preferred to choose clear examples of the isolated (not blended), relatively straight filaments on simple backgrounds
that would enable more accurate measurements of their properties and more reliable conclusions.

In practice, we used the simple (non-branching) global skeletons, produced by \emph{getsf} and measured the
11.7{\arcsec}-resolution surface densities of the component of filaments, separated from the sources and backgrounds. The contrast
$C$ of a source- and background-subtracted filament can be defined as
\begin{equation}
C = {N_{\rm H_{2},F}} / {N_{\rm H_{2},B}}
\end{equation}
where ${N_{\rm H_{2},F}}$ is the median value of the surface densities along the filament crest (skeleton) and ${N_{\rm H_{2},B}}$
the background surface density.
After a visual inspection (Fig.~\ref{filamentexample}), only the prominent filaments with $C
\gtrsim 0.5$ and length $L \gtrsim 0.4$ pc were used in the measurements of their profiles. We selected 68 filaments, in which at
least one side does not blend with the adjacent structures. A histogram of physical parameters of the 68 selected filamentary
filaments is shown in Fig.~\ref{filamentshist} and correlations of some parameters are displayed in Fig.~\ref{filamentcorrelation}.

Let us denote $W$ the median half-maximum width of the surface density profiles of a filament
along its length and $W_{\rm B}$ the width, measured on the broader side and $W_{\rm N}$ the width, measured on
the narrower side. When the two one-sided estimates are only moderately different (by less than 50\%), we choose their arithmetic
mean as the filament width $W$. Otherwise, we set the filament width $W = W_{\rm N}$. The widths are in the range of 0.15 to 0.63
pc (median of $0.3\pm 0.11$ pc) and they have no correlation with the filament temperature and surface density
(Fig.~\ref{filamentcorrelation}). The filament lengths $L$, defined as the length of the skeletons, are in the
range of 0.35 to 2.14 pc (median of 0.7 pc). The median filament dust temperatures are in the range of 12.6
to 26.6 K (median of 15.5 K), measured on the crest in the temperature map.

Three estimates of the linear density $\Lambda$ of each filament can be computed, from the average integrated
profile along the crest ($\Lambda^{\rm P}$), from the average width ($\Lambda^{\rm W}$), and from the filament mass 
($\Lambda^{\rm M}$),
\begin{eqnarray} 
\left.\begin{aligned}
\label{mlineP}
&\Lambda^{\rm P} = \mu_{\rm H_{2}} m_{\rm H} \,\langle \int N_{\rm H_{2}}(r,x)\,{\rm d}r \,\rangle_{x}, \\
&\Lambda^{\rm W} = \mu_{\rm H_{2}} m_{\rm H} \,N_{\rm H_{2}}^{0} \,\langle W \rangle_{x}, \\
&\Lambda^{\rm M} = M_{\rm F} / L,
\end{aligned}\right.
\end{eqnarray} 
where the averaging is done over the coordinate $x$ along the filament crest, $N_{\rm H_{2}}^{0}$ is the crest surface
density, and the filament mass $M_{\rm F}$ is obtained by integrating the surface densities over the filament footprint. The
linear densities span more than an order of magnitude and are clearly correlated with some other parameters
(Fig.~\ref{filamentcorrelation}). Table \ref{filamentpara} summarizes the derived physical parameters of the filaments that include their lengths, masses, linear densities, and widths.

 With the assumption that filaments can be described by the infinitely-long linear isothermal hydrostatic
non-magnetic cylinders \citep{Stodolkiewicz1963,Ostriker1964}, their critical linear density has a very simple expression
$\Lambda_{\rm cr} = 2\, c_{\rm s}^2/G$ that depends on the sound speed (gas temperature) in the filament. Observed filaments with
linear densities above the critical value are usually considered unstable, fragmenting into cores under their own gravity
\citep[e.g.,][]{Andre2014,Zhanggy2020}.

Applicability of the above instability criterion to the real observations is, however, completely unclear. This is
because the properties of the complex filaments observed with \emph{Herschel} (e.g., Figs.~\ref{nh2map}, \ref{coresinmap}, and
\ref{filamentexample}) show that the assumptions of the highly idealized model are invalid. Indeed, the filaments do not resemble a
linear cylinder, they are finite in lengths (actually relatively short) and non-isothermal, they have non-negligible magnetic
fields, and there is no clear indication that they are in a hydrostatic equilibrium. The observed filamentary structures appear
embedded in very complex, inhomogeneous environments (backgrounds), they are quite substantially curved, overlap with other nearby
structures (filaments and sources), they have temperature gradients toward their centers, as well as significantly variable radial
profiles and widths, hence linear densities, along their crests \citep[e.g., Figs.~6, 13, 19--23 in][]{Men2021method}. It is quite
reasonable to conclude, therefore, that the critical linear density $\Lambda_{\rm cr}$, based on the simplistic model assumptions,
must be considered only a rough (order-of-magnitude) indicator that they perhaps might be unstable.

In the absence of a more realistic criterion, we analyzed the linear densities of the extracted filaments in Vela
C, calculating the critical linear density using the median value of the dust temperature along each filament crest and assuming
that $\Lambda_{\rm cr}$ is uncertain within a factor of 3. The linear densities $\Lambda^{\rm W}$ are shown in
Fig.~\ref{coresinmap} for each each point along the crests of the entire network of the filamentary structures. A distribution of
filaments over the stability parameter $\Lambda^{\rm M} / \Lambda_{\rm cr}$ is also shown in Fig.~\ref{filamentshist}. The peak of
the distribution is contained within the uncertainties $0.33 < \Lambda^{\rm M} / \Lambda_{\rm cr} < 3$, which means that most of
the filaments (39 of 68) may well be both supercritical and subcritical. There is also a tail of 29 filaments with the parameter
values above $3$ may definitely be considered supercritical, in which are found 94 prestellar cores, 83 protostellar cores, and
only 1 unbound starless core. On the other hand, the filaments with the stability parameter above $1$ contain 210 prestellar cores,
127 protostellar cores, and 8 unbound starless cores. In other words, taking into account the uncertainties, the supercritical
filaments contain only prestellar and protostellar cores.

\section{Discussion}\label{sec:discussion}

\subsection{Extracted sources at different resolutions}

Insufficient angular resolutions are known to make extracted sources artificially wider and more massive
\citep{Louvet2021,Men2023}, because of the heavier blending with the structured backgrounds and incorporation of the background
into the extracted source, caused by the errors in background determination. Convergence tests at lower angular resolutions might
indicate, whether a given resolution is sufficiently high to produce accurate measurements \citep{Louvet2021}. To check for
possible resolution effects, we measured properties of the extracted cores in the available surface densities at the resolutions of
8.5, 11.7 and 18.2{\arcsec}, selecting only isolated sources for the distributions of source sizes in Fig.~\ref{coreR} to
avoid considerable inaccuracies, associated with deblending overlapping sources.

The starless (prestellar) cores are consistently well-resolved at all three resolutions (Fig.~\ref{starlesscore}),
whereas the protostellar cores appear partially-resolved at 18.2{\arcsec} and become progressively less resolved toward the highest
8.5{\arcsec} resolution (Fig.~\ref{protocore}). The differences are caused by the fact that the protostellar cores are the
accreting envelopes around protostars, generating accretion luminosity that heats up the inner opaque area of the envelopes to $T
\ga 1000$ K at radii $\theta \ll 1${\arcsec} \citep[cf. Fig.~3 in][]{Men2016fitfluxes}, making the compact hot zone appear as an
unresolved peak at 70 $\mu$m and a weaker peak at 160 $\mu$m (Fig.~\ref{protocore}). The strong temperature gradients around the
unresolved protostellar peaks make their derived surface densities inaccurate, especially the highest-resolution images
\citep[Appendix A in][]{Men2021method}.

The significantly overestimated unresolved peaks of the protostellar cores in the 11.7 and 8.5{\arcsec} resolution
surface densities provide a likely explanation, why their size distributions in Fig.~\ref{coreR} become increasingly unresolved at
higher resolutions. We note, however, that protostellar envelopes have power-law volume densities $\rho \propto r^{-2}$ (surface
densities $\sigma \propto \theta^{-1}$) and, therefore, their sizes $A$ must be proportional to the angular resolution $O$, if
their outer boundaries are sufficiently distant ($O \ll \Theta$). Simple numerical convolutions with Gaussian kernels confirm that
the surface density image of such a power-law structure obeys the relationship $A = 1.6 \times O$. The size distribution of
protostellar cores in Fig.~\ref{coreR} in the 18.2{\arcsec} resolution surface densities (much less affected by the inaccuracies of
pixel SED fitting) is indeed centered at sizes approximately $1.3 \times 18.2${\arcsec}.

\subsection{Star formation and the CMF}

Our analysis of the source extraction in Vela C resulted in the massive CMF that peaks at 5 $M_{\odot}$ and
extends to 200 $M_{\odot}$, with a slope of $-1.35 \pm 0.16$ above 20 $M_{\odot}$, which happened to have the same value as the one
proposed by \cite{Salpeter1955} for an initial mass function (IMF) of stars (from 0.4 to 10 $M_{\odot}$). The cores extracted in
this work are much more massive than the low-mass stars analyzed in the more abundant studies of the nearby regions of low-mass
star formation. We constructed mass functions for the starless, prestellar, and protostellar cores. The CMF of protostellar cores
can have a more complex shape, with a peak or plateau at intermediate masses, and a steep drop-off at high masses. This is because
the protostar formation process tends to concentrate mass in a smaller range of core masses, and can also be affected by feedback
mechanisms from the newly formed protostar, such as outflows and radiation \citep{McKee2002,Bate2003,Enoch2008}.

There is a debate in the literature of whether the IMF and CMF have the same distribution. Some observations from
the ground-based single-aperture telescopes \citep[e.g.,][]{Motte1998,Alves2007} and \emph{Herschel} \citep[e.g.,][]{Konyves2015}
have shown that there are similar mass distributions in the CMF and IMF in star-forming regions. There are also cases within the
Galaxy, where the CMF and IMF are dissimilar \citep[e.g.,][]{Li2007,Motte2018,Zhang2018}. The similarity of the two is often
interpreted as indicating that there may be a constant core-to-star conversion efficiency \citep{Alves2007}. A recent ALMA study of
the MM2 and MM3 clouds in W43 \citep{Pouteau2022} found an excess of high-mass cores and a significantly shallower CMF slope. There
are also indications that the CMF peak may shift to the lower masses with increasing angular resolution of the images
\citep{Louvet2021}.

The mechanisms of high-mass star formation are not very clear yet. One scenario is that the massive, dense cores
collapse gravitationally and form high-mass stars, because of their higher mass reservoirs and stronger gravitational potentials
\citep{McKee2003,Tan2014a}. Other scenarios include disk fragmentation, competitive accretion, and mergers of the lower-mass
protostars \citep[e.g.,][]{Bonnell2004,Bate2009,Kruijssen2012}. The similarity of the CMF and IMF slopes in the high-mass end
may be suggestive of a possible link between the formation of high-mass stars and the formation of their progenitors, the dense
cores. More studies are necessary to reach a reliable understanding of the CMFs and their relationship with the IMF, including
independent source and filament extractions and re-evaluation of the existing reach dataset of the \emph{Herschel} observations of
both low- and high-mass star-forming regions.

\subsection{Filamentary structures}

\emph{Herschel} discovered widespread networks of filamentary structures within the molecular clouds
\citep{Men2010}, proposed as the key actors in the formation of stars \citep{Andre2014}. The dense cores, formed by the
gravitational fragmentation of the supercritical filaments, are thought to be the progenitors of the next generations of stars
\citep[e.g.,][]{Zhanggy2020}.

The filamentary structures in Vela C have crest temperatures of 15.5 K with a dispersion of roughly 2 K
(Fig.~\ref{filamentshist}). Radial dust temperature profiles, created from the dust temperature map, reveal
shapes similar to an inverted surface density profile \citep{Zhanggy2020}, wherever the filament does not contain a hot source. The
higher the surface or linear densities, the lower the crest temperature (Fig.~\ref{filamentcorrelation}), because
of the increased shielding from the interstellar radiation field \citep{Arzoumanian2019}. The very common filament pattern is a
main filament or a ridge and a network of stripes or sub-filaments \citep[e.g.,][]{Kirk2013,Men2021method}. To facilitate
more accurate measurements, \emph{getsf} separated all branches of the skeleton network. The length of the 68
well-shaped filaments is approximately 0.5 pc, which is comparable to the lengths of the 599 filaments measured by
\citet{Arzoumanian2019}. Filamentary structures are interconnected with various nearby branches, and variations along the crest
introduce uncertainties in the width, mass, and profile measurements (Fig.~\ref{filamentexample}).

Studies of the nearby molecular clouds observed with \emph{Herschel} suggested that the widths of the filaments are independent of
the surface density and have a typical values of $\sim 0.1$ pc \citep{Arzoumanian2011, Arzoumanian2019}. Whether such a typical
width does actually exist or whether it is also applicable to the high-mass star-forming regions is still debated
\citep{Panopoulou2022,Andre2022}. The widths of the filamentary structures that we selected in this work range from 0.15 to 0.63
pc, with a median value of $0.3\pm 0.11$ pc. The angular resolution of the surface density map (11.7{\arcsec}) corresponds to $\sim
0.05$ pc at the distance of 905 pc adopted for Vela C, which means that all the filaments appear resolved. The
widths and linear masses of the filaments in our study are similar to those found in the Cygnus X high-mass star-forming region,
located at 1.4 kpc \citep{Hennemann2012}.

\section{Conclusions}\label{sec:conclusions}

Using the \emph{hires} algorithm to create high-resolution surface densities from the \emph{Herschel} observations and the
\emph{getsf} method to extract sources and filaments \citep{Men2021method}, we detected the sources and filamentary structures in
the Vela C molecular cloud and analyzed their physical properties. Our systematic study of the individual molecular cloud led
to the following main results.

The PDF$_{N}$ of the surface density map with a resolution of 11.7{\arcsec} shows a prominent power-law tail at
$A_{V}\ga 7$ mag with the power exponent of $-2.8 \pm 0.1$. The steep power law indicates that a major part of the molecular cloud
space is the low-density background, with the source and filamentary components comprising $40${\%} of the mass of the entire cloud
($4.6 \times 10^{4}$ $M_{\odot}$).


We extracted 570 reliable sources that were classified as 176 protostellar cores and 394 starless cores, which
were further interpreted as 291 self-gravitating prestellar cores and 103 unbound cores. The median values of the core masses and
dust temperatures, estimated from fitting the SEDs are 4.8 $M_{\odot }$ and 11.7 K, respectively. The starless and prestellar CMF
can be well fitted with ${\rm d}N/{\rm d}\log M\propto M^{-1.35\pm0.16}$ between 20 and 200 $M_{\odot}$, with an exponent identical
to that of the Salpeter's IMF.


We measured isolated sources on the surface density maps with angular resolutions of 8.5, 11.7 and 18.2{\arcsec}
and found that the prestellar and unbound starless cores are well resolved, especially at the highest resolutions. Their
distributions are centered on a half-maximum size of $30${\arcsec} ($0.13$ pc). Proportionality of the sizes of the protostellar
cores to the angular resolution is likely caused by both overestimated peaks in the derived surface densities and the power-law
radial density distribution of the collapsing envelopes.


We measured the 68 well-behaving, isolated filaments with contrasts $C \gtrsim 0.5$ and lengths $L \gtrsim 0.4$
pc, extracted in the surface density map of Vela C at 11.7{\arcsec} resolution. The filament widths are in the range of 0.15 to
0.63 pc, with a median of $0.3 \pm 0.11$ pc. The filament instability criterion $\Lambda > \Lambda_{\rm cr}$, based on the critical
surface density of the highly simplified model of an infinite isothermal hydrostatic cylinder, is considered in this work only a
rough, uncertain indicator. We analyzed the linear densities of the extracted filaments, assuming that the critical value is
uncertain within a factor of 3.

A majority of the filaments (39) may well be supercritical or subcritical, within the large uncertainties.
Only 29 filaments (with $\Lambda^{\rm M} > 3\,\Lambda_{\rm cr}$) may be definitely considered supercritical, in which we found 94
prestellar cores, 83 protostellar cores, and only 1 unbound starless core. Thus, the supercritical filaments contain only
prestellar and protostellar cores.
We found that $64$\% of the prestellar cores and $85${\%} of the protostellar cores are associated with the filamentary structures
of $\Lambda^{\rm M} > \Lambda_{\rm cr}$, whereas only $1$\% of the unbound starless cores are located on such filaments. The
observed close connection between the cores and filaments is consistent with the idea that the dense filamentary structures provide
the most favorable local environment for the formation of stars in molecular clouds.


\begin{acknowledgements}
We carried out this work at the China-Argentina Cooperation Station of NAOC/CAS. This work was supported by the Key Project of
International Cooperation of Ministry of Science and Technology of China through grant 2010DFA02710, and by the National Natural
Science Foundation of China through grants 11503035, 11573036, 11373009, 11433008, 11403040 and 11403041. G.Z. acknowledges support
from the Postdoctoral Science Foundation of China (No. 2021T140672), and the National Natural Science Foundation of China (No.
U2031118). Z.W. acknowledges support from the National Natural Science foundation of China (No. U1931203). The 
authors thank the anonymous referee for constructive comments that prompted improvements of this paper.
\end{acknowledgements}
\bibliographystyle{aa} 
\bibliography{ref.bib} 

\begin{thebibliography}{91}
\expandafter\ifx\csname natexlab\endcsname\relax\def\natexlab#1{#1}\fi

\bibitem[{{Alves} {et~al.}(2007){Alves}, {Lombardi}, \& {Lada}}]{Alves2007}
{Alves}, J., {Lombardi}, M., \& {Lada}, C.~J. 2007, \aap, 462, L17

\bibitem[{{Alves} {et~al.}(2001){Alves}, {Lada}, \& {Lada}}]{Alves2001}
{Alves}, J.~F., {Lada}, C.~J., \& {Lada}, E.~A. 2001, \nat, 409, 159

\bibitem[{{Andr{\'e}} {et~al.}(2014){Andr{\'e}}, {Di Francesco},
  {Ward-Thompson}, {Inutsuka}, {Pudritz}, \& {Pineda}}]{Andre2014}
{Andr{\'e}}, P., {Di Francesco}, J., {Ward-Thompson}, D., {et~al.} 2014, in
  Protostars and Planets VI, ed. H.~{Beuther}, R.~S. {Klessen}, C.~P.
  {Dullemond}, \& T.~{Henning}, 27

\bibitem[{{Andr{\'e}} {et~al.}(2010){Andr{\'e}}, {Men'shchikov}, {Bontemps},
  {K{\"o}nyves}, {Motte}, {Schneider}, {Didelon}, {Minier}, {Saraceno},
  {Ward-Thompson}, {di Francesco}, {White}, {Molinari}, {Testi}, {Abergel},
  {Griffin}, {Henning}, {Royer}, {Mer{\'\i}n}, {Vavrek}, {Attard},
  {Arzoumanian}, {Wilson}, {Ade}, {Aussel}, {Baluteau}, {Benedettini},
  {Bernard}, {Blommaert}, {Cambr{\'e}sy}, {Cox}, {di Giorgio}, {Hargrave},
  {Hennemann}, {Huang}, {Kirk}, {Krause}, {Launhardt}, {Leeks}, {Le Pennec},
  {Li}, {Martin}, {Maury}, {Olofsson}, {Omont}, {Peretto}, {Pezzuto}, {Prusti},
  {Roussel}, {Russeil}, {Sauvage}, {Sibthorpe}, {Sicilia-Aguilar}, {Spinoglio},
  {Waelkens}, {Woodcraft}, \& {Zavagno}}]{Andre2010}
{Andr{\'e}}, P., {Men'shchikov}, A., {Bontemps}, S., {et~al.} 2010, \aap, 518,
  L102

\bibitem[{{Andre} {et~al.}(2000){Andre}, {Ward-Thompson}, \&
  {Barsony}}]{Andre2000}
{Andre}, P., {Ward-Thompson}, D., \& {Barsony}, M. 2000, in Protostars and
  Planets IV, ed. V.~{Mannings}, A.~P. {Boss}, \& S.~S. {Russell}, 59

\bibitem[{{Andr{\'e}} {et~al.}(2022){Andr{\'e}}, {Palmeirim}, \&
  {Arzoumanian}}]{Andre2022}
{Andr{\'e}}, P.~J., {Palmeirim}, P., \& {Arzoumanian}, D. 2022, \aap, 667, L1

\bibitem[{{Arzoumanian} {et~al.}(2011){Arzoumanian}, {Andr{\'e}}, {Didelon},
  {K{\"o}nyves}, {Schneider}, {Men'shchikov}, {Sousbie}, {Zavagno}, {Bontemps},
  {di Francesco}, {Griffin}, {Hennemann}, {Hill}, {Kirk}, {Martin}, {Minier},
  {Molinari}, {Motte}, {Peretto}, {Pezzuto}, {Spinoglio}, {Ward-Thompson},
  {White}, \& {Wilson}}]{Arzoumanian2011}
{Arzoumanian}, D., {Andr{\'e}}, P., {Didelon}, P., {et~al.} 2011, \aap, 529, L6

\bibitem[{{Arzoumanian} {et~al.}(2019){Arzoumanian}, {Andr{\'e}},
  {K{\"o}nyves}, {Palmeirim}, {Roy}, {Schneider}, {Benedettini}, {Didelon}, {Di
  Francesco}, \& {Kirk}}]{Arzoumanian2019}
{Arzoumanian}, D., {Andr{\'e}}, P., {K{\"o}nyves}, V., {et~al.} 2019, \aap,
  621, A42

\bibitem[{{Baba} {et~al.}(2006){Baba}, {Sato}, {Nagashima}, {Nishiyama},
  {Kato}, {Haba}, {Nagata}, {Nagayama}, {Tamura}, \& {Sugitani}}]{Baba2006}
{Baba}, D., {Sato}, S., {Nagashima}, C., {et~al.} 2006, \aj, 132, 1692

\bibitem[{{Ballesteros-Paredes} {et~al.}(2020){Ballesteros-Paredes},
  {Andr{\'e}}, {Hennebelle}, {Klessen}, {Kruijssen}, {Chevance}, {Nakamura},
  {Adamo}, \& {V{\'a}zquez-Semadeni}}]{Javier2020}
{Ballesteros-Paredes}, J., {Andr{\'e}}, P., {Hennebelle}, P., {et~al.} 2020,
  \ssr, 216, 76

\bibitem[{{Ballesteros-Paredes} {et~al.}(2003){Ballesteros-Paredes}, {Klessen},
  \& {V{\'a}zquez-Semadeni}}]{Ballesteros2003}
{Ballesteros-Paredes}, J., {Klessen}, R.~S., \& {V{\'a}zquez-Semadeni}, E.
  2003, \apj, 592, 188

\bibitem[{{Ballesteros-Paredes} {et~al.}(2011){Ballesteros-Paredes},
  {V{\'a}zquez-Semadeni}, {Gazol}, {Hartmann}, {Heitsch}, \&
  {Col{\'\i}n}}]{Ballesteros2011}
{Ballesteros-Paredes}, J., {V{\'a}zquez-Semadeni}, E., {Gazol}, A., {et~al.}
  2011, \mnras, 416, 1436

\bibitem[{{Bate}(2009)}]{Bate2009}
{Bate}, M.~R. 2009, \mnras, 392, 590

\bibitem[{{Bate} {et~al.}(2003){Bate}, {Bonnell}, \& {Bromm}}]{Bate2003}
{Bate}, M.~R., {Bonnell}, I.~A., \& {Bromm}, V. 2003, \mnras, 339, 577

\bibitem[{{Bergin} \& {Tafalla}(2007)}]{Bergin2007}
{Bergin}, E.~A. \& {Tafalla}, M. 2007, \araa, 45, 339

\bibitem[{{Bernard} {et~al.}(2010){Bernard}, {Paradis}, {Marshall}, {Montier},
  {Lagache}, {Paladini}, {Veneziani}, {Brunt}, {Mottram}, {Martin},
  {Ristorcelli}, {Noriega-Crespo}, {Compi{\`e}gne}, {Flagey}, {Anderson},
  {Popescu}, {Tuffs}, {Reach}, {White}, {Benedettini}, {Calzoletti},
  {Digiorgio}, {Faustini}, {Juvela}, {Joblin}, {Joncas}, {Mivilles-Deschenes},
  {Olmi}, {Traficante}, {Piacentini}, {Zavagno}, \& {Molinari}}]{Bernard2010}
{Bernard}, J.~P., {Paradis}, D., {Marshall}, D.~J., {et~al.} 2010, \aap, 518,
  L88

\bibitem[{{Bohlin} {et~al.}(1978){Bohlin}, {Savage}, \& {Drake}}]{Bohlin1978}
{Bohlin}, R.~C., {Savage}, B.~D., \& {Drake}, J.~F. 1978, \apj, 224, 132

\bibitem[{{Bonnell} {et~al.}(2004){Bonnell}, {Vine}, \& {Bate}}]{Bonnell2004}
{Bonnell}, I.~A., {Vine}, S.~G., \& {Bate}, M.~R. 2004, \mnras, 349, 735

\bibitem[{{Bonnor}(1956)}]{Bonnor1956}
{Bonnor}, W.~B. 1956, \mnras, 116, 351

\bibitem[{{Chabrier}(2005)}]{Chabrier2005}
{Chabrier}, G. 2005, in Astrophysics and Space Science Library, Vol. 327, The
  Initial Mass Function 50 Years Later, ed. E.~{Corbelli}, F.~{Palla}, \&
  H.~{Zinnecker}, 41

\bibitem[{{Ebert}(1955)}]{Ebert1955}
{Ebert}, R. 1955, \zap, 37, 217

\bibitem[{{Enoch} {et~al.}(2008){Enoch}, {Evans}, {Sargent}, {Glenn},
  {Rosolowsky}, \& {Myers}}]{Enoch2008}
{Enoch}, M.~L., {Evans}, Neal~J., I., {Sargent}, A.~I., {et~al.} 2008, \apj,
  684, 1240

\bibitem[{{Federrath} {et~al.}(2008){Federrath}, {Klessen}, \&
  {Schmidt}}]{Federrath2008}
{Federrath}, C., {Klessen}, R.~S., \& {Schmidt}, W. 2008, \apjl, 688, L79

\bibitem[{{Gaia Collaboration} {et~al.}(2018){Gaia Collaboration}, {Brown},
  {Vallenari}, {Prusti}, {de Bruijne}, {Babusiaux}, {Bailer-Jones}, {Biermann},
  {Evans}, \& {Eyer}}]{Gaia2018}
{Gaia Collaboration}, {Brown}, A.~G.~A., {Vallenari}, A., {et~al.} 2018, \aap,
  616, A1

\bibitem[{{Galli} {et~al.}(2002){Galli}, {Walmsley}, \&
  {Gon{\c{c}}alves}}]{Galli2002}
{Galli}, D., {Walmsley}, M., \& {Gon{\c{c}}alves}, J. 2002, \aap, 394, 275

\bibitem[{{Giannini} {et~al.}(2012){Giannini}, {Elia}, {Lorenzetti},
  {Molinari}, {Motte}, {Schisano}, {Pezzuto}, {Pestalozzi}, {Di Giorgio},
  {Andr{\'e}}, {Hill}, {Benedettini}, {Bontemps}, {Di Francesco}, {Fallscheer},
  {Hennemann}, {Kirk}, {Minier}, {Nguyen Luong}, {Polychroni}, {Rygl},
  {Saraceno}, {Schneider}, {Spinoglio}, {Testi}, {Ward-Thompson}, \&
  {White}}]{Giannini2012}
{Giannini}, T., {Elia}, D., {Lorenzetti}, D., {et~al.} 2012, \aap, 539, A156

\bibitem[{{Griffin} {et~al.}(2010){Griffin}, {Abergel}, {Abreu}, {Ade},
  {Andr{\'e}}, {Augueres}, {Babbedge}, {Bae}, {Baillie}, {Baluteau}, {Barlow},
  {Bendo}, {Benielli}, {Bock}, {Bonhomme}, {Brisbin}, {Brockley-Blatt},
  {Caldwell}, {Cara}, {Castro-Rodriguez}, {Cerulli}, {Chanial}, {Chen},
  {Clark}, {Clements}, {Clerc}, {Coker}, {Communal}, {Conversi}, {Cox},
  {Crumb}, {Cunningham}, {Daly}, {Davis}, {de Antoni}, {Delderfield}, {Devin},
  {di Giorgio}, {Didschuns}, {Dohlen}, {Donati}, {Dowell}, {Dowell}, {Duband},
  {Dumaye}, {Emery}, {Ferlet}, {Ferrand}, {Fontignie}, {Fox}, {Franceschini},
  {Frerking}, {Fulton}, {Garcia}, {Gastaud}, {Gear}, {Glenn}, {Goizel},
  {Griffin}, {Grundy}, {Guest}, {Guillemet}, {Hargrave}, {Harwit}, {Hastings},
  {Hatziminaoglou}, {Herman}, {Hinde}, {Hristov}, {Huang}, {Imhof}, {Isaak},
  {Israelsson}, {Ivison}, {Jennings}, {Kiernan}, {King}, {Lange}, {Latter},
  {Laurent}, {Laurent}, {Leeks}, {Lellouch}, {Levenson}, {Li}, {Li},
  {Lilienthal}, {Lim}, {Liu}, {Lu}, {Madden}, {Mainetti}, {Marliani}, {McKay},
  {Mercier}, {Molinari}, {Morris}, {Moseley}, {Mulder}, {Mur}, {Naylor},
  {Nguyen}, {O'Halloran}, {Oliver}, {Olofsson}, {Olofsson}, {Orfei}, {Page},
  {Pain}, {Panuzzo}, {Papageorgiou}, {Parks}, {Parr-Burman}, {Pearce},
  {Pearson}, {P{\'e}rez-Fournon}, {Pinsard}, {Pisano}, {Podosek}, {Pohlen},
  {Polehampton}, {Pouliquen}, {Rigopoulou}, {Rizzo}, {Roseboom}, {Roussel},
  {Rowan-Robinson}, {Rownd}, {Saraceno}, {Sauvage}, {Savage}, {Savini},
  {Sawyer}, {Scharmberg}, {Schmitt}, {Schneider}, {Schulz}, {Schwartz},
  {Shafer}, {Shupe}, {Sibthorpe}, {Sidher}, {Smith}, {Smith}, {Smith},
  {Spencer}, {Stobie}, {Sudiwala}, {Sukhatme}, {Surace}, {Stevens}, {Swinyard},
  {Trichas}, {Tourette}, {Triou}, {Tseng}, {Tucker}, {Turner}, {Vaccari},
  {Valtchanov}, {Vigroux}, {Virique}, {Voellmer}, {Walker}, {Ward}, {Waskett},
  {Weilert}, {Wesson}, {White}, {Whitehouse}, {Wilson}, {Winter}, {Woodcraft},
  {Wright}, {Xu}, {Zavagno}, {Zemcov}, {Zhang}, \& {Zonca}}]{Griffin2010}
{Griffin}, M.~J., {Abergel}, A., {Abreu}, A., {et~al.} 2010, \aap, 518, L3

\bibitem[{{Hennemann} {et~al.}(2012){Hennemann}, {Motte}, {Schneider},
  {Didelon}, {Hill}, {Arzoumanian}, {Bontemps}, {Csengeri}, {Andr{\'e}},
  {Konyves}, {Louvet}, {Marston}, {Men'shchikov}, {Minier}, {Nguyen Luong},
  {Palmeirim}, {Peretto}, {Sauvage}, {Zavagno}, {Anderson}, {Bernard}, {Di
  Francesco}, {Elia}, {Li}, {Martin}, {Molinari}, {Pezzuto}, {Russeil}, {Rygl},
  {Schisano}, {Spinoglio}, {Sousbie}, {Ward-Thompson}, \&
  {White}}]{Hennemann2012}
{Hennemann}, M., {Motte}, F., {Schneider}, N., {et~al.} 2012, \aap, 543, L3

\bibitem[{{Heyer} \& {Dame}(2015)}]{Heyer2015}
{Heyer}, M. \& {Dame}, T.~M. 2015, \araa, 53, 583

\bibitem[{{Heyer} \& {Terebey}(1998)}]{Heyer1998}
{Heyer}, M.~H. \& {Terebey}, S. 1998, \apj, 502, 265

\bibitem[{{Hill} {et~al.}(2011){Hill}, {Motte}, {Didelon}, {Bontemps},
  {Minier}, {Hennemann}, {Schneider}, {Andr{\'e}}, {Men'shchikov}, {Anderson},
  {Arzoumanian}, {Bernard}, {di Francesco}, {Elia}, {Giannini}, {Griffin},
  {K{\"o}nyves}, {Kirk}, {Marston}, {Martin}, {Molinari}, {Nguyen Luong},
  {Peretto}, {Pezzuto}, {Roussel}, {Sauvage}, {Sousbie}, {Testi},
  {Ward-Thompson}, {White}, {Wilson}, \& {Zavagno}}]{Hill2011}
{Hill}, T., {Motte}, F., {Didelon}, P., {et~al.} 2011, \aap, 533, A94

\bibitem[{{Johnstone} {et~al.}(2000){Johnstone}, {Wilson}, {Moriarty-Schieven},
  {Joncas}, {Smith}, {Gregersen}, \& {Fich}}]{Johnstone2000}
{Johnstone}, D., {Wilson}, C.~D., {Moriarty-Schieven}, G., {et~al.} 2000, \apj,
  545, 327

\bibitem[{{Kainulainen} {et~al.}(2011){Kainulainen}, {Beuther}, {Banerjee},
  {Federrath}, \& {Henning}}]{Kainulainen2011}
{Kainulainen}, J., {Beuther}, H., {Banerjee}, R., {Federrath}, C., \&
  {Henning}, T. 2011, \aap, 530, A64

\bibitem[{{Kainulainen} {et~al.}(2009){Kainulainen}, {Beuther}, {Henning}, \&
  {Plume}}]{Kainulainen2009}
{Kainulainen}, J., {Beuther}, H., {Henning}, T., \& {Plume}, R. 2009, \aap,
  508, L35

\bibitem[{{Kauffmann} {et~al.}(2010){Kauffmann}, {Pillai}, {Shetty}, {Myers},
  \& {Goodman}}]{Kauffmann2010}
{Kauffmann}, J., {Pillai}, T., {Shetty}, R., {Myers}, P.~C., \& {Goodman},
  A.~A. 2010, \apj, 716, 433

\bibitem[{{Kirk} {et~al.}(2013){Kirk}, {Myers}, {Bourke}, {Gutermuth},
  {Hedden}, \& {Wilson}}]{Kirk2013}
{Kirk}, H., {Myers}, P.~C., {Bourke}, T.~L., {et~al.} 2013, \apj, 766, 115

\bibitem[{{Kirk} {et~al.}(2005){Kirk}, {Ward-Thompson}, \&
  {Andr{\'e}}}]{Kirk2005}
{Kirk}, J.~M., {Ward-Thompson}, D., \& {Andr{\'e}}, P. 2005, \mnras, 360, 1506

\bibitem[{{K{\"o}nyves} {et~al.}(2015){K{\"o}nyves}, {Andr{\'e}},
  {Men'shchikov}, {Palmeirim}, {Arzoumanian}, {Schneider}, {Roy}, {Didelon},
  {Maury}, {Shimajiri}, {Di Francesco}, {Bontemps}, {Peretto}, {Benedettini},
  {Bernard}, {Elia}, {Griffin}, {Hill}, {Kirk}, {Ladjelate}, {Marsh}, {Martin},
  {Motte}, {Nguy{\^e}n Luong}, {Pezzuto}, {Roussel}, {Rygl}, {Sadavoy},
  {Schisano}, {Spinoglio}, {Ward-Thompson}, \& {White}}]{Konyves2015}
{K{\"o}nyves}, V., {Andr{\'e}}, P., {Men'shchikov}, A., {et~al.} 2015, \aap,
  584, A91

\bibitem[{{Kramer} {et~al.}(1998){Kramer}, {Stutzki}, {Rohrig}, \&
  {Corneliussen}}]{Kramer1998}
{Kramer}, C., {Stutzki}, J., {Rohrig}, R., \& {Corneliussen}, U. 1998, \aap,
  329, 249

\bibitem[{{Kritsuk} {et~al.}(2011){Kritsuk}, {Norman}, \&
  {Wagner}}]{Kritsuk2011}
{Kritsuk}, A.~G., {Norman}, M.~L., \& {Wagner}, R. 2011, \apjl, 727, L20

\bibitem[{{Kruijssen} {et~al.}(2012){Kruijssen}, {Pelupessy}, {Lamers},
  {Portegies Zwart}, {Bastian}, \& {Icke}}]{Kruijssen2012}
{Kruijssen}, J.~M.~D., {Pelupessy}, F.~I., {Lamers}, H. J.~G.~L.~M., {et~al.}
  2012, \mnras, 421, 1927

\bibitem[{{Li} \& {Goldsmith}(2012)}]{Li2012}
{Li}, D. \& {Goldsmith}, P.~F. 2012, \apj, 756, 12

\bibitem[{{Li} {et~al.}(2007){Li}, {Velusamy}, {Goldsmith}, \&
  {Langer}}]{Li2007}
{Li}, D., {Velusamy}, T., {Goldsmith}, P.~F., \& {Langer}, W.~D. 2007, \apj,
  655, 351

\bibitem[{{Liseau} {et~al.}(1992){Liseau}, {Lorenzetti}, {Nisini}, {Spinoglio},
  \& {Moneti}}]{Liseau1992}
{Liseau}, R., {Lorenzetti}, D., {Nisini}, B., {Spinoglio}, L., \& {Moneti}, A.
  1992, \aap, 265, 577

\bibitem[{{Lorenzetti} {et~al.}(1993){Lorenzetti}, {Spinoglio}, \&
  {Liseau}}]{Lorenzetti1993}
{Lorenzetti}, D., {Spinoglio}, L., \& {Liseau}, R. 1993, \aap, 275, 489

\bibitem[{{Louvet} {et~al.}(2021){Louvet}, {Hennebelle}, {Men'shchikov},
  {Didelon}, {Ntormousi}, \& {Motte}}]{Louvet2021}
{Louvet}, F., {Hennebelle}, P., {Men'shchikov}, A., {et~al.} 2021, \aap, 653,
  A157

\bibitem[{{Massi} {et~al.}(2003){Massi}, {Lorenzetti}, \&
  {Giannini}}]{Massi2003}
{Massi}, F., {Lorenzetti}, D., \& {Giannini}, T. 2003, \aap, 399, 147

\bibitem[{{Massi} {et~al.}(2019){Massi}, {Weiss}, {Elia}, {Csengeri},
  {Schisano}, {Giannini}, {Hill}, {Lorenzetti}, {Menten}, {Olmi}, {Schuller},
  {Strafella}, {De Luca}, {Motte}, \& {Wyrowski}}]{Massi2019}
{Massi}, F., {Weiss}, A., {Elia}, D., {et~al.} 2019, \aap, 628, A110

\bibitem[{{May} {et~al.}(1988){May}, {Murphy}, \& {Thaddeus}}]{May1988}
{May}, J., {Murphy}, D.~C., \& {Thaddeus}, P. 1988, \aaps, 73, 51

\bibitem[{{McKee} \& {Tan}(2002)}]{McKee2002}
{McKee}, C.~F. \& {Tan}, J.~C. 2002, \nat, 416, 59

\bibitem[{{McKee} \& {Tan}(2003)}]{McKee2003}
{McKee}, C.~F. \& {Tan}, J.~C. 2003, \apj, 585, 850

\bibitem[{{Men'shchikov}(2013)}]{Men2013getfilaments}
{Men'shchikov}, A. 2013, \aap, 560, A63

\bibitem[{{Men'shchikov}(2016)}]{Men2016fitfluxes}
{Men'shchikov}, A. 2016, \aap, 593, A71

\bibitem[{{Men'shchikov}(2021{\natexlab{a}})}]{Men2021benchmark}
{Men'shchikov}, A. 2021{\natexlab{a}}, \aap, 654, A78

\bibitem[{{Men'shchikov}(2021{\natexlab{b}})}]{Men2021method}
{Men'shchikov}, A. 2021{\natexlab{b}}, \aap, 649, A89

\bibitem[{{Men'shchikov}(2023)}]{Men2023}
{Men'shchikov}, A. 2023, \aap, submitted

\bibitem[{{Men'shchikov} {et~al.}(2010){Men'shchikov}, {Andr{\'e}}, {Didelon},
  {K{\"o}nyves}, {Schneider}, {Motte}, {Bontemps}, {Arzoumanian}, {Attard},
  {Abergel}, {Baluteau}, {Bernard}, {Cambr{\'e}sy}, {Cox}, {di Francesco}, {di
  Giorgio}, {Griffin}, {Hargrave}, {Huang}, {Kirk}, {Li}, {Martin}, {Minier},
  {Miville-Desch{\^e}nes}, {Molinari}, {Olofsson}, {Pezzuto}, {Roussel},
  {Russeil}, {Saraceno}, {Sauvage}, {Sibthorpe}, {Spinoglio}, {Testi},
  {Ward-Thompson}, {White}, {Wilson}, {Woodcraft}, \& {Zavagno}}]{Men2010}
{Men'shchikov}, A., {Andr{\'e}}, P., {Didelon}, P., {et~al.} 2010, \aap, 518,
  L103

\bibitem[{{Men'shchikov} {et~al.}(2012){Men'shchikov}, {Andr{\'e}}, {Didelon},
  {Motte}, {Hennemann}, \& {Schneider}}]{Men2012getsources}
{Men'shchikov}, A., {Andr{\'e}}, P., {Didelon}, P., {et~al.} 2012, \aap, 542,
  A81

\bibitem[{{Minier} {et~al.}(2013){Minier}, {Tremblin}, {Hill}, {Motte},
  {Andr{\'e}}, {Lo}, {Schneider}, {Audit}, {White}, {Hennemann}, {Cunningham},
  {Deharveng}, {Didelon}, {Di Francesco}, {Elia}, {Giannini}, {Nguyen Luong},
  {Pezzuto}, {Rygl}, {Spinoglio}, {Ward-Thompson}, \& {Zavagno}}]{Minier2013}
{Minier}, V., {Tremblin}, P., {Hill}, T., {et~al.} 2013, \aap, 550, A50

\bibitem[{{Molinari} {et~al.}(2011){Molinari}, {Schisano}, {Faustini},
  {Pestalozzi}, {di Giorgio}, \& {Liu}}]{Molinari2011}
{Molinari}, S., {Schisano}, E., {Faustini}, F., {et~al.} 2011, \aap, 530, A133

\bibitem[{{Molinari} {et~al.}(2010){Molinari}, {Swinyard}, {Bally}, {Barlow},
  {Bernard}, {Martin}, {Moore}, {Noriega-Crespo}, {Plume}, {Testi}, {Zavagno},
  {Abergel}, {Ali}, {Andr{\'e}}, {Baluteau}, {Benedettini}, {Bern{\'e}},
  {Billot}, {Blommaert}, {Bontemps}, {Boulanger}, {Brand}, {Brunt}, {Burton},
  {Campeggio}, {Carey}, {Caselli}, {Cesaroni}, {Cernicharo}, {Chakrabarti},
  {Chrysostomou}, {Codella}, {Cohen}, {Compiegne}, {Davis}, {de Bernardis}, {de
  Gasperis}, {Di Francesco}, {di Giorgio}, {Elia}, {Faustini}, {Fischera},
  {Fukui}, {Fuller}, {Ganga}, {Garcia-Lario}, {Giard}, {Giardino}, {Glenn},
  {Goldsmith}, {Griffin}, {Hoare}, {Huang}, {Jiang}, {Joblin}, {Joncas},
  {Juvela}, {Kirk}, {Lagache}, {Li}, {Lim}, {Lord}, {Lucas}, {Maiolo},
  {Marengo}, {Marshall}, {Masi}, {Massi}, {Matsuura}, {Meny}, {Minier},
  {Miville-Desch{\^e}nes}, {Montier}, {Motte}, {M{\"u}ller}, {Natoli}, {Neves},
  {Olmi}, {Paladini}, {Paradis}, {Pestalozzi}, {Pezzuto}, {Piacentini},
  {Pomar{\`e}s}, {Popescu}, {Reach}, {Richer}, {Ristorcelli}, {Roy}, {Royer},
  {Russeil}, {Saraceno}, {Sauvage}, {Schilke}, {Schneider-Bontemps},
  {Schuller}, {Schultz}, {Shepherd}, {Sibthorpe}, {Smith}, {Smith},
  {Spinoglio}, {Stamatellos}, {Strafella}, {Stringfellow}, {Sturm}, {Taylor},
  {Thompson}, {Tuffs}, {Umana}, {Valenziano}, {Vavrek}, {Viti}, {Waelkens},
  {Ward-Thompson}, {White}, {Wyrowski}, {Yorke}, \& {Zhang}}]{Molinari2010}
{Molinari}, S., {Swinyard}, B., {Bally}, J., {et~al.} 2010, \pasp, 122, 314

\bibitem[{{Motte} {et~al.}(1998){Motte}, {Andre}, \& {Neri}}]{Motte1998}
{Motte}, F., {Andre}, P., \& {Neri}, R. 1998, \aap, 336, 150

\bibitem[{{Motte} {et~al.}(2018){Motte}, {Nony}, {Louvet}, {Marsh}, {Bontemps},
  {Whitworth}, {Men'shchikov}, {Nguyen Luong}, {Csengeri}, {Maury}, {Gusdorf},
  {Chapillon}, {K{\"o}nyves}, {Schilke}, {Duarte-Cabral}, {Didelon}, \&
  {Gaudel}}]{Motte2018}
{Motte}, F., {Nony}, T., {Louvet}, F., {et~al.} 2018, Nature Astronomy, 2, 478

\bibitem[{{Motte} {et~al.}(2010){Motte}, {Zavagno}, {Bontemps}, {Schneider},
  {Hennemann}, {di Francesco}, {Andr{\'e}}, {Saraceno}, {Griffin}, {Marston},
  {Ward-Thompson}, {White}, {Minier}, {Men'shchikov}, {Hill}, {Abergel},
  {Anderson}, {Aussel}, {Balog}, {Baluteau}, {Bernard}, {Cox}, {Csengeri},
  {Deharveng}, {Didelon}, {di Giorgio}, {Hargrave}, {Huang}, {Kirk}, {Leeks},
  {Li}, {Martin}, {Molinari}, {Nguyen-Luong}, {Olofsson}, {Persi}, {Peretto},
  {Pezzuto}, {Roussel}, {Russeil}, {Sadavoy}, {Sauvage}, {Sibthorpe},
  {Spinoglio}, {Testi}, {Teyssier}, {Vavrek}, {Wilson}, \&
  {Woodcraft}}]{Motte2010}
{Motte}, F., {Zavagno}, A., {Bontemps}, S., {et~al.} 2010, \aap, 518, L77

\bibitem[{{Murphy} \& {May}(1991)}]{Murphy1991}
{Murphy}, D.~C. \& {May}, J. 1991, \aap, 247, 202

\bibitem[{{Netterfield} {et~al.}(2009){Netterfield}, {Ade}, {Bock}, {Chapin},
  {Devlin}, {Griffin}, {Gundersen}, {Halpern}, {Hargrave}, {Hughes}, {Klein},
  {Marsden}, {Martin}, {Mauskopf}, {Olmi}, {Pascale}, {Patanchon}, {Rex},
  {Roy}, {Scott}, {Semisch}, {Thomas}, {Truch}, {Tucker}, {Tucker}, {Viero}, \&
  {Wiebe}}]{Netterfield2009}
{Netterfield}, C.~B., {Ade}, P. A.~R., {Bock}, J.~J., {et~al.} 2009, \apj, 707,
  1824

\bibitem[{{Ostriker}(1964)}]{Ostriker1964}
{Ostriker}, J. 1964, \apj, 140, 1056

\bibitem[{{Panopoulou} {et~al.}(2022){Panopoulou}, {Clark}, {Hacar}, {Heitsch},
  {Kainulainen}, {Ntormousi}, {Seifried}, \& {Smith}}]{Panopoulou2022}
{Panopoulou}, G.~V., {Clark}, S.~E., {Hacar}, A., {et~al.} 2022, \aap, 657, L13

\bibitem[{{Passot} \& {V{\'a}zquez-Semadeni}(1998)}]{Passot1998}
{Passot}, T. \& {V{\'a}zquez-Semadeni}, E. 1998, \pre, 58, 4501

\bibitem[{{Piazzo} {et~al.}(2015){Piazzo}, {Calzoletti}, {Faustini},
  {Pestalozzi}, {Pezzuto}, {Elia}, {di Giorgio}, \& {Molinari}}]{Piazzo2015}
{Piazzo}, L., {Calzoletti}, L., {Faustini}, F., {et~al.} 2015, \mnras, 447,
  1471

\bibitem[{{Planck Collaboration} {et~al.}(2014){Planck Collaboration},
  {Abergel}, {Ade}, {Aghanim}, {Alves}, {Aniano}, {Armitage-Caplan}, {Arnaud},
  {Ashdown}, {Atrio-Barandela}, {Aumont}, {Baccigalupi}, {Banday}, {Barreiro},
  {Bartlett}, {Battaner}, {Benabed}, {Beno{\^\i}t}, {Benoit-L{\'e}vy},
  {Bernard}, {Bersanelli}, {Bielewicz}, {Bobin}, {Bock}, {Bonaldi}, {Bond},
  {Borrill}, {Bouchet}, {Boulanger}, {Bridges}, {Bucher}, {Burigana}, {Butler},
  {Cardoso}, {Catalano}, {Chamballu}, {Chary}, {Chiang}, {Chiang},
  {Christensen}, {Church}, {Clemens}, {Clements}, {Colombi}, {Colombo},
  {Combet}, {Couchot}, {Coulais}, {Crill}, {Curto}, {Cuttaia}, {Danese},
  {Davies}, {Davis}, {de Bernardis}, {de Rosa}, {de Zotti}, {Delabrouille},
  {Delouis}, {D{\'e}sert}, {Dickinson}, {Diego}, {Dole}, {Donzelli},
  {Dor{\'e}}, {Douspis}, {Draine}, {Dupac}, {Efstathiou}, {En{\ss}lin},
  {Eriksen}, {Falgarone}, {Finelli}, {Forni}, {Frailis}, {Fraisse},
  {Franceschi}, {Galeotta}, {Ganga}, {Ghosh}, {Giard}, {Giardino},
  {Giraud-H{\'e}raud}, {Gonz{\'a}lez-Nuevo}, {G{\'o}rski}, {Gratton},
  {Gregorio}, {Grenier}, {Gruppuso}, {Guillet}, {Hansen}, {Hanson}, {Harrison},
  {Helou}, {Henrot-Versill{\'e}}, {Hern{\'a}ndez-Monteagudo}, {Herranz},
  {Hildebrandt}, {Hivon}, {Hobson}, {Holmes}, {Hornstrup}, {Hovest},
  {Huffenberger}, {Jaffe}, {Jaffe}, {Jewell}, {Joncas}, {Jones}, {Juvela},
  {Keih{\"a}nen}, {Keskitalo}, {Kisner}, {Knoche}, {Knox}, {Kunz},
  {Kurki-Suonio}, {Lagache}, {L{\"a}hteenm{\"a}ki}, {Lamarre}, {Lasenby},
  {Laureijs}, {Lawrence}, {Leonardi}, {Le{\'o}n-Tavares}, {Lesgourgues},
  {Levrier}, {Liguori}, {Lilje}, {Linden-V{\o}rnle}, {L{\'o}pez-Caniego},
  {Lubin}, {Mac{\'\i}as-P{\'e}rez}, {Maffei}, {Maino}, {Mandolesi}, {Maris},
  {Marshall}, {Martin}, {Mart{\'\i}nez-Gonz{\'a}lez}, {Masi}, {Massardi},
  {Matarrese}, {Matthai}, {Mazzotta}, {McGehee}, {Melchiorri}, {Mendes},
  {Mennella}, {Migliaccio}, {Mitra}, {Miville-Desch{\^e}nes}, {Moneti},
  {Montier}, {Morgante}, {Mortlock}, {Munshi}, {Murphy}, {Naselsky}, {Nati},
  {Natoli}, {Netterfield}, {N{\o}rgaard-Nielsen}, {Noviello}, {Novikov},
  {Novikov}, {Osborne}, {Oxborrow}, {Paci}, {Pagano}, {Pajot}, {Paladini},
  {Paoletti}, {Pasian}, {Patanchon}, {Perdereau}, {Perotto}, {Perrotta},
  {Piacentini}, {Piat}, {Pierpaoli}, {Pietrobon}, {Plaszczynski},
  {Pointecouteau}, {Polenta}, {Ponthieu}, {Popa}, {Poutanen}, {Pratt},
  {Pr{\'e}zeau}, {Prunet}, {Puget}, {Rachen}, {Reach}, {Rebolo}, {Reinecke},
  {Remazeilles}, {Renault}, {Ricciardi}, {Riller}, {Ristorcelli}, {Rocha},
  {Rosset}, {Roudier}, {Rowan-Robinson}, {Rubi{\~n}o-Mart{\'\i}n}, {Rusholme},
  {Sandri}, {Santos}, {Savini}, {Scott}, {Seiffert}, {Shellard}, {Spencer},
  {Starck}, {Stolyarov}, {Stompor}, {Sudiwala}, {Sunyaev}, {Sureau}, {Sutton},
  {Suur-Uski}, {Sygnet}, {Tauber}, {Tavagnacco}, {Terenzi}, {Toffolatti},
  {Tomasi}, {Tristram}, {Tucci}, {Tuovinen}, {T{\"u}rler}, {Umana},
  {Valenziano}, {Valiviita}, {Van Tent}, {Verstraete}, {Vielva}, {Villa},
  {Vittorio}, {Wade}, {Wandelt}, {Welikala}, {Ysard}, {Yvon}, {Zacchei}, \&
  {Zonca}}]{Planck2014}
{Planck Collaboration}, {Abergel}, A., {Ade}, P.~A.~R., {et~al.} 2014, \aap,
  571, A11

\bibitem[{{Poglitsch} {et~al.}(2010){Poglitsch}, {Waelkens}, {Geis},
  {Feuchtgruber}, {Vandenbussche}, {Rodriguez}, {Krause}, {Renotte}, {van
  Hoof}, {Saraceno}, {Cepa}, {Kerschbaum}, {Agn{\`e}se}, {Ali}, {Altieri},
  {Andreani}, {Augueres}, {Balog}, {Barl}, {Bauer}, {Belbachir}, {Benedettini},
  {Billot}, {Boulade}, {Bischof}, {Blommaert}, {Callut}, {Cara}, {Cerulli},
  {Cesarsky}, {Contursi}, {Creten}, {De Meester}, {Doublier}, {Doumayrou},
  {Duband}, {Exter}, {Genzel}, {Gillis}, {Gr{\"o}zinger}, {Henning},
  {Herreros}, {Huygen}, {Inguscio}, {Jakob}, {Jamar}, {Jean}, {de Jong},
  {Katterloher}, {Kiss}, {Klaas}, {Lemke}, {Lutz}, {Madden}, {Marquet},
  {Martignac}, {Mazy}, {Merken}, {Montfort}, {Morbidelli}, {M{\"u}ller},
  {Nielbock}, {Okumura}, {Orfei}, {Ottensamer}, {Pezzuto}, {Popesso},
  {Putzeys}, {Regibo}, {Reveret}, {Royer}, {Sauvage}, {Schreiber}, {Stegmaier},
  {Schmitt}, {Schubert}, {Sturm}, {Thiel}, {Tofani}, {Vavrek}, {Wetzstein},
  {Wieprecht}, \& {Wiezorrek}}]{Poglitsch2010}
{Poglitsch}, A., {Waelkens}, C., {Geis}, N., {et~al.} 2010, \aap, 518, L2

\bibitem[{{Pouteau} {et~al.}(2022){Pouteau}, {Motte}, {Nony},
  {Galv{\'a}n-Madrid}, {Men'shchikov}, {Bontemps}, {Robitaille}, {Louvet},
  {Ginsburg}, {Herpin}, {L{\'o}pez-Sepulcre}, {Dell'Ova}, {Gusdorf},
  {Sanhueza}, {Stutz}, {Brouillet}, {Thomasson}, {Armante}, {Baug}, {Bonfand},
  {Busquet}, {Csengeri}, {Cunningham}, {Fern{\'a}ndez-L{\'o}pez}, {Liu},
  {Olguin}, {Towner}, {Bally}, {Braine}, {Bronfman}, {Joncour}, {Gonz{\'a}lez},
  {Hennebelle}, {Lu}, {Menten}, {Moraux}, {Tatematsu}, {Walker}, \&
  {Whitworth}}]{Pouteau2022}
{Pouteau}, Y., {Motte}, F., {Nony}, T., {et~al.} 2022, \aap, 664, A26

\bibitem[{{Salpeter}(1955)}]{Salpeter1955}
{Salpeter}, E.~E. 1955, \apj, 121, 161

\bibitem[{{Sanders} {et~al.}(1985){Sanders}, {Scoville}, \&
  {Solomon}}]{Sanders1985}
{Sanders}, D.~B., {Scoville}, N.~Z., \& {Solomon}, P.~M. 1985, \apj, 289, 373

\bibitem[{{Schisano} {et~al.}(2020){Schisano}, {Molinari}, {Elia},
  {Benedettini}, {Olmi}, {Pezzuto}, {Traficante}, {Brescia}, {Cavuoti}, {di
  Giorgio}, {Liu}, {Moore}, {Noriega-Crespo}, {Riccio}, {Baldeschi},
  {Becciani}, {Peretto}, {Merello}, {Vitello}, {Zavagno}, {Beltr{\'a}n},
  {Cambr{\'e}sy}, {Eden}, {Li Causi}, {Molinaro}, {Palmeirim}, {Sciacca},
  {Testi}, {Umana}, \& {Whitworth}}]{Schisano2020}
{Schisano}, E., {Molinari}, S., {Elia}, D., {et~al.} 2020, \mnras, 492, 5420

\bibitem[{{Sousbie}(2011)}]{Sousbie2011}
{Sousbie}, T. 2011, \mnras, 414, 350

\bibitem[{{Stod{\'o}lkiewicz}(1963)}]{Stodolkiewicz1963}
{Stod{\'o}lkiewicz}, J.~S. 1963, \actaa, 13, 30

\bibitem[{{Stutzki} \& {Guesten}(1990)}]{Stutzki1990}
{Stutzki}, J. \& {Guesten}, R. 1990, \apj, 356, 513

\bibitem[{{Tan} {et~al.}(2014){Tan}, {Beltr{\'a}n}, {Caselli}, {Fontani},
  {Fuente}, {Krumholz}, {McKee}, \& {Stolte}}]{Tan2014a}
{Tan}, J.~C., {Beltr{\'a}n}, M.~T., {Caselli}, P., {et~al.} 2014, in Protostars
  and Planets VI, ed. H.~{Beuther}, R.~S. {Klessen}, C.~P. {Dullemond}, \&
  T.~{Henning}, 149--172

\bibitem[{{Tassis} {et~al.}(2010){Tassis}, {Christie}, {Urban}, {Pineda},
  {Mouschovias}, {Yorke}, \& {Martel}}]{Tassis2010}
{Tassis}, K., {Christie}, D.~A., {Urban}, A., {et~al.} 2010, \mnras, 408, 1089

\bibitem[{{Vazquez-Semadeni}(1994)}]{Vazquez1994}
{Vazquez-Semadeni}, E. 1994, \apj, 423, 681

\bibitem[{{Ward-Thompson} {et~al.}(2007{\natexlab{a}}){Ward-Thompson},
  {Andr{\'e}}, {Crutcher}, {Johnstone}, {Onishi}, \& {Wilson}}]{{Ward2007}
{Ward-Thompson}, D., {Andr{\'e}}, P., {Crutcher}, R., {et~al.}
  2007{\natexlab{a}}, in Protostars and Planets V, ed. B.~{Reipurth},
  D.~{Jewitt}, \& K.~{Keil}, 33

\bibitem[{{Ward-Thompson} {et~al.}(2007{\natexlab{b}}){Ward-Thompson},
  {Andr{\'e}}, {Crutcher}, {Johnstone}, {Onishi}, \& {Wilson}}]{Ward2007}
{Ward-Thompson}, D., {Andr{\'e}}, P., {Crutcher}, R., {et~al.}
  2007{\natexlab{b}}, in Protostars and Planets V, ed. B.~{Reipurth},
  D.~{Jewitt}, \& K.~{Keil}, 33

\bibitem[{{Wenger} {et~al.}(2000){Wenger}, {Ochsenbein}, {Egret}, {Dubois},
  {Bonnarel}, {Borde}, {Genova}, {Jasniewicz}, {Lalo{\"e}}, {Lesteven}, \&
  {Monier}}]{Wenger2000}
{Wenger}, M., {Ochsenbein}, F., {Egret}, D., {et~al.} 2000, \aaps, 143, 9

\bibitem[{{Williams} {et~al.}(1994){Williams}, {de Geus}, \&
  {Blitz}}]{Williams1994}
{Williams}, J.~P., {de Geus}, E.~J., \& {Blitz}, L. 1994, \apj, 428, 693

\bibitem[{{Yamaguchi} {et~al.}(1999){Yamaguchi}, {Mizuno}, {Saito},
  {Matsunaga}, {Mizuno}, {Ogawa}, \& {Fukui}}]{Yamaguchi1999}
{Yamaguchi}, N., {Mizuno}, N., {Saito}, H., {et~al.} 1999, \pasj, 51, 775

\bibitem[{{Zhang} {et~al.}(2022){Zhang}, {Zhang}, {Li}, \&
  {Li}}]{2022RAA....22e5012Z}
{Zhang}, C., {Zhang}, G.-Y., {Li}, J.-Z., \& {Li}, X.-M. 2022, Research in
  Astronomy and Astrophysics, 22, 055012

\bibitem[{{Zhang} {et~al.}(2020){Zhang}, {Andr{\'e}}, {Men'shchikov}, \&
  {Wang}}]{Zhanggy2020}
{Zhang}, G.-Y., {Andr{\'e}}, P., {Men'shchikov}, A., \& {Wang}, K. 2020, \aap,
  642, A76

\bibitem[{{Zhang} {et~al.}(2018){Zhang}, {Xu}, {Vasyunin}, {Semenov}, {Wang},
  {Dib}, {Liu}, {Liu}, {Zhang}, {Liu}, {Wang}, {Li}, {Wu}, {Yuan}, {Li}, \&
  {Gao}}]{Zhang2018}
{Zhang}, G.-Y., {Xu}, J.-L., {Vasyunin}, A.~I., {et~al.} 2018, \aap, 620, A163

\bibitem[{{Zucker} {et~al.}(2020){Zucker}, {Speagle}, {Schlafly}, {Green},
  {Finkbeiner}, {Goodman}, \& {Alves}}]{Zucker2020}
{Zucker}, C., {Speagle}, J.~S., {Schlafly}, E.~F., {et~al.} 2020, \aap, 633,
  A51

\end{thebibliography}

\begin{appendix}

\section{Catalogs of dense cores and filaments identified with \emph{Herschel} in Vela C}

Our extraction in the \emph{Herschel} SPIRE and PACS images of the Vela C molecular cloud identified a total of
570 dense cores, including 421 starless cores and 149 protostellar cores. A template of our online catalog, listing the observed properties of the extracted \emph{Herschel} cores, is given in Table~\ref{coreobs}. A template of our online catalog of the derived properties for each core (physical radius, mass, SED dust temperature, etc.) is given in Table~\ref{corederive}. A template of our online catalog of the physical parameters of 68 filaments (lengths, masses, linear densities, widths, etc.) is given in Table~\ref{filamentpara}. 

\longtab[1]{
\begin{landscape}
\tiny
\setlength{\tabcolsep}{1.0mm}{
\begin{longtable}{c|ccccccc}
\caption{\label{coreobs} Catalog of 570 reliable cores identified in the multiwavelength \emph{Herschel} maps of the Vela C molecular cloud (template, full catalog only provided online).}\\
  \hline\noalign{\smallskip}
$n$ & RA & Dec & $f$ & $\Gamma$ & $\Xi$ & Core type & SIMBAD \\
 & (J2000, $\degr$) & (J2000, $\degr$) &  &  &  &  &  \\
(1) & (2) & (3) & (4) & (5) & (6) & (7) & (8) \\
  \hline\noalign{\smallskip}
1 & 134.8590106 & -43.4369271 & 1 & 7.831E+03 & 1.491E+06 & protostellar & [MHL2007] G264.9032+01.6584 1 \\
2 & 135.2813467 & -45.0079501 & 0 & 5.121E+03 & 5.271E+05 & protostellar & [MHL2007] G266.2856+00.8533 1	MSX6C G266.2856+00.8533 \\
3 & 135.1585678 & -43.9928924 & 0 & 4.561E+03 & 5.021E+05 & protostellar & MSX6C G265.4642+01.4561  IRAS 08588-4347 \\
  $\cdots$ & $\cdots$ \\
  \noalign{\smallskip}\hline
\end{longtable}
\begin{longtable}{ccccccccccccccccc}
  \hline\noalign{\smallskip}
$f_{70}$ & $\Gamma_{70}$ & $\Xi_{70}$ & $F_{\textrm{P},70}$ & $\sigma_{\textrm{P},70}$ & $F_{\textrm{T},70}$ & $\sigma_{\textrm{T},70}$ & $F_{\textrm{G},70}$ & $S_{j_{\textrm{F}},70}$& $A_{70}$ & $B_{70}$ & $A_{\textrm{M},70}$ & $B_{\textrm{M},70}$ & $\omega_{\textrm{M},70}$ & $\phi_{70}$ & $A_{\textrm{F},70}$ & $B_{\textrm{F},70}$ \\
(9) & (10) & (11) & (12) & (13) & (14) & (15) & (16) & (17) & (18) & (19) & (20) & (21) & (22) & (23) & (24) & (25) \\   
  \hline\noalign{\smallskip}
  \noalign{\smallskip}\hline
  
$f_{160}$ & $\Gamma_{160}$ & $\Xi_{160}$ & $F_{\textrm{P},160}$ & $\sigma_{\textrm{P},160}$ & $F_{\textrm{T},160}$ & $\sigma_{\textrm{T},160}$ & $F_{\textrm{G},160}$ & $S_{j_{\textrm{F}},160}$ & $A_{160}$ & $B_{160}$ & $A_{\textrm{M},160}$ & $B_{\textrm{M},160}$ & $\omega_{\textrm{M},160}$ & $\phi_{160}$ & $A_{\textrm{F},160}$ & $B_{\textrm{F},160}$ \\
 &  &  & (Jy/beam) & (Jy/beam) & (Jy) & (Jy) & (Jy) & (\arcsec) & (\arcsec) & (\arcsec) & (\arcsec) & (\arcsec) & (\arcsec) & ($\degr$) & (\arcsec) & (\arcsec) \\
(26) & (27) & (28) & (29) & (30) & (31) & (32) & (33) & (34) & (35) & (36) & (37) & (38) & (39) & (40) & (41) & (42) \\
  \hline\noalign{\smallskip}
 0 & 9.654E+02 & 2.549E+04 & 2.462E+01 & 8.177E-02 & 3.648E+01 & 1.323E-01 & 2.787E+01 & 1.170E+01 & 1.300E+01 & 1.191E+01 & 1.798E+01 & 1.670E+01 & 1.186E+02 & -4.644E+00 & 5.931E+01 & 5.433E+01 \\ 
 0 & 1.297E+03 & 4.218E+04 & 3.095E+01 & 6.114E-02 & 3.737E+01 & 1.347E-01 & 3.128E+01 & 1.170E+01 & 1.263E+01 & 1.096E+01 & 1.829E+01 & 1.678E+01 & 8.827E+01 & -6.158E+00 & 8.303E+01 & 7.205E+01 \\      
 0 & 1.323E+03 & 2.972E+04 & 3.596E+01 & 1.203E-01 & 4.212E+01 & 1.491E-01 & 3.463E+01 & 1.170E+01 & 1.302E+01 & 1.052E+01 & 1.584E+01 & 1.273E+01 & 1.154E+02 & -3.342E+00 & 4.839E+01 & 3.910E+01 \\  
   $\cdots$ \\
  \hline\noalign{\smallskip}
  \noalign{\smallskip}\hline

$f_{250}$ & $\Gamma_{250}$ & $\Xi_{250}$ & $F_{\textrm{P},250}$ & $\sigma_{\textrm{P},250}$ & $F_{\textrm{T},250}$ & $\sigma_{\textrm{T},250}$ & $F_{\textrm{G},250}$ & $S_{j_{\textrm{F}},250}$ & $A_{250}$ & $B_{250}$ & $A_{\textrm{M},250}$ & $B_{\textrm{M},250}$ & $\omega_{\textrm{M},250}$ & $\phi_{250}$ & $A_{\textrm{F},250}$ & $B_{\textrm{F},250}$ \\
(43) & (44) & (45) & (46) & (47) & (48) & (49) & (50) & (51) & (52) & (53) & (54) & (55) & (56) & (57) & (58) & (59) \\ 
  \hline\noalign{\smallskip}
  \noalign{\smallskip}\hline

$f_{350}$ & $\Gamma_{350}$ & $\Xi_{350}$ & $F_{\textrm{P},350}$ & $\sigma_{\textrm{P},350}$ & $F_{\textrm{T},350}$ & $\sigma_{\textrm{T},350}$ & $F_{\textrm{G},350}$ & $S_{j_{\textrm{F}},350}$ & $A_{350}$ & $B_{350}$ & $A_{\textrm{M},350}$ & $B_{\textrm{M},350}$ & $\omega_{\textrm{M},350}$ & $\phi_{350}$ & $A_{\textrm{F},350}$ & $B_{\textrm{F},350}$ \\
(60) & (61) & (62) & (63) & (64) & (65) & (66) & (67) & (68) & (69) & (70) & (71) & (72) & (73) & (74) & (75) & (76) \\ 
  \hline\noalign{\smallskip}
  \noalign{\smallskip}\hline

$f_{500}$ & $\Gamma_{500}$ & $\Xi_{500}$ & $F_{\textrm{P},500}$ & $\sigma_{\textrm{P},500}$ & $F_{\textrm{T},500}$ & $\sigma_{\textrm{T},500}$ & $F_{\textrm{G},500}$ & $S_{j_{\textrm{F}},500}$ & $A_{500}$ & $B_{500}$ & $A_{\textrm{M},500}$ & $B_{\textrm{M},500}$ & $\omega_{\textrm{M},500}$ & $\phi_{500}$ & $A_{\textrm{F},500}$ & $B_{\textrm{F},500}$ \\
(77) & (78) & (79) & (80) & (81) & (82) & (83) & (84) & (85) & (86) & (87) & (88) & (89) & (90) & (91) & (92) & (93) \\
  \hline\noalign{\smallskip}
  \noalign{\smallskip}\hline

$f_{8.5{\arcsec}}$ & $\Gamma_{8.5{\arcsec}}$ & $\Xi_{8.5{\arcsec}}$ & $F_{\textrm{P},8.5{\arcsec}}$ & $\sigma_{\textrm{P},8.5{\arcsec}}$ & $F_{\textrm{T},8.5{\arcsec}}$ & $\sigma_{\textrm{T},8.5{\arcsec}}$ & $F_{\textrm{G},8.5{\arcsec}}$ & $S_{j_{\textrm{F}},8.5{\arcsec}}$ & $A_{8.5{\arcsec}}$ & $B_{8.5{\arcsec}}$ & $A_{\textrm{M},8.5{\arcsec}}$ & $B_{\textrm{M},8.5{\arcsec}}$ & $\omega_{\textrm{M},8.5{\arcsec}}$ & $\phi_{8.5{\arcsec}}$ & $A_{\textrm{F},8.5{\arcsec}}$ & $B_{\textrm{F},8.5{\arcsec}}$ \\
(94) & (95) & (96) & (97) & (98) & (99) & (100) & (101) & (102) & (103) & (104) & (105) & (106) & (107) & (108) & (109) & (110) \\
  \hline\noalign{\smallskip}
  \noalign{\smallskip}\hline

$f_{11.7{\arcsec}}$ & $\Gamma_{11.7{\arcsec}}$ & $\Xi_{11.7{\arcsec}}$ & $F_{\textrm{P},11.7{\arcsec}}$ & $\sigma_{\textrm{P},11.7{\arcsec}}$ & $F_{\textrm{T},11.7{\arcsec}}$ & $\sigma_{\textrm{T},11.7{\arcsec}}$ & $F_{\textrm{G},11.7{\arcsec}}$ & $S_{j_{\textrm{F}},11.7{\arcsec}}$ & $A_{11.7{\arcsec}}$ & $B_{11.7{\arcsec}}$ & $A_{\textrm{M},11.7{\arcsec}}$ & $B_{\textrm{M},11.7{\arcsec}}$ & $\omega_{\textrm{M},11.7{\arcsec}}$ & $\phi_{11.7{\arcsec}}$ & $A_{\textrm{F},11.7{\arcsec}}$ & $B_{\textrm{F},11.7{\arcsec}}$ \\
 &  &  & $(\textrm{cm}^{-2})$ & $(\textrm{cm}^{-2})$ & $(\textrm{cm}^{-2})$ & $(\textrm{cm}^{-2})$ & $(M_{\odot})$ & (\arcsec) & (\arcsec) & (\arcsec) & (\arcsec) & (\arcsec) & ($\degr$) & (\arcsec) & (\arcsec) & (\arcsec) \\
(111) & (112) & (113) & (114) & (115) & (116) & (117) & (118) & (119) & (120) & (121) & (122) & (123) & (124) & (125) & (126) & (127) \\
  \hline\noalign{\smallskip}
 0 & 8.125E+02 & 2.715E+03 & 1.160E+25 & 3.324E+23 & 1.314E+25 & 2.745E+23 & 2.722E+00 & 3.256E-01 & 1.236E+01 & 1.108E+01 & 1.215E+01 & 1.155E+01 & 4.561E+01 & -2.453E+00 & 2.926E+01 & 2.870E+01 \\ 
 0 & 1.444E+03 & 1.028E+04 & 2.898E+25 & 3.574E+23 & 3.871E+25 & 3.844E+23 & 6.537E+00 & 1.356E-01 & 1.228E+01 & 1.115E+01 & 1.470E+01 & 1.354E+01 & 1.597E+02 & -3.166E+00 & 3.847E+01 & 3.704E+01 \\      
 0 & 1.004E+03 & 5.867E+03 & 2.054E+25 & 3.203E+23 & 2.531E+25 & 2.705E+23 & 4.340E+00 & 2.042E-01 & 1.220E+01 & 1.122E+01 & 1.216E+01 & 1.147E+01 & 1.353E+02 & -2.502E+00 & 3.003E+01 & 2.927E+01 \\ 
 $\cdots$ \\
  \hline\noalign{\smallskip}
  \noalign{\smallskip}\hline

$f_{18.2{\arcsec}}$ & $\Gamma_{18.2{\arcsec}}$ & $\Xi_{18.2{\arcsec}}$ & $F_{\textrm{P},18.2{\arcsec}}$ & $\sigma_{\textrm{P},18.2{\arcsec}}$ & $F_{\textrm{T},18.2{\arcsec}}$ & $\sigma_{\textrm{T},18.2{\arcsec}}$ & $F_{\textrm{G},18.2{\arcsec}}$ & $S_{j_{\textrm{F}},18.2{\arcsec}}$ & $A_{18.2{\arcsec}}$ & $B_{18.2{\arcsec}}$ & $A_{\textrm{M},18.2{\arcsec}}$ & $B_{\textrm{M},18.2{\arcsec}}$ & $\omega_{\textrm{M},18.2{\arcsec}}$ & $\phi_{18.2{\arcsec}}$ & $A_{\textrm{F},18.2{\arcsec}}$ & $B_{\textrm{F},18.2{\arcsec}}$ \\
(128) & (129) & (130) & (131) & (132) & (133) & (134) & (135) & (136) & (137) & (138) & (139) & (140) & (141) & (142) & (143) & (144) \\     
  \hline\noalign{\smallskip}
\end{longtable}
}
\tablefoot{Catalog entries are as follows: 
(1) Source running number; 
(2) and (3): Centroid equatorial coordinates;
(4) Flag describing global properties over all wavelengths (global flag), 0: source is not blended with any other source in any waveband; 1: source's footprints intersect by more than 20\% in at least one waveband; 2: source's footprint area contains at least one other source; 3: source is causing a larger source to be sub-structured; 
(5) The detection significance over all wavelengths; 
(6) The monochromatic goodness (combining detection significance and signal-to-noise ratio); 
(7) Core type: protostellar, prestellar, unbound starless, and unresolved starless;
(8) SIMBAD infrared source counterparts within a radius of 6{\arcsec} from the centroid position of the \emph{Herschel} source; 
(9), (26), (43), (60), (77), (94), (111), (128): Wavelength-dependent flag; 
(10), (27), (44), (61), (78), (95), (112), (129): The detection significance from monochromatic single scales; 
(11), (28), (45), (62), (79), (96), (113), (130): The monochromatic goodness (combining detection significance and signal-to-noise ratio); 
(12), (29), (46), (63), (80), (97), (114), (131): The peak intensity; 
(13), (30), (47), (64), (81), (98), (115), (132): The error of $F_{\textrm{P},\lambda}$; 
(14), (31), (48), (65), (82), (99), (116), (133): The total flux; 
(15), (32), (49), (66), (83), (100), (117), (134): The error of $F_{\textrm{T},\lambda}$; 
(16), (33), (50), (67), (84): The Gaussian flux; 
(101), (118), (135): The mass derived from the surface density image; 
(17), (34), (51), (68), (85), (102), (119), (136): The characteristic size of sources; 
(18)-(19), (35)-(36), (53)-(54), (69)-(70), (86)-(87), (103)-(104), (120)-(121), (137)-(138): The major and minor sizes at half-maximum of sources; 
(20)-(21), (37)-(38), (55)-(56), (71)-(72), (88)-(89), (105)-(106), (122)-(123), (139)-(140): The major and minor size from intensity moments of sources;  
(22), (39), (57), (73), (90), (107), (124), (141): The position angle; 
(23), (40), (58), (74), (91), (108), (125), (142): The footprint factor;
(24)-(25), (41)-(42), (59)-(60), (75)-(76), (92)-(93), (109)-(110), (126)-(127), (143)-(144): The major and minor sizes at footprint axes of sources. 
}
\end{landscape}
}
\longtab[2]{
\begin{landscape}
\setlength{\tabcolsep}{1.0mm}{
\begin{longtable}{c|ccccccccccccccccccc}
\caption{\label{corederive} Derived properties of 570 reliable cores identified identified in the multiwavelength \emph{Herschel} maps of the Vela C molecular cloud (template, full table only provided online).}\\
  \hline\noalign{\smallskip}
  \noalign{\smallskip}\hline
  $n$ & RA & Dec & $T$ & $\sigma_{T}$ & $M$ & $\sigma_{M}$ & $L$ & $\sigma_{L}$ & $R_{8.5 \arcsec}$ & $R_{11.7 \arcsec}$ & $R_{18.2 \arcsec}$ & Core type \\
   & (J2000, $\degr$) & (J2000, $\degr$) & (K) & (K) & $(M_\odot)$ & $(M_\odot)$ & $(L_\odot)$ & $(L_\odot)$ & (\arcsec) & (\arcsec) & (\arcsec) &  \\
  (1) & (2) & (3) & (4) & (5) & (6) & (7) & (8) & (9) & (10) & (11) & (12) & (13) \\
    \hline\noalign{\smallskip}
  1 & 134.8590106 & -43.4369271 & 1.934E+01 & 1.126E+00 & 4.141E+00 & 1.852E+00 & 2.087E+01 & 5.903E+00 & 8.500E+00 & 1.170E+01 & 2.117E+01 & protostellar \\
  2 & 135.2813467 & -45.0079501 & 1.668E+01 & 7.853E-01 & 8.763E+00 & 3.839E+00 & 1.817E+01 & 5.138E+00 & 9.202E+00 & 1.170E+01 & 2.003E+01 & protostellar \\
  3 & 135.1585678 & -43.9928924 & 1.977E+01 & 1.208E+00 & 4.317E+00 & 1.944E+00 & 2.477E+01 & 7.007E+00 & 8.500E+00 & 1.170E+01 & 1.990E+01 & protostellar \\
  $\cdots$ & $\cdots$ \\
  \hline\noalign{\smallskip}
\end{longtable}
}
\tablefoot{Catalog entries are as follows: 
(1) Source running number; 
(2) and (3): Centroid equatorial coordinates; 
(4) Derived dust temperature from SED fitting; 
(5) Uncertainty of derived temperature; 
(6) Derived total mass (gas and dust) from SED fitting; 
(7) Uncertainty of derived total mass; 
(8) Derived bolometric luminosity from SED fitting; 
(9) Uncertainty of derived luminosity; 
(10), (11), (12): Geometrical average between the major and minor FWHM size measured at 8.5 ,11.7 and 18.2{\arcsec} resolution; 
(13) Core type: protostellar, prestellar, unbound starless, and unresolved starless.
}

\tiny
\setlength{\tabcolsep}{1.0mm}{
\begin{longtable}{c|ccccccccccccccccc}
\caption{\label{filamentpara} Catalog of the 68 selected filaments identified in the multiwavelength \emph{Herschel} maps of the Vela C molecular cloud (template, full catalog only provided online).}\\
  \hline\noalign{\smallskip}
  \noalign{\smallskip}\hline
  $n$ & RA & Dec & $W$ & $L$ & $W$ & $W_{\alpha}$ & $W_{\beta}$ & $\overline{W}$ & $\overline{W}_{\alpha}$ & $\overline{W}_{\beta}$ & $\varsigma _{\overline{W}}$  & $\varsigma _{\overline{W}_{\alpha}}$ & $\varsigma _{\overline{W}_{\beta}}$ & $N_{\rm W}$ & $\Omega_{\overline{W}}$ & $\Omega_{\overline{W}_{\alpha}}$ & $\Omega_{\overline{W}_{\beta}}$ \\
   & (J2000, $\degr$) & (J2000, $\degr$) & (\arcsec) & (pc) & (pc) & (pc) & (pc) & (pc) & (pc) & (pc) & (pc) & (pc) & (pc) &  &  \\
  (1) & (2) & (3) & (4) & (5) & (6) & (7) & (8) & (9) & (10) & (11) & (12) & (13) & (14) & (15) & (16) & (17) & (18) \\
    \hline\noalign{\smallskip}
  1 & 135.1401131 & -45.1953743 & 6.775E+01 & 5.432E-01 & 2.973E-01 & 2.487E-01 & 3.552E-01 & 7.143E-01 & 2.614E-01 & 1.952E+00 & 4.964E-01 & 4.875E-02 & 3.450E+00 &  56 & 1.439E+00 & 5.361E+00 & 5.658E-01 \\
  4 & 135.1348641 & -45.1181584 & 7.461E+01 & 2.014E+00 & 3.274E-01 & 3.245E-01 & 3.302E-01 & 4.593E-01 & 5.432E-01 & 3.883E-01 & 1.933E-01 & 6.344E-01 & 2.733E-01 & 195 & 2.376E+00 & 8.561E-01 & 1.421E+00 \\         
  6 & 135.1244612 & -45.0512890 & 6.876E+01 & 1.685E+00 & 3.017E-01 & 2.848E-01 & 3.195E-01 & 6.570E-01 & 8.204E-01 & 5.261E-01 & 4.348E-01 & 1.126E+00 & 5.126E-01 & 165 & 1.511E+00 & 7.284E-01 & 1.026E+00 \\        
  17 & 135.0980823 & -44.9982407 & 1.124E+02 & 5.151E-01 & 4.932E-01 & 5.596E-01 & 4.347E-01 & 5.778E-01 & 7.804E-01 & 4.278E-01 & 1.828E-01 & 5.036E-01 & 8.486E-02 &  54 & 3.161E+00 & 1.550E+00 & 5.042E+00 \\       
  $\cdots$ & $\cdots$ \\
  \hline\noalign{\smallskip}
 \end{longtable}
 
 \begin{longtable}{cccccccccccccccccc}  
  \hline\noalign{\smallskip}
  $N_{\textrm{H}_{2}}$ & $\varsigma _{N_{\textrm{H}_{2}}}$ & $\Lambda^{\textrm{P}}$ & $\Lambda^{\textrm{P}}_{\alpha}$ & $\Lambda^{\textrm{P}}_{\beta}$ & $\Lambda^{\textrm{M}}$ & $\Lambda^{\textrm{M}}_{\alpha}$ & $\Lambda^{\textrm{M}}_{\beta}$ & $M$ & $M_{\alpha}$ & $M_{\beta}$ & $\varsigma _{\Lambda^{\textrm{P}}}$ & $\Omega_{\Lambda^{\textrm{P}}}$ & $T$ & $\varsigma _{T}$ & $C$ & $\varsigma _{C}$ \\
   $(\textrm{cm}^{-2})$ & $(\textrm{cm}^{-2})$ & $(M_{\odot}/\textrm{pc})$ & $(M_{\odot}/\textrm{pc})$ & $(M_{\odot}/\textrm{pc})$ & $(M_{\odot}/\textrm{pc})$ & $(M_{\odot}/\textrm{pc})$ & $(M_{\odot}/\textrm{pc})$ & $(M_{\odot})$ & $(M_{\odot})$ & $(M_{\odot})$ & $(M_{\odot}/\textrm{pc})$ &  & (K) & (K) &  & \\
  (19) & (20) & (21) & (22) & (23) & (24) & (25) & (26) & (27) & (28) & (29) & (30) & (31) & (32) & (33) & (34) & (35) \\
  \hline\noalign{\smallskip}
  3.815E+21 & 4.963E+20 & 3.226E+01 & 2.607E+01 & 3.598E+01 & 5.206E+01 & 2.578E+01 & 7.835E+01 & 2.828E+01 & 1.401E+01 & 4.256E+01 & 1.050E+01 & 3.073E+00 & 1.549E+01 & 7.833E-02 & 7.059E-01 & 2.329E-01 \\
  1.379E+22 & 3.999E+21 & 1.204E+02 & 1.127E+02 & 1.217E+02 & 1.677E+02 & 2.069E+02 & 1.285E+02 & 3.377E+02 & 4.167E+02 & 2.588E+02 & 5.638E+01 & 2.136E+00 & 1.391E+01 & 5.460E-01 & 2.291E+00 & 7.993E-01 \\
  1.652E+22 & 5.429E+21 & 1.976E+02 & 1.958E+02 & 1.861E+02 & 3.076E+02 & 3.227E+02 & 2.926E+02 & 5.185E+02 & 5.438E+02 & 4.931E+02 & 4.743E+01 & 4.167E+00 & 1.366E+01 & 4.703E-01 & 2.520E+00 & 1.096E+00 \\
  1.304E+22 & 3.913E+21 & 1.927E+02 & 2.048E+02 & 1.845E+02 & 2.486E+02 & 2.150E+02 & 2.821E+02 & 1.280E+02 & 1.108E+02 & 1.453E+02 & 5.081E+01 & 3.793E+00 & 1.387E+01 & 3.380E-01 & 1.698E+00 & 7.485E-01 \\
  $\cdots$ \\
  \hline\noalign{\smallskip}
\end{longtable}
}
\tablefoot{Filaments are measured on 11.7\arcsec-resolution surface density map. Catalog entries are as follows: 
(1) Filament number; 
(2) and (3): Equatorial coordinates of the middle position; 
(4) and (6): The median of filament's FWHM measured from both side in radial direction (different units); 
(5): Filament (skeleton) length; 
(7) and (8): The median of filament's FWHM measured from the left and right side in radial direction; 
(9): The mean value of filament's FWHM measured from both side in radial direction; 
(10) and (11): The mean value of filament's FWHM measured from the left and right side in radial direction; 
(12), (13), (14): Standard deviation about $\overline{W}$, $\overline{W}_{\alpha}$ and $\overline{W}_{\beta}$; 
(15) Number of points used for $\overline{W}$ and $\varsigma _{\overline{W}}$; 
(16) Signal-to-noise ratio of $\overline{W}$/$\varsigma _{\overline{W}}$; 
(17) and (18): Signal-to-noise ratio of $\overline{W}_{\alpha}$/$\varsigma _{\overline{W}_{\alpha}}$ and $\overline{W}_{\beta}$/$\varsigma _{\overline{W}_{\beta}}$; 
(19) Filament crest value of surface density (on skeleton); 
(20) Standard deviation about $N_{H_{2}}$; 
(21) The linear density of filament's measured from median integrated profile in radial direction; 
(22) and (23): The linear density of filament's measured from the left and right median integrated profile in radial direction; 
(24) The linear density of filament's measured from $M/L$; 
(25) and (26): The linear density of filament's measured from $M_{\alpha}/L$ and $M_{\beta}/L$; 
(27) Filament mass derived from both sides; 
(28) and (29): Filament mass derived from the left and right side; 
(30) Standard deviation about $\Lambda^{P}$; 
(31) Signal-to-noise ratio of $\Lambda^{P}$/$\varsigma _{\Lambda^{P}}$; 
(32) Filament crest value of temperature (on skeleton);
(33) Standard deviation about $T$; 
(34) Filament crest value of contrast (on skeleton).
(35) Standard deviation about $C$.
}
\end{landscape}
}
\end{appendix}
\end{document}